\documentclass[fleqn,usenatbib]{mnras}

\usepackage{newtxtext,newtxmath}
\usepackage[T1]{fontenc}
\usepackage{ae,aecompl}
\usepackage{placeins}

\usepackage{graphicx}	 
\usepackage{amsmath}	
\usepackage{color}
\usepackage{gensymb}
\usepackage{longtable}

\usepackage{scalerel,tikz}
\usetikzlibrary{svg.path}
\definecolor{orcidlogocol}{HTML}{A6CE39}
\tikzset{orcidlogo/.pic={
 \fill[orcidlogocol] svg{M256,128c0,70.7-57.3,128-128,128C57.3,256,0,198.7,0,128C0,57.3,57.3,0,128,0C198.7,0,256,57.3,256,128z};
 \fill[white] svg{M86.3,186.2H70.9V79.1h15.4v48.4V186.2z}
 svg{M108.9,79.1h41.6c39.6,0,57,28.3,57,53.6c0,27.5-21.5,53.6-56.8,53.6h-41.8V79.1z M124.3,172.4h24.5c34.9,0,42.9-26.5,42.9-39.7c0-21.5-13.7-39.7-43.7-39.7h-23.7V172.4z}
 svg{M88.7,56.8c0,5.5-4.5,10.1-10.1,10.1c-5.6,0-10.1-4.6-10.1-10.1c0-5.6,4.5-10.1,10.1-10.1C84.2,46.7,88.7,51.3,88.7,56.8z};
}}
\newcommand\orcidicon[1]{\href{https://orcid.org/#1}{\mbox{\scalerel*{
\begin{tikzpicture}[yscale=-1,transform shape]
\pic{orcidlogo};
\end{tikzpicture}
}{|}}}}

\newcommand{\aref}[1]{\hyperref[#1]{Appendix~\ref{#1}}}

\usepackage[normalem]{ulem}

\definecolor{darkgreen}{rgb}{0.13, 0.55, 0.13}
\definecolor{brown}{rgb}{0.65, 0.16, 0.16}

\title[Evolution of the IMF with metallicity]{When did the initial mass function become bottom-heavy?}

\author[Sharda and Krumholz]{Piyush Sharda$^{\orcidicon{0000-0003-3347-7094}\,1,2}$\thanks{piyush.sharda@anu.edu.au (PS)}
and Mark R. Krumholz$^{\orcidicon{0000-0003-3893-854X}\,1,2}$\thanks{mark.krumholz@anu.edu.au (MRK)}\\
$^{1}$Research School of Astronomy and Astrophysics, Australian National University, Canberra, ACT 2611, Australia\\
$^{2}$Australian Research Council Centre of Excellence for All Sky Astrophysics in 3 Dimensions (ASTRO 3D), Australia\\
}

\date{Accepted 2021 October 07. Received 2021 October 05; in original form 2021 July 19}

\pubyear{2021}

\begin{document}
\label{firstpage}
\pagerange{\pageref{firstpage}--\pageref{lastpage}}
\maketitle

\begin{abstract}
The characteristic mass that sets the peak of the stellar initial mass function (IMF) is closely linked to the thermodynamic behaviour of interstellar gas, which controls how gas fragments as it collapses under gravity. As the Universe has grown in metal abundance over cosmic time, this thermodynamic behaviour has evolved from a primordial regime dominated by the competition between compressional heating and molecular hydrogen cooling to a modern regime where the dominant process in dense gas is protostellar radiation feedback, transmitted to the gas by dust-gas collisions. In this paper we map out the primordial-to-modern transition by constructing a model for the thermodynamics of collapsing, dusty gas clouds at a wide range of metallicities. We show the transition from the primordial regime to the modern regime begins at metallicity $Z\sim 10^{-4} \rm{Z_\odot}$, passes through an intermediate stage where metal line cooling is dominant at $Z \sim 10^{-3}\,\rm{Z_{\odot}}$, and then transitions to the modern dust- and feedback-dominated regime at $Z\sim 10^{-2} \rm{Z_\odot}$. In low pressure environments like the Milky Way, this transition is accompanied by a dramatic change in the characteristic stellar mass, from $\sim 50\,\rm{M_\odot}$ at $Z \sim 10^{-6}\,\rm{Z_{\odot}}$ to $\sim 0.3\,\rm{M_\odot}$ once radiation feedback begins to dominate, which marks the appearance of the modern bottom-heavy Milky Way IMF. In the high pressure environments typical of massive elliptical galaxies, the characteristic mass for the modern, dust-dominated regime falls to $\sim 0.1\,\rm{M_{\odot}}$, thus providing an explanation for the more bottom-heavy IMF observed in these galaxies. We conclude that metallicity is a key driver of variations in the characteristic stellar mass, and by extension, the IMF.
\end{abstract}

\begin{keywords}
stars: formation -- stars: mass function -- ISM: dust -- ISM: abundances -- ISM: clouds -- stars: Population III
\end{keywords}



\section{Introduction}
\label{s:introduction}
The thermodynamics of modern (metal-rich) star formation is largely determined by the combined action of dust and stellar radiation feedback. At gas number densities $\lesssim 10^{4}-10^{5}$ cm$^{-3}$, dust and gas temperatures can differ, but once the density in a collapsing molecular cloud core exceeds this threshold, efficient gas-dust collisional coupling forces the gas and dust temperatures to track one another closely \citep[e.g.,][]{Masunaga98a, Masunaga00a, Goldsmith01a}. This means that, prior to the formation of stellar sources capable of heating it, dust is the main coolant as molecular clouds collapse and compress on their way to star formation. In this regime the gas temperature is set by a competition between adiabatic compression of the gas due to collapse (possibly supplemented by cosmic ray heating) and cooling due to dust thermal emission (possibly supplemented by metal line cooling). The cooling processes initially keep the gas close to isothermal, allowing it to fragment to ever-smaller masses as the Jeans mass decreases with rising density \citep{2016MNRAS.458..673G, Guszejnov18a}, until the system reaches the so-called opacity limit for fragmentation \citep{Low76a, Rees76a}. At this point a collapsing cloud becomes opaque and can no longer radiate away its gravitational potential energy on a free-fall timescale, preventing the gas from fragmenting any further \citep{2000ApJ...534..809O}. This transition occurs when the collapsed mass reaches $\sim 10^{-2}$ M$_\odot$.\footnote{Throughout this manuscript, we use `$\approx$' to imply `approximately equal to' and `$\sim$' to imply `of the order'.}, and early simulations using an equation of state that stiffens at high density to mimic the effects of opacity found a mass spectrum that is essentially flat between $0.01-0.5\,\rm{M_{\odot}}$ \citep{Bate03a,2005MNRAS.363..363B}.

Once hydrostatic objects form the thermodynamic regime radically changes: the gravitational potential energy liberated as mass accretes onto $\sim 10^{-2}$ M$_\odot$ ``seed'' protostars is transformed into heat and then radiatively transferred outward. In the highly-opaque environment of a collapsing cloud, this radiation is in turn absorbed by and heats the surrounding dust, which then heats the gas to temperatures far higher than those that prevail prior to hydrostatic core formation. This suppresses the formation of small objects and shifts the peak in the stellar mass distribution to $\sim 0.2$ M$_\odot$ \citep{2009ApJ...703..131O, Bate09a, Bate12a, 2011ApJ...743..110K,  2011ApJ...740...74K, Krumholz12b,Myers14a, Bate15a, 2016MNRAS.458..673G, 2017JPhCS.837a2007F, 2018MNRAS.476..771C, Mathew20a,2021arXiv210606521M}. This characteristic mass is effectively set by the ``sphere of influence'' that each hydrostatic object creates around itself by heating the gas therein to the point where it is unable to fragment \citep{2016MNRAS.460.3272K}.

The thermodynamic behaviour of collapsing gas at much lower metallicities, characteristic of the formation of the first stars (Population III) and the generation that immediately followed them (Population II), is quite different. Stellar feedback in the form of heating (as opposed to ionisation or dissociation) is unimportant, since there is no dust to absorb the radiation and transmit it to the gas. In the absence of metals, molecular hydrogen is the dominant cooling agent that competes against compressional heating \citep{1984ApJ...280..465L,1998A&A...335..403G}. Nonetheless, because H$_2$ is a poor coolant, the characteristic gas temperature is much higher than during modern, dust-mediated star formation, even considering the elevated dust temperatures that prevail once hydrostatic objects form. Thus while some low-mass objects can form due to disc fragmentation even in the absence of metals \citep[e.g.,][]{Clark11a, 2011ApJ...727..110C, Greif11a,2012MNRAS.422..290S,2014ApJ...781...60H,2015MNRAS.449...77L,2019ApJ...877...99S, 2020MNRAS.497..336S,2021MNRAS.503.2014S}, the typical mass of a star formed under these thermodynamic conditions is far larger than in the present-day case \citep{2009Natur.459...49B,2013RPPh...76k2901B,2014ApJ...792...32S,2019ffbh.book...67K}.

However, the transition between the two extremes of modern (metal-rich) and primordial (metal-poor) star formation, and in particular the role of dust coupling and stellar radiation feedback at low metallicity, has thus far received limited exploration. \citet{2011ApJ...743..110K} present analytic models for radiation feedback and predict a weak scaling of IMF peak mass with metallicity, while \citet{2011ApJ...735...49M} and \citet{2014MNRAS.442..285B,2019MNRAS.484.2341B} carry out radiation-hydrodynamic simulations of star formation over a metallicity range from $0.01-3\,Z_{\rm{\odot}}$ and find negligible effects on gas fragmentation. However, these studies do not explore lower metallicities, despite available evidence for the existence of a low-metallicity ISM in the past through the discovery of stars with metallicities as low as $10^{-4}\,\rm{Z_{\odot}}$ \citep{2011Natur.477...67C,2018MNRAS.481.3838S}, as well as several others with $\rm{[Fe/H]} < -5$ \citep{2004ApJ...603..708C,2014Natur.506..463K,2015ApJ...810L..27F,2017A&A...604A...9A,2018ApJ...854L..34A,2019MNRAS.488L.109N,2019ApJ...876...97E}. Coming from the opposite direction, \citet{2001MNRAS.328..969B} and \citet{2005ApJ...626..627O,2010ApJ...722.1793O} consider the thermodynamics of gas of increasing metallicity, and find that dust and metal line cooling permits fragmentation to reach masses $\lesssim 1$ M$_\odot$ only once the metallicity exceeds $\sim 10^{-3.5} Z_{\rm{\odot}}$. Dust is a more efficient coolant than metal lines, and permits fragmentation to lower masses at lower metallicity \citep[e.g.,][]{2014ApJ...783...75M,2020arXiv200806107C,2021arXiv210206312S}, but exactly by how much depends on the poorly-known distribution of dust grain sizes in the early Universe \citep{2006MNRAS.369.1437S, 2012MNRAS.419.1566S, 2010ApJ...722.1793O, 2010MNRAS.402..429S, 2015MNRAS.446.2659C}. However, the early Universe star formation models are fundamentally misanalogous to the modern ones that consider decreasing metallicity, in that the early Universe models consider dust solely as a coolant that enables fragmentation, whereas the modern ones assign it a more nuanced role, as both a source of cooling and later, once stellar feedback begins, a source of heating -- a changeover that seems crucial to explaining why the IMF in the present-day Universe peaks at $\sim 0.2\,\rm{M_{\odot}}$ rather than $\sim 10^{-2}\,\rm{M_{\odot}}$ \citep{2001MNRAS.322..231K,2003PASP..115..763C,2005ASSL..327...41C}.

Our goal in this paper is to explore the thermodynamics of star-forming gas, and the implications of those thermodynamics for fragmentation, across a wide range of metallicity, from near zero to super-Solar. In particular, we aim to study the transition in the peak of the IMF as a function of metallicity, and figure out when the IMF became bottom-heavy as the metal content of the Universe increased. Crucially, and in contrast with earlier work, we consider the evolving role of stellar radiation feedback, which is perhaps sub-dominant in the primordial Universe but evolves to a dominant effect in the present. We arrange the rest of the paper as follows: \autoref{s:theory} describes the theoretical framework that we use to construct our models, \autoref{s:results} describes the resulting dust and gas temperatures and the characteristic stellar mass, and \autoref{s:discussion} presents a discussion of the robustness of the results. \autoref{s:discussions_evolutionIMF} looks at the evolution of the IMF in various stellar systems in the context of our models. We then use our models to explore implications for the cosmic star formation history in \autoref{s:implications}. Finally, we present a summary of our work in \autoref{s:conclusions}.

\section{Theoretical Framework}
\label{s:theory}
The basic system we consider is a spherical cloud core shortly after its centre has collapsed and produced a first hydrostatic object of mass $\sim 10^{-2}$ M$_\odot$. We are interested in determining how much of the gas around this object is available to accrete onto it -- thereby increasing the object's mass above the minimum imposed by the opacity limit for fragmentation and shifting the stellar mass distribution higher -- and how much is likely to undergo independent collapse and form other objects. Since both analytic models and simulations show that fragmentation is closely linked to the temperature structure of the gas, we address this question by first computing the expected gas temperature structure, which requires balancing heating against cooling, and then examining the implications of that temperature structure for fragmentation. For simplicity, we collect all the major parameters we use in this work and list them in \autoref{tab:tab1}.

Before proceeding further, however, we note an important caveat: the basic premise of our model is that the location of the IMF peak is set by thermal fragmentation, and, while this proposition has significant theoretical and numerical support (as discussed in \autoref{s:introduction}), this is not the only possible explanation. For example, several early authors proposed that the IMF peak might be set by the feedback provided by protostellar outflows \citep{1996ApJ...464..256A, Shu04a}, and more recently several have proposed that it is imposed by tidal forces that appear when a first \citet{1969MNRAS.145..271L} core forms  \citep{2018A&A...611A..89L, 2019ApJ...883..140H,2020MNRAS.492.4727C, Hennebelle20a}. Our calculation will yield correct results only to the extent that the IMF peak is controlled by gas thermodynamics rather than outflows, tidal forces, or some other process that is insensitive to the temperature and pressure of the collapsing gas.

\begin{table}
\centering
\caption{List of the main parameters used in this work.}
\begin{tabular}{|l|l|l|}
\hline
Parameter & Description & Reference \\
\hline
&Physical model & \autoref{s:physicalsetting}\\
\hline
$P$ & Cloud pressure & \autoref{eq:rho0} \\
$\sigma_{\rm{v}}$ & Cloud velocity dispersion \\
$\rho_0$ & Density at the cloud edge & \autoref{eq:rho0} \\
$\Sigma$ & Surface density at the cloud edge & \autoref{eq:Sigma} \\
$R$ & Cloud radius & \autoref{eq:R} \\
$M$ & Cloud mass & \autoref{eq:M}\\
\hline
&Chemical model & \autoref{s:metallicity_chemicalmodel}\\
\hline
$Z$ & Metallicity & \ldots \\
$\mathcal{Z}$ & $\log_{10}\,Z/\rm{Z_{\odot}}$ & \ldots \\
$\delta$ & Dust to gas ratio & \ldots \\
$n_{\rm{H}}$ & Number density per H nuclei & \ldots \\
$n$ & Free particle number density & \ldots \\
$\mu$ & Mean mass per free particle & \ldots \\
$\mu_{\rm{H}}$ & Mean mass per H nucleon & \ldots \\
\hline
&Dust temperature profile & \autoref{s:dusttemperature}\\
\hline
$L$ & Luminosity & \autoref{eq:L}\\
$T_{\rm{d0}}$ & Dust temperature at cloud edge & \autoref{eq:Td(r)} \\
$T_{\rm{d}}(r)$ & Dust temperature at radius $r$ & \autoref{eq:Td(r)}\\
$\kappa$ & Dust opacity & \autoref{eq:tdust} \\
\hline
&Gas temperature profile & \autoref{s:gastemperature}\\
\hline
$T_{\rm{g}}(r)$ & Gas temperature at radius $r$ & \ldots\\
$\Gamma_{\rm{c}}$ & Compressional heating & \autoref{eq:compress}\\
$\Psi_{\rm{gd}}$ & Dust-gas energy exchange & \autoref{eq:gd}\\
$\Lambda_{\rm{M}}$ & Metal line cooling & \autoref{eq:C_O_cool}\\
$\Lambda_{\rm{H_2}}$ & Molecular hydrogen cooling & \autoref{eq:h2}\\
$\Lambda_{\rm{HD}}$ & Hydrogen deuteride cooling & \ldots\\
$\Gamma_{\rm{H_2,3b}}$ & 3-body H$_2$ formation heating & \autoref{eq:gammaH23b}\\
$\Gamma_{\rm{H_2,d}}$ & Heating due to H$_2$ formation on dust & \autoref{eq:dustH2}\\
$\Gamma_{\rm{CR}}$ & Cosmic ray heating & \autoref{eq:cosmicrays}\\
$\mathcal{C}(\eta)$ & Relative contribution of a process $\eta$ & \autoref{eq:term_contribution}\\
&to gas heating/cooling&\\ 
\hline
&Characteristic stellar mass & \autoref{s:charmass}\\
\hline
$M_{\rm{enc}}(r)$ & Enclosed mass as a function of $r$ & \autoref{eq:Menclosed} \\
& around a protostar&\\
$M_{\rm{BE}}(r)$ & Bonnor-Ebert mass as a function of $r$ & \autoref{eq:BEmass} \\
& around a protostar&\\
$n_{\rm{crit}}$ & Critical density where $M_{\rm{enc}}=M_{\rm{BE}}$  & \ldots\\
$M_{\rm{ch}}$ & Characteristic stellar mass & \ldots\\
\hline
\end{tabular}
\label{tab:tab1}
\end{table}

\subsection{Physical model}
\label{s:physicalsetting}
Following numerical simulations of star formation at various metallicities \citep[e.g.,][]{2011ApJ...740...74K,2013ApJ...763...51F,2016MNRAS.463.2781C,2019MNRAS.490..513S} as well as observations \citep{1995ApJ...446..665C,2000ApJ...537..283V,2002A&A...389..908J,2002ApJS..143..469M,2009ARep...53.1127P,2015A&A...578A..29S,2021A&A...648A..66G}, we assume that the volume density of the cloud has a radial profile $\rho(r) = \rho_0 (r/R)^{-k_\rho}$ that can be fully described by three parameters; one of these is $k_{\rho}$, which both observations and simulations find is always in the range $k_\rho = 1-2$, and we therefore fix to $1.5$. The other two can be the edge density $\rho_0$ and outer radius $R$, but we can equally well specify any two of the total mass ($M$), the mean surface density ($\Sigma$), or anything else related to these. For the purposes of exploring the parameter space, it is convenient to choose the two parameters to be the pressure at the cloud edge $P$ and velocity dispersion at the cloud edge $\sigma_{\rm v}$ -- the former because the pressure in a molecular cloud core will be at a minimum bounded below by the mean interstellar pressure in the galaxy wherein it resides, and the latter because observed velocity dispersions in molecular cloud cores span a relatively narrow range. The edge density and radius are related to these by\footnote{Note that the relationship between $\sigma_{\rm{v}}$, $M$, and $R$ in this system is the same as that in \citet[equation~14]{2011ApJ...735...49M} within 20 per cent.}
\begin{eqnarray}
\rho_0  & = & \frac{P}{\sigma_{\rm v}^2}
\label{eq:rho0} \\
\Sigma & = & \sqrt{\frac{20P}{3\pi\alpha_{\rm{vir}}G}}
\label{eq:Sigma} \\
R & = & \frac{\left(3-k_{\rho}\right)\Sigma}{4\rho_0}
\label{eq:R} \\
M & = & \pi\Sigma R^2
\label{eq:M} 
\end{eqnarray}
where $\alpha_{\rm{vir}} \approx 2$ is the virial ratio for a collapsing molecular cloud. Here, we have ignored the effects of magnetic pressure, which, if dominant (e.g., in high mass star formation regions), can lead to $\alpha_{\rm{vir}} \ll 1$ \citep{2011A&A...530A.118P,2013ApJ...779..185K}. At least for modern star formation, \citet{2016MNRAS.460.3272K} find that magnetic forces are unimportant on the small scales where fragmentation sets the IMF, even in simulations where the magnetic field is dynamically very strong on larger scales.

We vary the pressure between $10^4\,k_{\rm{B}}\,\rm{K\,cm^{-3}}$ (typical of Milky Way molecular clouds - \citealt{2001ApJ...547..792D,2017ApJ...834...57M}) and $10^8\,k_{\rm{B}}\,\rm{K\,cm^{-3}}$ (typical of molecular clouds in starburst environments - \citealt{2000ApJ...532L.109T,2008ApJ...677L...5V,2008ApJ...686..948B}, or super star clusters - \citealt{2006A&A...448..881B}). Observed values of $\sigma_{\rm{v}}$ for molecular cloud cores are between $0.5-5\,\rm{km\,s^{-1}}$ \citep[see also,][]{1997MNRAS.288..145P,2011ApJ...735...49M,2014ApJ...796...75C}.  Note that not all combinations of $P$ and $\sigma_{\rm v}$ are physically plausible, and some are plausible but not very probable. For example, a molecular cloud with a very high $P$ in general cannot have a very high $\sigma_{\rm{v}}$ since high $P$ corresponds to a high surface density, and the volume density is proportional to $1/\sigma^2_{\rm{v}}$; high $P$ and high $\sigma_{\rm v}$ therefore corresponds to the implausible combination a very high surface density but a very low volume density.

\subsection{Metallicity and chemical model}
\label{s:metallicity_chemicalmodel}
We parameterize the chemical composition of the gas in terms of the metallicity $Z$, which we vary between $10^{-6}\,\rm{Z_{\odot}}$ (extremely metal-poor or primordial-like) and $2\,\rm{Z_{\odot}}$ (super solar-like). This is of course a significant oversimplification. In reality, the chemical composition of a cloud changes according to the density, metallicity, and temperature, and may or may not be in steady state \citep[e.g.,][]{2009ASPC..417...71L,Krumholz11a,2021arXiv210303889H,2021arXiv210501681S}. This matters for the thermodynamics because, even for fixed abundances of elements such as C and O, the rate of cooling depends on whether these are mostly in the form of molecules such as CO, neutral atoms such as \ion{C}{i}, or ions such as \ion{C}{ii}. Capturing this complexity in detail is possible only with the aid of a fully time-dependent chemodynamic simulation \citep[e.g.,][]{2016MNRAS.463.2781C}. However, we will see that the generic result we obtain below regarding the existence of different thermodynamic regimes and the locations of transitions between them is not qualitatively dependent on details of the chemical composition.

For the purposes of making a simple model, we will limit ourselves to considering the following chemical species, which are responsible for the great majority of cooling in neutral gas: \ion{H}{i}, H$_2$\footnote{In places where we make use of atomic data that distinguishes ortho-H$_2$ from para-H$_2$, we adopt a fixed ortho-H$_2$ to para-H$_2$ ratio of 3:1 - variations in this ratio as observed in clouds at different metallicities \citep[e.g.,][]{1999ApJ...516..371S,2000MNRAS.316..901F,2000A&A...356..695R} make little difference to the results.}, \ion{C}{i}, \ion{C}{ii}, \ion{O}{i}, and CO. We define $x_{\rm{S}}$ as the ratio of the number density of species $S$ to the number density of H nuclei, where we distinguish between the abundance of the neutral state of a particular species and all atoms of that species by denoting the former with $\textsc{i}$, i.e., $x_{\ion{C}{i}}$ is the number of neutral carbon atoms present per H nucleus, while $x_{\rm C}$ is the number of all carbon atoms per H nucleus in all chemical and ionisation states -- ionised, free neutral, and bound into CO. We set total atomic abundances as a function of metallicity simply by scaling from local ISM gas-phase abundances, i.e., we set $x_{\rm C} = x_{\rm (C,MW)} (Z/\rm{Z_\odot})$, and similarly for O. Here $x_{\rm (C,MW)}$ is the gas-phase abundance of C atoms in the local ISM; for the purposes of this paper, we adopt $x_{\rm (C,MW)} = 1.4\times 10^{-4}$ and $x_{\rm (O,MW)} = 3.0\times 10^{-4}$ following \citet[table 23.1]{2011piim.book.....D}. The abundances of the various possible chemical states are then related by
\begin{eqnarray}
    1 & = & x_{\ion{H}{i}} + 2 x_{\rm H_2} \\
    x_{\rm C} & = & x_{\ion{C}{i}} + x_{\ion{C}{ii}} + x_{\rm CO} \\
    x_{\rm O} & = & x_{\ion{O}{i}} + x_{\rm CO},
\end{eqnarray}
which is equivalent to stating that all (gas-phase) O atoms are in the form of \ion{O}{i} or CO, C atoms are in the form of \ion{C}{i}, \ion{C}{ii}, or CO, and H atoms are \ion{H}{i} or H$_2$.

\begin{figure}
\includegraphics[width=1.0\columnwidth]{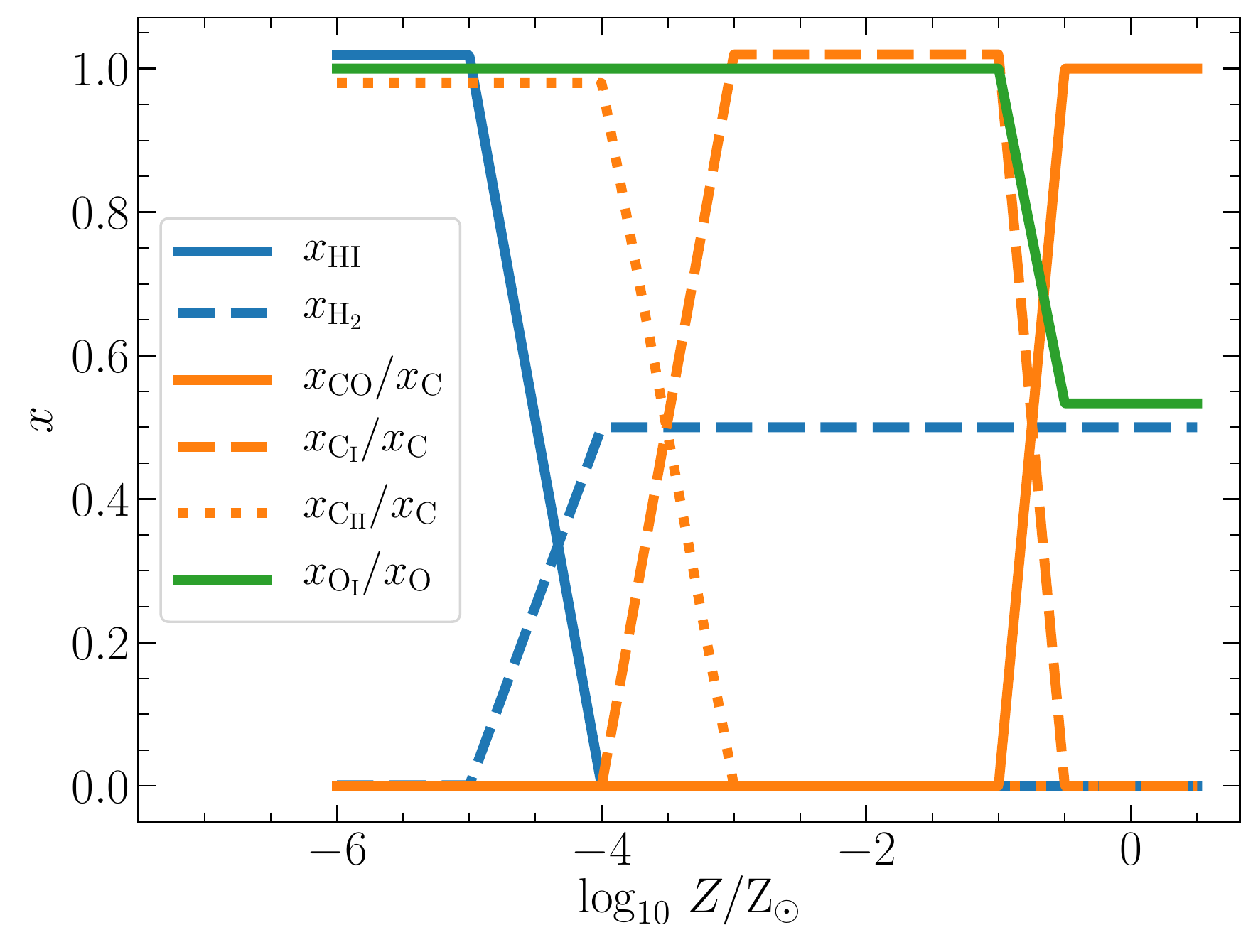}
\caption{Number fraction of hydrogen, carbon and oxygen atoms in various forms -- neutral (\ion{H}{i}, \ion{C}{i}, \ion{O}{i}), ionized (\ion{C}{ii}), and molecular (H$_2$, CO), as a function of metallicity $Z$ for the fiducial model (see \autoref{eq:chemicalcomposition_CO} and \autoref{eq:chemicalcomposition_C}). The curves for $x_{\ion{H}{i}}$ and $x_{\ion{O}{i}}/x_{\rm{O}}$ have been shifted by $\pm2$ per cent, respectively, for better visibility.}
\label{fig:dummy}
\end{figure}

In keeping with our simple approach, as a fiducial case we will adopt plausible scalings of $x_{\rm H_2}$, $x_{\ion{C}{i}}$, and $x_{\rm CO}$ with metallicity, guided by a combination of simulations and observations; the abundances of the remaining states can then be deduced from conservation. We therefore adopt

\begin{equation}
x_{\ion{H}{i}} = \left\{ \begin{array}{ll}
  1 - 2x_{\rm{H_2,\,min}}, & \mathcal{Z} < -5\\
  (1-2x_{\rm{H_2,\,min}}) + {} \\
  \quad 2(\mathcal{Z}+5)\left(x_{\rm{H_2,\,min}}-x_{\rm{H_2,\,max}}\right), & -5 \leq \mathcal{Z} < -4\\
  1 - 2x_{\rm{H_2,\,max}}, & \mathcal{Z} \geq -4\,,
\end{array} \right.
\label{eq:chemicalcomposition_hydrogen}
\end{equation}
where $\mathcal{Z}\equiv \log_{10}(Z/Z_\odot)$, $x_{\rm{H_2,max}} = 0.5$ for a fully molecular composition, and we set $x_{\rm{H_2,min}} \approx 0.0007$ based on the results from Population III star formation simulations \citep{2019MNRAS.490..513S,2020MNRAS.497..336S}. For CO, we adopt
\begin{equation}
\frac{x_{\rm{CO}}}{x_{\rm{C}}} = \left\{ \begin{array}{ll}
  0, & \mathcal{Z} < -1\\
  2\left(\mathcal{Z} + 1\right), & -1 \leq \mathcal{Z} < -0.5\\
  1,  & \mathcal{Z} \geq -0.5\,,
\end{array} \right.
\label{eq:chemicalcomposition_CO}
\end{equation}
and for \ion{C}{i}, we use
\begin{equation}
\frac{x_{\ion{C}{i}}}{x_{\rm{C}}} = \left\{ \begin{array}{ll}
  0, & \mathcal{Z} < -4\\
  4 + \mathcal{Z}, & -4 \leq \mathcal{Z} < -3\\
  1, & -3 \leq \mathcal{Z} < -1\\
  -\left(1+2\mathcal{Z}\right), & -1 \leq \mathcal{Z} < -0.5\\
  0, & \eta \geq -0.5
\end{array} \right.
\label{eq:chemicalcomposition_C}
\end{equation}
We plot our adopted fiducial chemical abundances as a function of metallicity in \autoref{fig:dummy}.

Qualitatively, these scalings describe a sequence of chemical states through which star-forming gas passes, starting with a purely atomic, moderately ionised composition (mostly \ion{H}{i}, \ion{C}{ii}, \ion{O}{i}) at the lowest metallicity, changing to one where the H converts to H$_2$ at higher metallicity (mostly H$_2$, \ion{C}{ii}, \ion{O}{i}), then allowing the C to start transitioning to neutral as dust shielding increases (H$_2$, \ion{C}{i}, \ion{O}{i}), and finally reaching the CO-dominated composition observed to characterise modern star formation (H$_2$, CO). While our chemical setup is consistent with both theoretical \citep{2005ApJ...626..627O,2009ApJ...693..216K,2010ApJ...709..308M,2012MNRAS.426..377G,2014ApJ...790...10S,2015MNRAS.446.2659C,2015MNRAS.450.4424B,2016MNRAS.456.3596G,2020arXiv200806107C,2021arXiv210501681S,2021arXiv210303889H} and observational results \citep{2009ASPC..417...71L,2017ApJ...839..107P,2017ApJ...835..278S,2018ApJ...853..111J,2020A&A...643A.141M}, it is of course a simplification since the chemical composition also depends on cloud density and temperature as we describe above. We will show later in \autoref{s:discussions_chemicalcompositions} that the choice of the chemical composition does not significantly impact our results on the transition of the IMF from primordial to modern-day.

Since one of the main goals of this paper is to account for the effects of dust, as a fiducial model we also assume that the dust abundance scales linearly with total metallicity. Specifically, we adopt a dust to gas mass ratio $\delta_{\odot} = 1/162$ at $Z = \rm{Z_{\odot}}$ \citep{2004ApJS..152..211Z}, and at other metallicities adopt a mass ratio $\delta = \delta_\odot (Z/Z_\odot)$. As with gas-phase chemical abundances, we also consider alternative scalings of $\delta$ with $Z$ below in \autoref{s:discussions_metalscaling}, and show that our qualitative results are not sensitive to the particular scaling that we adopt.

Finally, at various points in the following discussion, we will require conversions between mass density and number density, which depend on chemical composition. Assuming that H and He nuclei always dominate the mass, and the usual cosmic He abundance $x_{\rm He}=0.1$, the mean mass per H nucleus, measured in units of $m_{\rm H}$, is $\mu_{\rm H}=1 + 4 x_{\rm He} = 1.4$. The number density of H nuclei is related to the mass density by $n_{\rm H}=\rho/\mu_{\rm H} m_{\rm H}$. By contrast, the number of free particles per H nucleus is $1 - x_{\rm H_2} + x_{\rm He}$, so the mean mass per free particle is $\mu = \mu_{\rm H}/(1 - x_{\rm H_2} + x_{\rm He})$, and the number density of free particles is $n = \rho/\mu m_{\rm H}$.

\subsection{Dust temperature profile}
\label{s:dusttemperature}
Once a protostar appears in a star-forming molecular cloud, the temperature of the dust around the protostar is mainly governed by radiation feedback from the protostar. Thus, radiation feedback plays a key role in the thermodynamics of dust in such regions. During the early phases of star formation, this feedback is powered primarily by accretion, rather than by nuclear burning, and we calculate the accretion luminosity following \citet{2011ApJ...743..110K}:
\begin{equation}
    L =\epsilon_{\rm{L}}\epsilon_{\rm{M}} \mathcal{A} M \sqrt{\frac{3G\rho_0}{3-k_{\rho}}}.
    \label{eq:L}
\end{equation}
Here the factor under the square root is simply the inverse of the mean free-fall time ($t_{\rm{ff}}$) in the cloud, so this expression amounts to a statement that the luminosity is proportional to an accretion rate of order $M/t_{\rm ff}$. The other factors appearing in the proportionality are $\epsilon_{\rm{M}} \approx 0.5$, the fraction of the mass falling onto the protostar that is accreted rather than ejected from the inner disc in a wind \citep{2000ApJ...545..364M,2007A&A...462L..17A,2008ApJ...684.1240E,2012ApJ...761..156F,2014ApJ...790..128F}, $\epsilon_{\rm{L}} \approx 0.75$, the fraction of accretion power that contributes to accretion luminosity rather than being used to drive the wind \citep{2003ApJ...585..850M}, and $\mathcal{A} = GM_{\star}/R_{\star}$, the energy per unit mass released by accretion onto a protostar of mass $M_\star$ and radius $R_\star$. \citet[Figure~3a]{2011ApJ...743..110K} point out that, due to deuterium burning during the star formation process, all protostars have similar, nearly-linear mass-radius relations, which yields an approximately constant value $\mathcal{A} = 2.5\times10^{14}\,\rm{erg\,g^{-1}}$ \citep[see also,][]{1980ApJ...241..637S,2011ApJ...738..140H}. We discuss how $\epsilon_{\rm{L}},\,\epsilon_{\rm{M}}$ and $\mathcal{A}$ impact the final results in \autoref{s:discussions_otherparams}.

To obtain the dust temperature from the luminosity and the physical structure of the cloud core, we use the analytic model developed by \citet{2005ApJ...631..792C} to produce spectral energy distributions (SEDs) of non-overlapping\footnote{Non-overlapping means that the thermal influence zone of one source does not overlap with another, which is a valid approximation as long as the star formation efficiency per freefall time is less than 10 per cent \citep{2011ApJ...740...74K}, observed in almost all star-forming environments \citep{2018MNRAS.477.4380S,2019MNRAS.487.4305S}.} spherically-symmetric sources. This model has been shown to reproduce the SEDs in a wide variety of dusty environments, ranging from protostellar sources to high-redshift sub-mm galaxies \citep{2008ApJ...683..693C,2013ApJ...773..113C}. \citeauthor{2005ApJ...631..792C} show that, for a dusty gas cloud with a powerlaw density profile such as we have assumed, the dust temperature profile also assumes an approximately powerlaw form, given by
\begin{equation}
T_{\rm{d}}(r) = T_{\rm{d0}} \left(\frac{r}{R}\right)^{-k_{\rm{T}}},
\label{eq:Td(r)}
\end{equation}
where, for a given wavelength-dependent dust opacity, the index $k_{\rm{T}}$ and the outer dust temperature $T_{\rm{d0}}$ are determined entirely by two parameters: the cloud luminosity-to-mass ratio $L/M$, and the surface density $\Sigma$. The index and outer dust temperature are in turn given by
\begin{eqnarray}
    k_{\rm{T}} & \approx & \frac{0.48k^{0.005}_{\rho}}{\tilde{R}^{0.02k^{1.09}_{\rho}}} + \frac{0.10k^{5.5}_{\rho}}{\tilde{R}^{0.70k^{1.09}_{\rho}}} \\
    T^\gamma_{\rm{d0}} & \approx & \left(\frac{L/M}{6.4\sigma_{\rm{SB}}\tilde{R}^{0.1}}\right)^{k_\rho - 1 +\beta k_{\rm{T}}}\left[\frac{(3-k_\rho)\delta \kappa_0}{4(k_\rho-1)T^{\beta}_0}\right]^{4k_{\rm{T}}-2}
    \nonumber\\
    & & \qquad {}\times \Sigma^{(4+\beta)k_{\rm{T}} + k_\rho - 3}
\label{eq:tdust}
\end{eqnarray}
where $\gamma = 2\beta + 4(k_\rho-1)$, $\sigma_{\rm{SB}}$ is the Stefan-Boltzmann constant, and the parameters $\beta$, $\delta$, $\kappa_0$, and $T_0$ describe the dust opacity as a function of wavelength in the infrared, $\kappa = \delta \kappa_0 (\lambda_0/\lambda)^\beta$, where $\lambda_0 = hc/k_B T_0$. We adopt $\beta=2$ and $\kappa_0=0.27\,\rm{cm^2\,g^{-1}}$ at $\lambda_0=100\,\rm{\mu m}$ from the dust opacity model of \cite{2001ApJ...548..296W}, which gives $T_0=144\,\rm{K}$.\footnote{Note that the dust opacity model we adopt does not take the effects of grain growth via coagulation into account. However, \cite{2005ApJ...631..792C} show that for typical cloud densities, the coagulation timescale is two orders of magnitude more than the freefall time at which the cloud collapses. Additionally, dust coagulation changes $\beta$ by at most 10 per cent \citep{1994A&A...291..943O}, which does not lead to any appreciable difference in $T_{\rm{d}}$. Thus, we do not expect dust coagulation to play a significant role in our work.} Other dust opacity models such as those of \cite{1994ApJ...421..615P} and \cite{2003A&A...410..611S} have been shown to make little difference in this calculation \citep{2011ApJ...735...49M}. Finally, the dimensionless constant $\tilde{R}$ is given by
\begin{equation}
\tilde{R}  = \left[\frac{(L/M)(M/\pi R^2)^{(4+\beta)/\beta}}{6.4 \sigma_{\rm{SB}}\tilde{R}^{0.1}}\left(\frac{\delta \kappa_0 \left(3-k_{\rho}\right)}{4 \left(k_{\rho}-1\right)T_0^{\beta}}\right)^{4/\beta}\right]^{-\beta/\gamma}\,.
\label{eq:tildeR}
\end{equation}
Note that the dust opacity model we incorporate is only valid as long as most of the emission is longward of $30\,\mu \rm{m}$, implying that dust opacities cannot be written down as a powerlaw in frequency for $T_{\rm{d}} \gtrsim 480\,\rm{K}$. We show below that $T_{\rm{d}} < 480\,\rm{K}$ in the regime where dust matters for the characteristic stellar mass. Furthermore, we check that the application of the \citeauthor{2005ApJ...631..792C} model for our work remains valid across the entire parameter space by ensuring $\tilde{R} > 1$.

\subsection{Gas temperature profile}
\label{s:gastemperature}
The next step in our calculation is to determine the gas temperature $T_{\rm g}$ as a function of radius within the cloud. In thermal equilibrium, we can solve for $T_{\rm g}$ by balancing gas heating and cooling,
\begin{equation}
    \Gamma_{\mathrm{c}} + \Psi_{\mathrm{gd}} + \Lambda_{\mathrm{M}} + \Lambda_{\mathrm{H_2}} + \Lambda_{\mathrm{HD}} = 0
\label{eq:equate_heat_cool}
\end{equation}
where the $\Gamma$ term represents heating processes (expressed as rate per unit mass, with units of energy per unit mass per unit time), $\Lambda$ terms represent cooling processes, and $\Psi$ is a term representing processes that can either heat or cool the gas depending on the circumstances. From left to right, the terms appearing in \autoref{eq:equate_heat_cool} represent heating due to adiabatic compression, heating/cooling due to dust-gas energy exchange, cooling due to metal lines, cooling due to molecular hydrogen, and cooling due to hydrogen deuteride, respectively. While this list of processes is by no means exhaustive, our list does cover the processes that dominate gas thermodynamics. In addition to these, we also consider the effects of H$_2$ formation heating and cosmic rays on gas thermodynamics later in \autoref{s:discussion}. We next express all of the heating, exchange, and cooling terms in terms of the local gas density $\rho$ and gas temperature $T_{\rm g}$; since we have specified $\rho(r)$ (as well as the dust temperature $T_{\rm d}(r)$), doing so allow us to solve \autoref{eq:equate_heat_cool} to obtain the gas temperature profile $T_{\rm g}(r)$.

\subsubsection{Heating processes}
\label{s:gatemp_heating}
\textit{Compressional heating.} Heating due to compression in a free-falling gas occurs at a rate \citep[e.g.,][]{Masunaga98a, Masunaga00a}
\begin{equation}
    \Gamma_{\mathrm{c}} \approx \frac{k_{\mathrm{B}}T_{\mathrm{g}}}{\mu} \sqrt{\frac{32G \mu_{\rm{H}}n_{\rm{H}}}{3\pi m_{\rm{H}}}}.
\label{eq:compress}
\end{equation}
This expression follows simply from the first law of thermodynamics, which requires that the work per unit mass done on the gas by compression be $P\, d(1/\rho)/dt$. \autoref{eq:compress} follows immediately from this by applying the ideal gas law to write $P$ in terms of $\rho$ and $T_{\rm g}$, and taking the time derivative to be of order of the free-fall time $1/t_{\rm ff} = \sqrt{32G\rho/3\pi}$ \citep[e.g.,][]{2007ARA&A..45..565M}. We discuss some additional heating mechanisms that could potentially contribute to gas heating in some parts of the parameter space (for example, cosmic rays and H$_2$ formation heating) in \autoref{s:discussion}, showing that they do not qualitatively alter our results on the trends in characteristic stellar mass as a function of metallicity.

\subsubsection{Gas-dust energy exchange}
\label{s:gatemp_psigd}
The energy exchange per unit mass between the dust and the gas due to collisions is given by \citep{1979ApJS...41..555H,2000ApJ...534..809O}
\begin{equation}
    \Psi_{\mathrm{gd}} = 2\alpha_{\rm{gd}} \delta \mathcal{S}_{\rm{gd}} n_{\rm{H}} \sqrt{\frac{8k_{\mathrm{B}}T_{\mathrm{g}}}{\pi m_{\mathrm{H}}}} k_{\mathrm{B}}\left(T_{\mathrm{d}}-T_{\mathrm{g}}\right)\times \sum_{\rm{S}} \frac{x_{\rm{S}}}{\sqrt{\mu_{\rm{S}}}}\,,
\label{eq:gd}
\end{equation}
where the sum is over the species \ion{H}{i}, H$_2$, and He, $\mu_{\rm{S}}$ is the species mass in units of $m_{\rm H}$ (1, 2, and 4 for \ion{H}{i}, H$_2$, and He, respectively), $\alpha_{\rm{gd}}$ is a factor between 0 and 1 that describes the inelasticity of dust-gas collisions \citep{1983ApJ...265..223B}, $\delta$ is the dust to gas ratio that we defined in \autoref{s:metallicity_chemicalmodel}, and $\mathcal{S}_{\rm gd}$ is the dust cross section per unit dust mass. Following results from experiments \citep{1967rgd1.conf..155T,1974PrSS....5..261G,1967JChPh..46.2376G}, we set $\alpha_{\rm{gd}} \sim 0.5$. In the absence of low metallicity measurements, we fix $\mathcal{S}_{\rm{gd}}=10^5\,\rm{cm^2\,g^{-1}}$ (based on observations of interstellar dust in the Milky Way - \citealt{2012MNRAS.419.1566S}) at all $Z$, cautioning that this is very uncertain, since $\mathcal{S}_{\rm{gd}}$ depends primarily on the poorly-understood grain size distribution \citep{2016MNRAS.457.1842S}. We discuss the effects of uncertainties in $\alpha_{\rm{gd}}$ and $\mathcal{S}_{\rm{gd}}$ on the final results in \autoref{s:discussions_otherparams}.

It is clear from \autoref{eq:gd} that $\Psi_{\rm{gd}}$ acts as a heating source if $T_{\rm{d}}>T_{\rm{g}}$, and a cooling source when $T_{\rm d}<T_{\rm g}$. In the analysis that follows later, we will split $\Psi_{\rm{gd}}$ into its positive (heating) and negative (cooling) terms, and denote them by $\Gamma_{\rm{gd}}$ and $\Lambda_{\rm{gd}}$, respectively. Our approach thus enables us to gauge the role of dust-gas energy exchange in the presence of stellar feedback by using $T_{\rm{d}}$ from the \cite{2005ApJ...631..792C} model.

\subsubsection{Cooling processes}
\label{s:gatemp_cooling}
The most important metal coolants in the ISM are carbon and oxygen, both of which can be present in different chemical forms (\ion{C}{ii}, \ion{C}{i}, \ion{O}{i}, CO). In atomic form, the most important coolants are the fine structure lines \ion{C}{ii} $158\,\rm{\mu m}$, \ion{O}{i} $63$ and $145\,\rm{\mu m}$, \ion{C}{i} $230$ and $610\,\rm{\mu m}$; when C and O are in molecular form, cooling is dominated by the low-$J$ levels of CO. Based on the chemical composition we describe in \autoref{eq:chemicalcomposition_CO} and \autoref{eq:chemicalcomposition_C}, we can express $\Lambda_{\rm{M}}$ as the sum of cooling provided by C and O in their various chemical forms.

\textit{\ion{C}{ii} cooling.} The ground spectroscopic term of \ion{C}{ii} is a $^2P$ doublet. In statistical equilibrium between the excited and the ground fine structure states, the cooling rate per unit mass is given by\footnote{Note that this calculation and the analogous ones that follow for \ion{C}{i} and \ion{O}{i} are only valid as long as the cooling lines are optically thin. We check that this is indeed the case by estimating the optical depth of each of the transitions of these species. The general result is that these species never become optically thick within the parameter space we cover in this work (at all $Z$); this is primarily because of the extremely low oscillator strengths of the fine structure transitions.}
\begin{equation}
    \Lambda_{\ion{C}{ii}} = -\frac{x_{\ion{C}{ii},1}A_{\ion{C}{ii},10}k_{\rm{B}}T_{\ion{C}{ii}}}{\mu_{\rm{H}} m_{\rm{H}}}\,
\label{eq:Cpluscooling}
\end{equation}
where $x_{\ion{C}{ii},1}$ is the level population in the excited level of \ion{C}{ii}, $A_{\ion{C}{ii},10}$ is the \ion{C}{ii} Einstein $A$ coefficient for radiative de-excitation from the excited to the ground level, and $T_{\ion{C}{ii}}$ is the energy per $k_{\rm{B}}$ for the excited level of \ion{C}{ii}. We use the Leiden Atomic and Molecular Database \citep[LAMDA,][]{2005A&A...432..369S} to obtain $A_{\ion{C}{ii},10} = 2.3\times10^{-6}\,\rm{s^{-1}}$ and $T_{\ion{C}{ii}} \approx 91\,\rm{K}$. The density in the upper level $x_{\ion{C}{ii},1}$ can then be expressed as
\begin{equation}
\begin{split}
    x_{\ion{C}{ii},1} = x_{\ion{C}{ii}} \times \left[1 +    \frac{n_{\rm{H}} \overline{k}_{\ion{C}{ii},10} + A_{\ion{C}{ii},10}}{n_{\rm{H}} \overline{k}_{\ion{C}{ii},01}}\right]^{-1}\,.
\end{split}
\label{eq:Ctwolevelsystemnew}
\end{equation}
Here, $\overline{k}_{\rm{\ion{C}{ii},10}}$ is the number fraction-weighted rate of collisional de-excitation from the excited to the ground level due to collisions with H, ortho-H$_2$, para-H$_2$, and He at a given $T_{\rm{g}}$, 
\begin{equation}
\begin{split}
\overline{k}_{\ion{C}{ii},10} = x_{\ion{H}{i}}k_{\ion{C}{ii},10-\rm{H}} + \frac{3}{4}x_{\rm{H_2}}k_{\ion{C}{ii},10-\rm{oH_2}} + \frac{1}{4}x_{\rm{H_2}}k_{\ion{C}{ii},10-\rm{pH_2}} + \\
x_{\rm{He}}k_{\ion{C}{ii},10-\rm{He}}\,.
\end{split}
\label{eq:weightedk}
\end{equation}
We adopt the collisional coefficient rates for collisions between \ion{C}{ii} and other species from the LAMDA database \citep{2005ApJ...620..537B,2013JChPh.138t4314L,2014ApJ...780..183W}.\footnote{The collisional rate coefficient for collisions between \ion{C}{ii} and He, $k_{\ion{C}{ii},10-\rm{He}}$, is not provided in the LAMDA database. So, we simply approximate $k_{\ion{C}{ii},10-\rm{He}} \approx \sqrt{2}k_{\ion{C}{ii},10-\rm{oH_2}}$, where the factor $\sqrt{2}$ accounts for the mass difference between He and ortho-H$_2$.} $\overline{k}_{\rm{\ion{C}{ii},01}}$ is the number fraction-weighted rate of collisional excitation from the ground to the excited state, given by
\begin{equation}
    \overline{k}_{\rm{\ion{C}{ii},01}} = \frac{g_{\ion{C}{ii},1}}{g_{\ion{C}{ii},0}}\,\overline{k}_{\rm{\ion{C}{ii},10}} e^{-T_{\ion{C}{ii}}/T_{\rm{g}}}
\label{eq:Ck10k01}
\end{equation}
where $g_{\ion{C}{ii},1}=4$ and $g_{\ion{C}{ii},0}=2$ are the statistical weights of the excited and the ground states of \ion{C}{ii}, respectively.

\textit{\ion{C}{i} and \ion{O}{i} cooling.} Both \ion{C}{i} and \ion{O}{i} have similar atomic shell configurations since the former has 2 filled and the latter has 2 empty 2$p$ shells. For both, cooling at low temperatures comes from fine structure transitions between the sub-states of the lowest spectroscopic term, which is a $^3P$ triplet. In statistical equilibrium, the level populations in the three states can be implicitly expressed as
\begin{equation}
\label{eq:CIrateqn}
    \left(n_{\rm{H}} \mathsf{\overline{K}} + \mathsf{A}\right) \mathbf{x}_{\ion{C}{i}} = \mathbf{0},
\end{equation}
where
\begin{equation}
    \mathsf{\overline{K}} = \left(
    \begin{array}{ccc}
    \overline{k}_{\ion{C}{i},01} + \overline{k}_{\ion{C}{i},02} & 
    -\overline{k}_{\ion{C}{i},10} & 
    -\overline{k}_{\ion{C}{i},20} \\
    -\overline{k}_{\ion{C}{i},01} &
    \overline{k}_{\ion{C}{i},10} + \overline{k}_{\ion{C}{i},12} &
    -\overline{k}_{\ion{C}{i},21} \\
    -\overline{k}_{\ion{C}{i},02} &
    -\overline{k}_{\ion{C}{i},12} &
    \overline{k}_{\ion{C}{i},20} + \overline{k}_{\ion{C}{i},21}
    \end{array}
    \right)
\end{equation}
is the matrix of weighted collisional transition rates,
\begin{equation}
    \mathsf{A} =  
    \left(
    \begin{array}{ccc}
        0 & -A_{\ion{C}{i},10} & -A_{\ion{C}{i},20} \\
        0 & A_{\ion{C}{i},10} & -A_{\ion{C}{i},21} \\
        0 & 0 & A_{\ion{C}{i},20} + A_{\ion{C}{i},21}
    \end{array}
    \right)
\end{equation}
is the matrix of radiative transition rates, $\mathbf{x}_{\ion{C}{i}} = (x_{\ion{C}{i},0}, x_{\ion{C}{i},1}, x_{\ion{C}{i},2})$ is a vector of the number fractions in the three fine structure sub-states (whose sum is constrained by $x_{\ion{C}{i},0}+ x_{\ion{C}{i},1}+ x_{\ion{C}{i},2} = x_{\ion{C}{i}}$), and $\mathbf{0}$ is the zero vector. The $\overline{k}_{\ion{C}{i},nm}$ and $A_{\ion{C}{i},nm}$ terms are the number fraction-weighted collisional rate coefficients and Einstein $A$ coefficients for transition from state $n$ to state $m$, respectively. The expression for \ion{O}{i} is analogous. We again use temperature-dependent collision rates for collisions of \ion{C}{i} or \ion{O}{i} with H, He and ortho/para-H$_2$ taken from the LAMDA database \citep{1977A&A....56..289L,1991JPhB...24.2487S,1991JPhB...24.2343S,2018MNRAS.474.2313L}. Once the level populations are known from the solution to \autoref{eq:CIrateqn}, the cooling rate per unit mass can then be written as
\begin{equation}
\Lambda_{\ion{C}{i}} = -\frac{k_{\rm B}}{\mu_{\rm{H}} m_{\rm{H}}} \sum_{i=1}^2 x_{\ion{C}{i},i} \sum_{j=0}^{i-1} A_{\ion{C}{i},i,j} T_{\ion{C}{i},i,j}.
\label{eq:C_O_cool}
\end{equation}
The procedure for computing $\Lambda_{\ion{O}{i}}$ is analogous.

\textit{CO cooling.} CO cooling is significantly more complex, because we cannot assume for it, as we do for the C and O fine structure lines, that the optical depth is small. We therefore handle CO using the software library \texttt{DESPOTIC} to estimate cooling rates and optical depths for molecular clouds of finite optical depth using the escape probability approximation \citep{2014MNRAS.437.1662K}. For this purpose, we create a 4D grid in $n_{\rm{H}}$, $T_{\rm{g}}$, CO column density $N_{\rm{CO}}$, and $\sigma_{\rm{v}}$, and interpolate across the grid to find the cooling rate due to CO (from all $J$ states) for a given molecular cloud. We estimate the column density of CO for our model clouds as $N_{\rm{CO}} = x_{\rm{CO}} \Sigma / (\mu_{\rm{H}} m_{\rm{H}})$, with $x_{\rm{CO}}$ given by \autoref{eq:chemicalcomposition_CO}.

\textit{H$_2$ cooling.} To implement H$_2$ cooling, we write the cooling rate per unit mass as
\begin{equation}
    \Lambda_{\mathrm{H_2}} = -\Lambda_{\mathrm{H_2,thin}}\frac{x_{\rm{H_2}}}{n_{\rm{H_2}} \mu_{\rm{H}} m_{\rm{H}}}    \times \mathrm{min}\,\left[1,\left(\frac{n}{8\times 10^9\,\mathrm{cm^{-3}}}\right)^{-0.45}\right]\,,
\label{eq:h2}
\end{equation}
where $\Lambda_{\mathrm{H_2,thin}}$ is the cooling rate in the optically thin regime assuming local thermodynamic equilibrium (LTE) that we obtain from \citet[equation~9]{2014MNRAS.439.2386G}. The factor in the square parentheses in \autoref{eq:h2} accounts for a reduction in H$_2$ cooling due to optical depth effects at high densities, following \cite{2004MNRAS.348.1019R}. Note that the optical depth correction we use is only approximate, and can deviate from a more accurate implementation especially at very high densities \citep{2006ApJ...652....6Y,2015ApJ...799..114H}. This is not a problem; as we show below, the densities in which we are interested are $\lesssim 10^{9}\,\rm{cm^{-3}}$ for all $P$, $\sigma_{\rm{v}}$ and $Z$.

\textit{HD cooling.} We follow \cite{2005MNRAS.361..850L} to calculate cooling due to HD as a function of $T_{\rm{g}}$, $n_{\rm{H}}$ and the number fraction of HD molecules that we approximate as $x_{\rm{HD}} = \mathrm{[D/H]}f_{\rm{HD}}$, where $\rm{[D/H]} \approx 10^{-5}$ \citep{1996Natur.381..207T,2001ApJ...552..718O,2014ApJ...781...31C}. Here, $f_{\rm{HD}}$ represents the fraction of deuterium that exists in the form of HD. Typically, $f_{\rm{HD}} \sim 10^{-3}$ at $Z < 10^{-5}\,\rm{Z_{\odot}}$ but it can rise to 10 per cent at $Z \sim 10^{-4}\,\rm{Z_{\odot}}$ \citep[see figure 4 of][]{2005ApJ...626..627O}. While such elevated HD abundances do not significantly impact the characteristic mass at low metallicities where HD cooling can be important, we nonetheless fix $f_{\rm{HD}} = 0.1$, thus including the maximum possible contribution of HD cooling at all $Z$.

\textit{Consistency check.} As a necessary check, we verify that the ratio of luminosity due to cooling radiation by all these processes to the luminosity due to blackbody cooling radiation never exceeds unity. This ratio is
\begin{equation}
    \mathcal{L} = -\frac{4 \pi r^3 \rho(r) \left(\Lambda_{\rm{M}}+\Lambda_{\rm{H_2}}+\Lambda_{\rm{HD}}-\min(\Psi_{\rm{gd}},0)\right)}{4 \pi r^2 \sigma_{\rm{SB}} T^4_{\rm{g}}(r)}
\label{eq:lumiratio}
\end{equation}
where the minimum operator in the final term is to ensure that we include the contribution of dust only when it is a source of cooling, not a source of heating. Below we compute a critical radius that sets the characteristic mass, and we find that $\mathcal{L} < 0.004$ at this radius and everywhere outside it throughout our model grid for all models we discuss in this work.

\subsection{Characteristic Stellar Mass}
\label{s:charmass}
Now that we have found the gas temperature profile, we are in a position to estimate how fragmentation will proceed, and what characteristic stellar mass, $M_{\rm ch}$, it will produce. We do so following the \textit{ansatz} proposed by \citet{2011ApJ...743..110K}, and which direct tests against radiation-magnetohydrodynamic simulations \citep{2016MNRAS.460.3272K,2018MNRAS.476..771C} have shown is reasonably accurate: we consider a spherical region of radius $r$ around the forming protostar, and within this region we compute both the mass enclosed
\begin{equation}
    M_{\rm{enc}}(r) = \frac{8 \pi \rho^2_0 R^3}{3\rho(r)}\,,
\label{eq:Menclosed}
\end{equation}
and the minimum mass required for the gas to be unstable against gravitational collapse, which is given by the Bonnor-Ebert mass \citep{1955ZA.....37..217E,1956MNRAS.116..351B,1957MNRAS.117..104B}
\begin{equation}
    M_{\rm{BE}}(r) = 1.18\times \epsilon_{\rm M} \sqrt{\frac{\left(k_{\rm{B}} T_{\rm{g}}(r)/G\mu m_{\rm{H}}\right)^3}{\rho(r)}}\,,
\label{eq:BEmass}    
\end{equation}
where the factor of $\epsilon_{\rm M}$ is to account for the fraction of the mass that survives being blown out by protostellar jets. The enclosed mass scales as $M_{\rm enc}\propto \rho^{-1}$, while the Bonnor-Ebert mass scales as $T_{\rm g}^{3/2}\rho^{-1/2}$. So, for small radii, where $\rho$ and $T_{\rm g}$ are both large, we have $M_{\rm enc} \ll M_{\rm BE}$. The physical meaning of this condition is that close to a protostar, the gas is prevented from fragmenting because its mass is unable to overcome thermal pressure support. It can accrete onto the already-existing protostar (or ejected as part of an outflow), but it cannot collapse to produce a new protostar. As one considers larger and larger regions, $\rho$ and $T_{\rm g}$ both fall, such that there is a critical radius at which $M_{\rm enc} = M_{\rm BE}$, meaning that the enclosed mass is large enough to be unstable. Our \textit{ansatz} is that this transition gives the approximate characteristic mass at which fragmentation can occur. Consequently, the characteristic mass $M_{\rm{ch}}$ can be read off at the location where $M_{\rm{enc}}=M_{\rm{BE}}$. Since we have specified $\rho(r)$ and $T_{\rm g}(r)$, we can solve for $M_{\rm ch}$ from any cloud of specified physical parameters -- $\sigma_{\rm v}$ and $P$ -- and chemical composition -- as parameterised by $Z$. Our goal in this work is to study how $M_{\rm{ch}}$ changes as we vary these parameters, particularly $Z$, and to understand which physical processes are responsible for driving these changes. 

It is important to highlight how the formation of binary stars fits into our cloud configuration. It is well known that binaries form via two main modes -- core fragmentation \citep[e.g.,][]{1991MNRAS.249..588C,2004A&A...414..633G,2004ApJ...600..769F,2009ApJ...703..131O,2010ApJ...725.1485O} and disc fragmentation \citep[e.g.,][]{1989ApJ...347..959A,1994MNRAS.269L..45B,2006MNRAS.373.1563K,2009Sci...323..754K,2017MNRAS.468.4093G}. However, it is not yet clear which mode of binary formation dominates near the characteristic stellar mass that sets the peak of the Galactic IMF (see, for example, \citealt{2010ApJ...725.1485O} versus \citealt{2020MNRAS.491.5158T}), with currently available observations providing support for both scenarios \citep{2011ApJ...731....8K,2016ApJ...818...73T,2017NatAs...1E.172L}. The same conundrum also holds at extremely low or zero metallicity, where simulations find a ubiquitous presence of binary systems due to both modes of fragmentation \citep{2009MNRAS.399.1255M,2015MNRAS.448.1405M,2019MNRAS.488.2658C,2019MNRAS.490..513S,2020MNRAS.497..336S,2020ApJ...892L..14S,2021arXiv210304997C,2021MNRAS.501..643L}, with a slight preference for disc fragmentation over core fragmentation \citep{2020arXiv200806107C}. Our model naturally incorporates the core fragmentation mode for binary formation but not the disc fragmentation mode. Thus, to the extent that core fragmentation dominates, we are able to predict the characteristic stellar mass of single star IMF. On the other hand, our predictions apply for the system IMF in the case disc fragmentation dominates binary star formation.

\begin{figure}
\includegraphics[width=1.0\columnwidth]{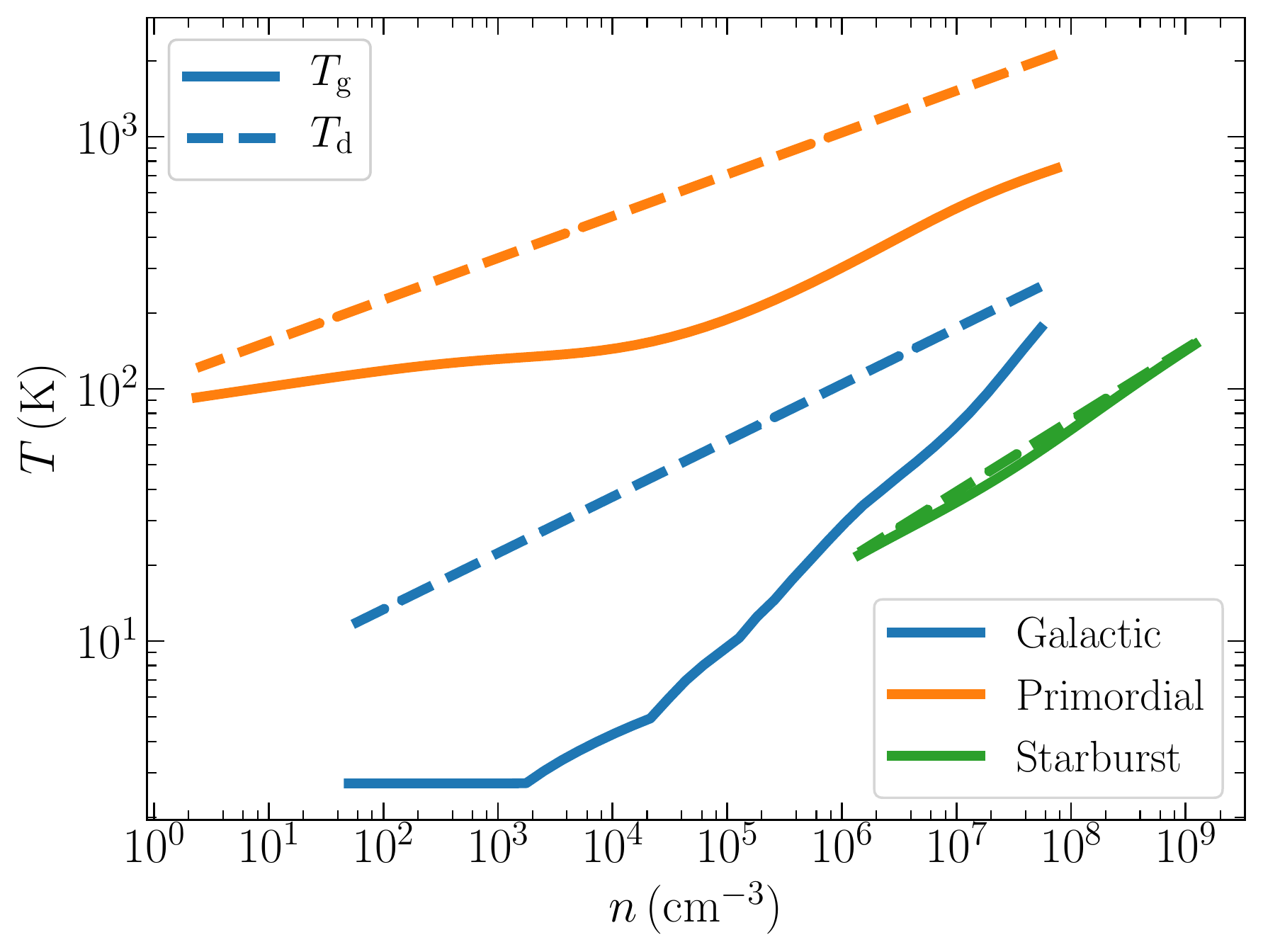}
\caption{Gas and dust temperatures ($T_{\rm{g}}-$ solid, and $T_{\rm{d}}-$ dashed) as a function of free particle number density $n$ in a molecular cloud in three different environments: Galactic (pressure $P/k_{\rm{B}}=10^4\,\rm{K\,cm^{-3}}$, velocity dispersion $\sigma_{\rm{v}} = 5\,\rm{km\,s^{-1}}$, metallicity $Z = \rm{Z_{\odot}}$), primordial (same as Galactic but with $Z = 10^{-6}\,\rm{Z_{\odot}}$), and starburst ($P/k_{\rm{B}}=10^8\,\rm{K\,cm^{-3}}$, $\sigma_{\rm{v}} = 0.5\,\rm{km\,s^{-1}}$, $Z = \rm{Z_{\odot}}$). Dust acts as a heating source for the gas if $T_{\rm{d}} > T_{\rm{g}}$, and vice-versa; it is unimportant for setting gas temperature in the primordial case due to the near-zero dust abundance.}
\label{fig:radialprofile_TgasTdust}
\end{figure}

\begin{figure}
\includegraphics[width=1.0\columnwidth]{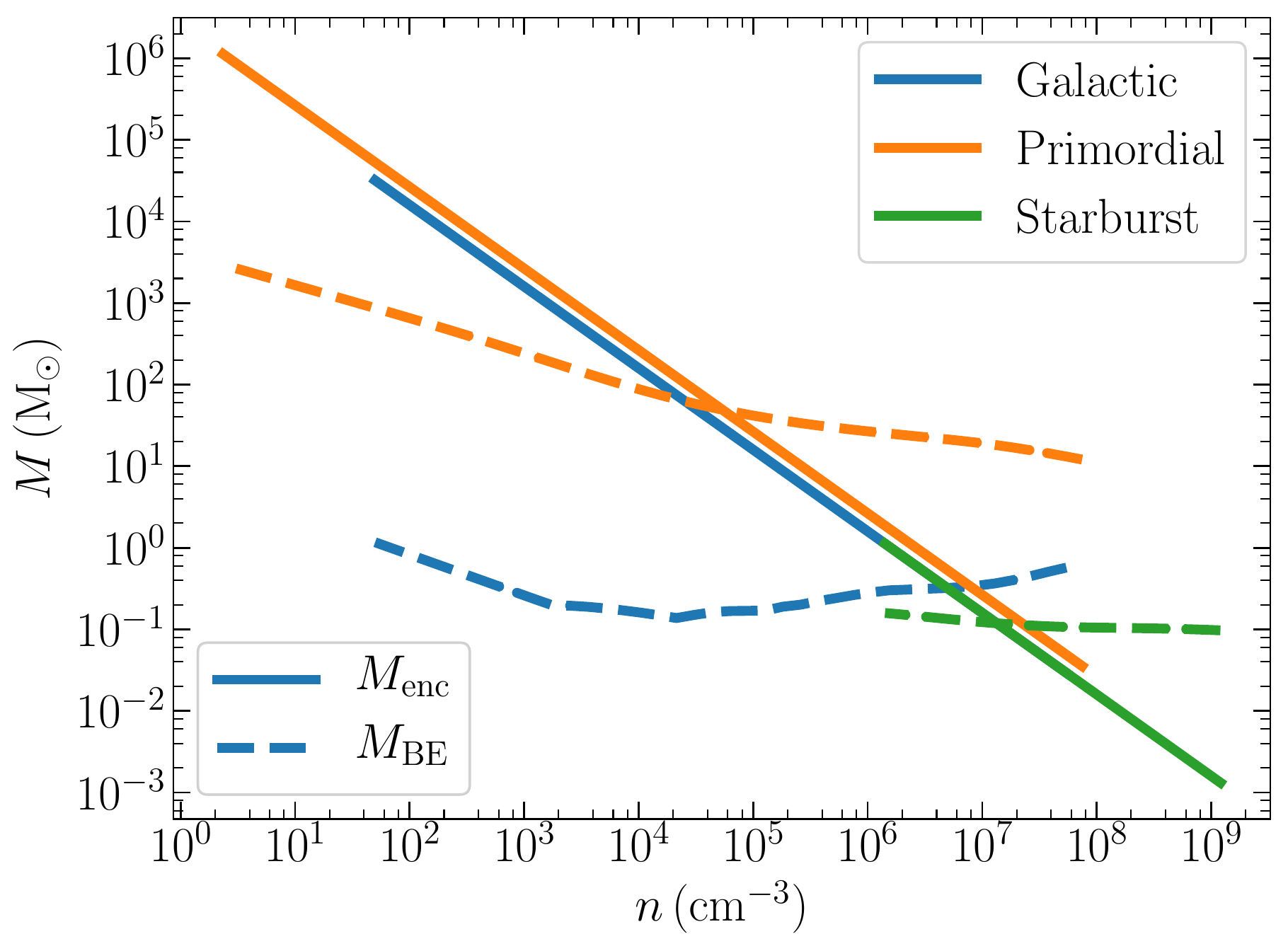}
\caption{Mass enclosed around a protostar ($M_{\rm{enc}}$, \autoref{eq:Menclosed}) and the Bonnor-Ebert mass ($M_{\rm{BE}}$, \autoref{eq:BEmass}) as a function of free particle number density $n$ in a typical molecular cloud in different environments (Galactic, primordial, starburst) as in \autoref{fig:radialprofile_TgasTdust}. The characteristic stellar mass $M_{\rm{ch}}$ can be read off at the critical density $n_{\rm{crit}}$ where $M_{\rm{enc}}=M_{\rm{BE}}$.}
\label{fig:radialprofile_Mch_n}
\end{figure}

\section{Results}
\label{s:results}
With the procedure complete, we can now look at the resulting models. We begin in \autoref{s:results_radialprofiles} by walking through the procedure using some example models, to orient the reader and provide some intuition for the workings of the model. We then explore variations in the gas and dust temperatures across parameter space in \autoref{s:results_gasdusttemperature}, the physical processes driving these variations in \autoref{s:results_gasthermalbalance}, and finally the consequences for the characteristic stellar mass and the IMF in \autoref{s:results_IMF}. We remind the reader that in this work we only focus on the characteristic stellar mass that sets the peak of the IMF. This is not sufficient by itself to fully describe the IMF since it consists of several other features (for example, the slopes at the low and the high-mass end, and the truncation mass) that we do not investigate in this work.

\subsection{Profiles}
\label{s:results_radialprofiles}

We begin by presenting results for three example cases: a typical Galactic molecular cloud of pressure $P/k_{\rm{B}} = 10^{4}\,\rm{K\,cm^{-3}}$, effective velocity dispersion $\sigma_{\rm{v}} = 5\,\rm{km\,s^{-1}}$, and metallicity $Z = \rm{Z_{\odot}}$, a near-primordial cloud with the same $P$ and $\sigma_{\rm v}$ but $Z = 10^{-6}Z_\odot$, and a very compact cloud in a high-pressure starburst environment, with $P/k_{\rm{B}} = 10^{8}\,\rm{K\,cm^{-3}}$, $\sigma_{\rm v} = 0.5$ km s$^{-1}$, and $Z=Z_\odot$. In all cases we use our fiducial choice of chemical composition for the corresponding metallicity, though, as we will show below, the qualitative results are insensitive to this.

\autoref{fig:radialprofile_TgasTdust} shows the profile of gas and the dust temperature as a function of density of free particles $n$ that we obtain in each of the three cases following the theoretical framework we outline in \autoref{s:theory}. For the Galactic cloud, dust acts as a heating source throughout the cloud as it has been warmed by the central protostar; however, the gas and dust do not reach close to equality until the density rises to almost $10^6$ cm$^{-3}$. By contrast, in the primordial case, the role of dust in gas thermodynamics becomes negligible, so even though $T_{\rm{d}} > 480\,\rm{K}$ across some part of the cloud (likely invalidating the \citealt{2005ApJ...631..792C} model, since its assumption of a powerlaw dust opacity fails at such high temperatures), this makes no difference to the gas temperature, which is entirely set by H$_2$/HD cooling. Finally, in the starburst case, the density is so high that the dust and gas temperatures are very well coupled to each other throughout the cloud (shown by the overlapping solid and dashed green curves in \autoref{fig:radialprofile_TgasTdust}), and dust completely dominates gas thermodynamics. However, note that dust properties (e.g., surface area per unit mass) likely vary at least somewhat with physical conditions and metallicity, and we have ignored this effect thus far. We discuss this further in \autoref{s:discussions_otherparams}.

\autoref{fig:radialprofile_Mch_n} shows the radial profiles of $M_{\rm{BE}}$ and $M_{\rm{enc}}$ for the three example cases. As expected, we observe that in all cases the mass enclosed at high densities (corresponding to small radii) is much too small to collapse close to the protostar, i.e., $M_{\rm enc} \ll M_{\rm BE}$, but becomes unstable to collapse as one moves outward (away from the protostar). The density and mass at which this changeover occurs is different in the three examples. For the Galactic case, we find $M_{\rm enc} = M_{\rm BE}$ at $n\approx 6\times10^6$ cm$^{-3}$. The resulting characteristic stellar mass in this case is $M_{\rm{ch}} = 0.31\,\rm{M_{\odot}}$, within $\approx 10$ per cent of the observed the peak of the Milky Way stellar IMF \citep{2001MNRAS.322..231K,2003PASP..115..763C}. Our model therefore naturally results in a characteristic stellar mass that is sub-solar at $Z=\rm{Z_{\odot}}$, thereby resulting in a bottom-heavy Milky Way stellar IMF. This exercise also demonstrates that protostellar feedback is important to reproduce the peak of the IMF, at least at Solar metallicity and low pressure, consistent with findings from several numerical simulations \citep{2009ApJ...703..131O,2010ApJ...713.1120K,2010ApJ...710.1343U,2014MNRAS.439.3420M,2021arXiv210606521M}. 

\begin{figure}
\includegraphics[width=1.0\columnwidth]{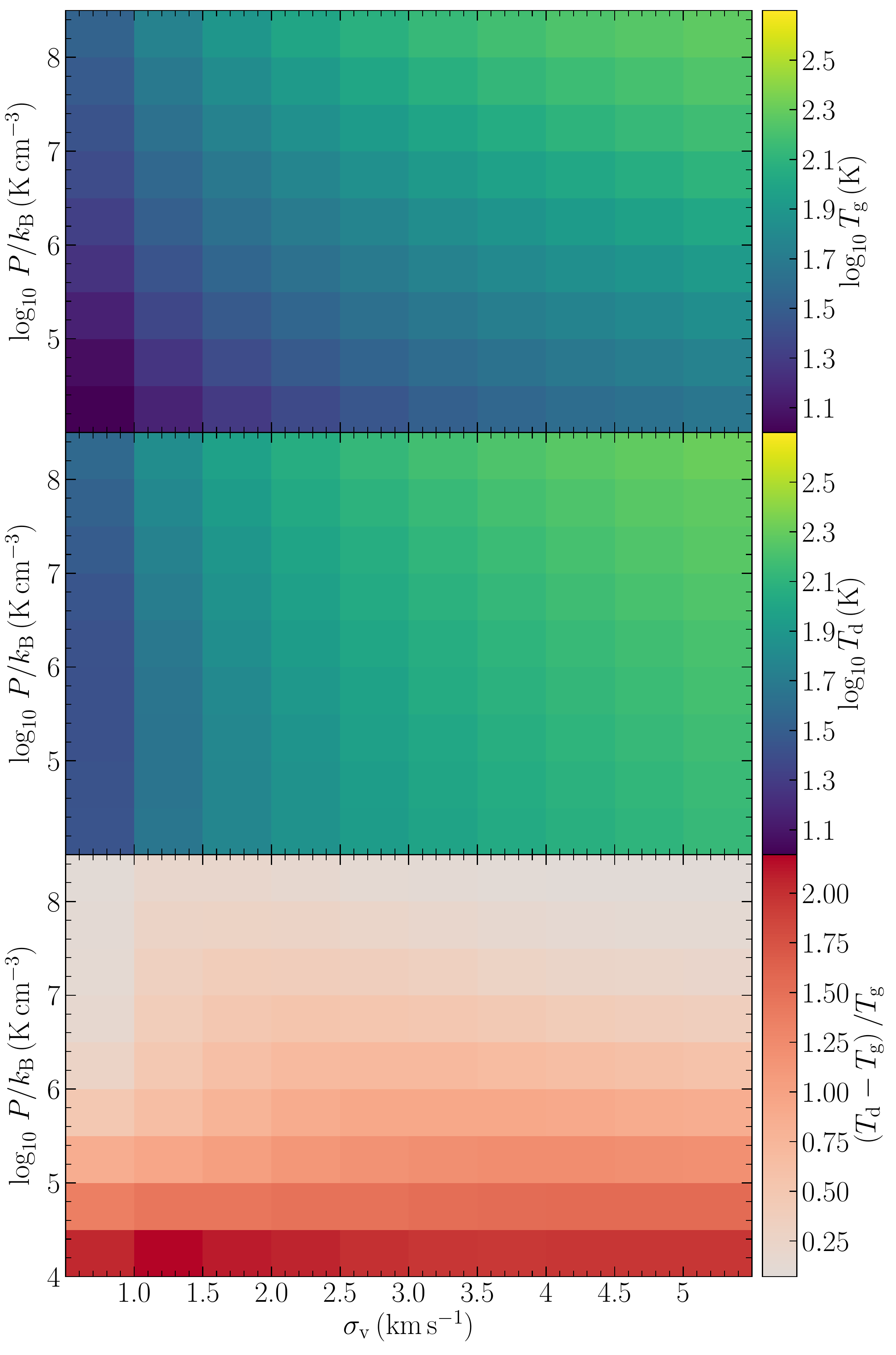}
\caption{2D plots of the gas and dust temperatures at the critical location that sets the characteristic mass as a function of cloud pressure $P$ and velocity dispersion $\sigma_{\rm{v}}$ for $Z=\rm{Z_{\odot}}$. Dust plays a crucial role in setting the gas temperature in this case, and is well-coupled to the gas.}
\label{fig:gdt_2D}
\end{figure}

\begin{figure}
\includegraphics[width=1.0\columnwidth]{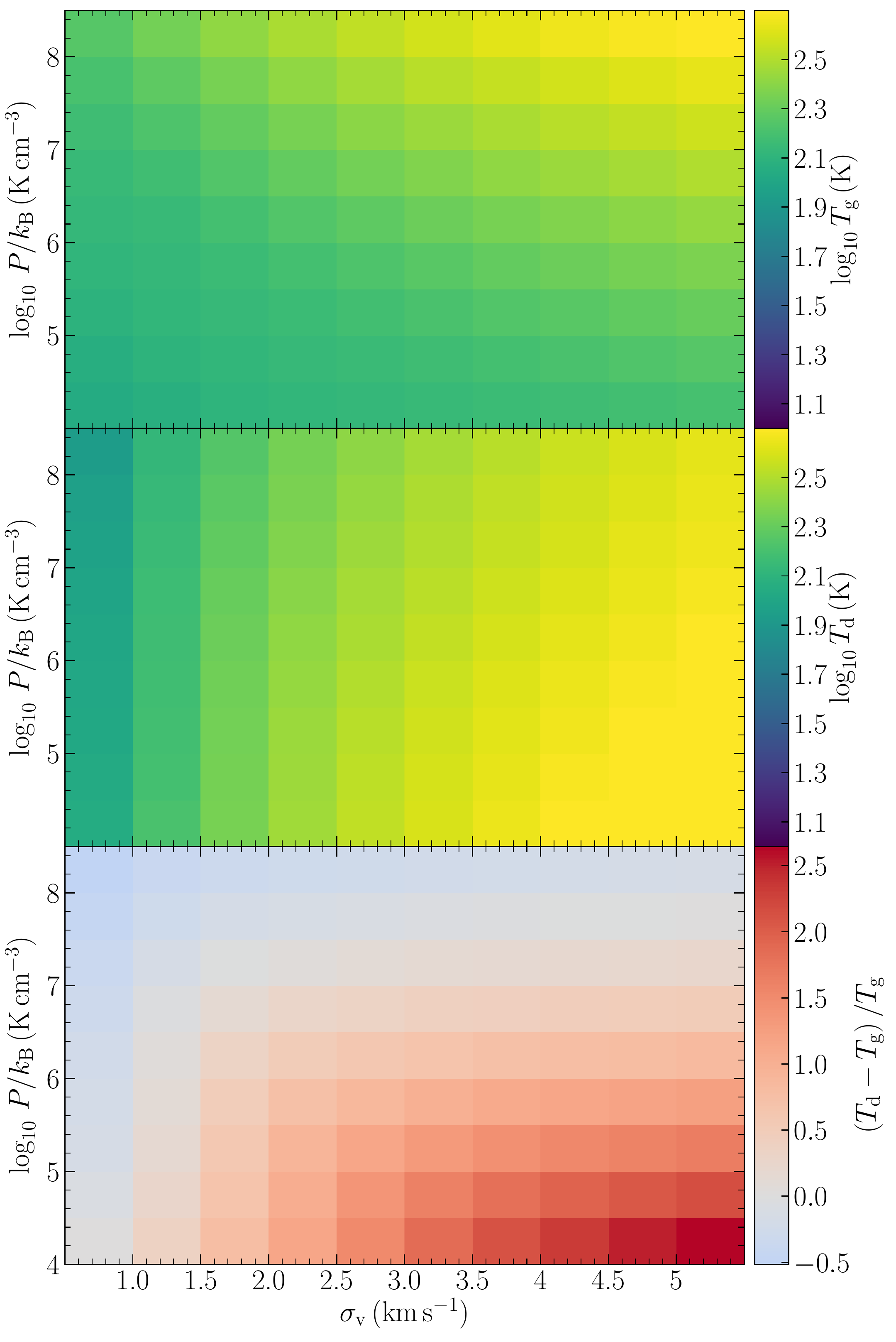}
\caption{Same as \autoref{fig:gdt_2D} but for an extremely metal-poor environment ($Z = 10^{-6}\,\rm{Z_{\odot}}$). Here, dust can act as both a heating and a cooling source for the gas, but is generally unimportant. Note the difference in colour scale in the bottom panel here as compared to \autoref{fig:gdt_2D}.}
\label{fig:gdt_2D_lowZ}
\end{figure}

For the primordial case, due to the high gas temperature, the enclosed mass only becomes sufficient to collapse by itself much farther out in the cloud (at low $n$), yielding $M_{\rm{ch}}=60\,\rm{M_{\odot}}$. Thus, primordial-like star formation naturally gives rise to a super-solar characteristic mass, thereby hinting at the formation of a top-heavy IMF in primordial-like environments. Finally, for the starburst case, $M_{\rm BE}$ is nearly independent of $n$ due to the combination of feedback and strong dust-gas coupling; feedback yields a dust temperature, and thus a gas temperature, that falls with radius in such a way as to keep $\sqrt{T^3/\rho}\propto M_{\rm BE}$ nearly constant. The resulting $M_{\rm{ch}}$ is $0.11\,\rm{M_{\odot}}$, close to the Galactic case but smaller by a factor of $\approx 3$. We revisit this finding in the context of the IMF in the centres of massive elliptical galaxies in \autoref{s:discussions_earlytypeSF}.

\begin{figure}
\includegraphics[width=1.0\columnwidth]{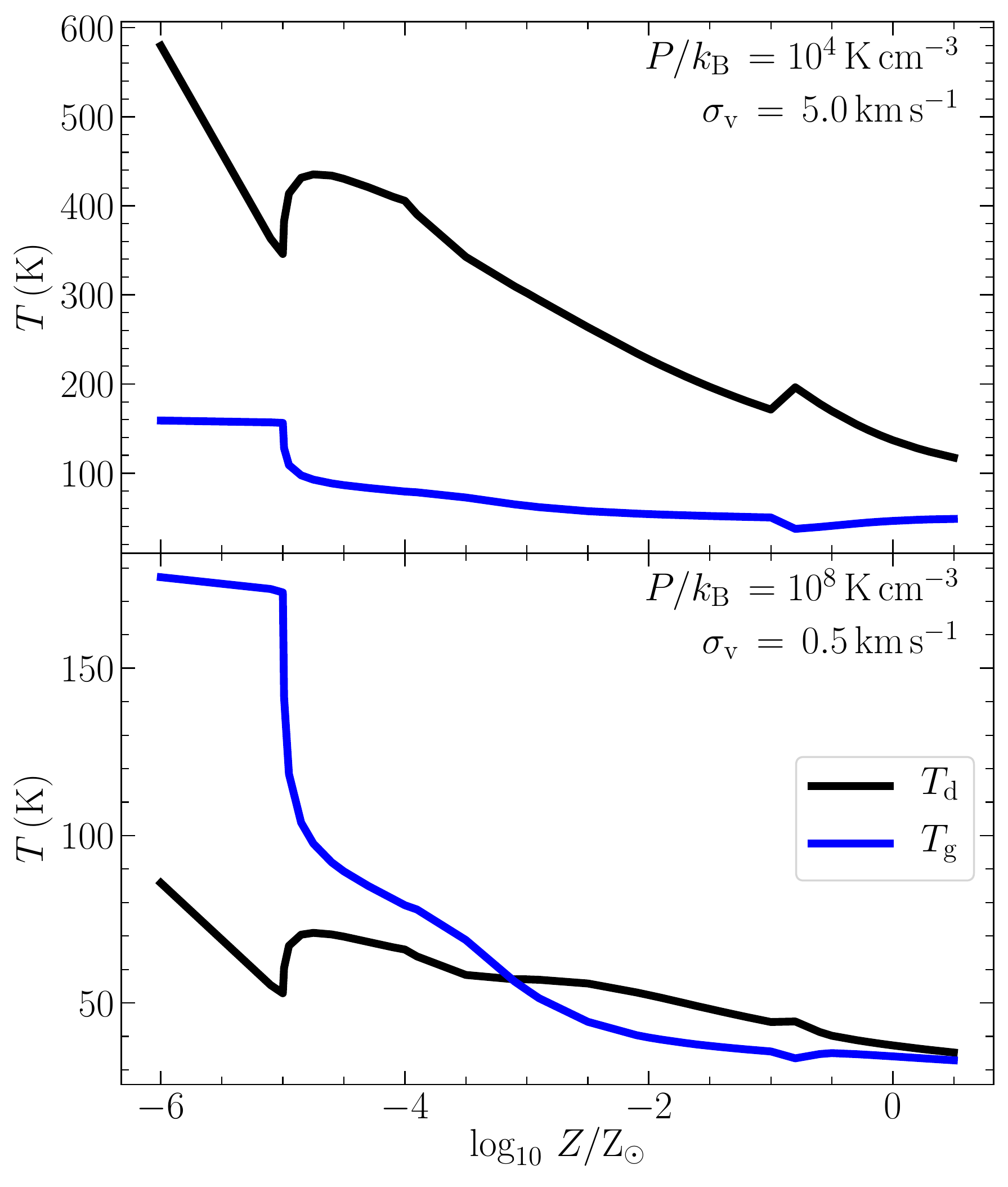}
\caption{\textit{Top panel:} Gas and dust temperatures ($T_{\rm{g}}$ and $T_{\rm{d}}$) at the location of the characteristic stellar mass $M_{\rm{ch}}$ as a function of metallicity $Z$ for a Galactic-type environment with pressure $P/k_{\rm{B}}=10^4\,\rm{K\,cm^{-3}}$ and velocity dispersion $\sigma_{\rm{v}}=5\,\rm{km\,s^{-1}}$. \textit{Bottom panel:} Same as the top panel but for a starburst environment with $P/k_{\rm{B}}=10^8\,\rm{K\,cm^{-3}}$ and velocity dispersion $\sigma_{\rm{v}}=0.5\,\rm{km\,s^{-1}}$.}
\label{fig:gdt_Zprofile}
\end{figure}

\subsection{Gas and dust temperatures}
\label{s:results_gasdusttemperature}

Having built some intuition for the functioning of the models, we now begin to examine results across parameter space. We first seek to answer two related questions: how well-coupled are the dust and the gas? And where in parameter space is dust a source of cooling versus a source of heating, or where is it unimportant? Since our parameter space is three-dimensional \{$P,\sigma_{\rm{v}},Z$\}, we address these questions by making slices through it. We first look at the 2D space in \{$P,\,\sigma_{\rm{v}}$\} for the metallicities $Z=Z_\odot$ and $Z=10^{-6}Z_\odot$. For each point in this parameter space, we find the location where the enclosed mass equals the Bonnor-Ebert mass (i.e., the location at which $M_{\rm ch}$ is determined) and measure the dust and gas temperatures there. \autoref{fig:gdt_2D} and \autoref{fig:gdt_2D_lowZ} show the results. At $Z=\rm{Z_{\odot}}$, we see that the gas and dust temperatures are well coupled to each other for all $P$ and $\sigma_{\rm{v}}$, consistent with our findings above that dust plays a crucial role in setting the gas temperature at high $Z$. Both the temperatures increase in tandem if either $P$ or $\sigma_{\rm{v}}$ is increased. On the other hand, the gas temperatures are much less sensitive to $P$ or $\sigma_{\rm{v}}$ for $Z=10^{-6}\,\rm{Z_{\odot}}$. The dust temperatures do change, and dust can both heat and cool the gas in this case, but its impact on the gas temperature is negligible.

We next explore variation with metallicity. \autoref{fig:gdt_Zprofile} plots the dust and gas temperatures as a function of $Z$ at two choices of fixed $P$ and $\sigma_{\rm{v}}$, one corresponding to a low-density, Galactic-type case ($P/k_{\rm B} = 10^4$ K cm$^{-3}$, $\sigma_{\rm v}=5$ km s$^{-1}$, and one to a dense, starburst-type environment ($P/k_{\rm B} = 10^8$ K cm$^{-3}$, $\sigma_{\rm v} = 0.5$ km s$^{-1}$). For the former, dust acts as a heating source for the gas since $T_{\rm{d}} > T_{\rm{g}}$ at all $Z$; however, this heating source becomes more and more feeble as we go to lower $Z$, such that the dust and gas temperatures begin to diverge. For the latter case, we see a more non-linear behavior. \autoref{fig:gdt_Zprofile} shows that dust and gas temperatures are very well-coupled at high metallicities ($Z > 10^{-0.5}\,\rm{Z_{\odot}}$), but $T_{\rm{g}}<T_{\rm{d}}$ between $10^{-3.3} \leq Z/\rm{Z_{\odot}} \leq 10^{-0.5}$, implying that dust begins to act as a heating source for these metallicities. At $Z < 10^{-3.3}\,\rm{Z_{\odot}}$, dust begins to act as a cooling source for the gas. However the effects of dust cooling at such low metallicities are rather limited compared to metal and H$_2$/HD cooling, a topic we explore below in \autoref{s:results_gasthermalbalance}.

\subsection{Gas thermal balance}
\label{s:results_gasthermalbalance}

We have already seen that, depending on the metallicity and environment, dust can be both a source of heating and of cooling, and the gas and dust temperatures can be closely locked or widely divergent. We now seek to place dust in the broader context of other heating and cooling mechanisms. To quantify the relative importance of all the various heating and cooling terms, we define 
\begin{equation}
\mathcal{C}(\eta) = \frac{2 \eta}{|\Gamma_{\rm{c}}| + |\Gamma_{\rm{gd}}| + |\Lambda_{\rm gd}| + |\Lambda_{\rm{M}}| + |\Lambda_{\rm{H_2}}| + |\Lambda_{\rm{HD}}|}
\label{eq:term_contribution}
\end{equation}
where $\mathcal{C}(\eta)$ represents the relative contribution of each term $\eta$ that heats or cools the gas, evaluated at the location in the model cloud where $M_{\rm enc} = M_{\rm BE}$, i.e., at the point that determines how the gas fragments. Thus, for example, $\mathcal{C}({\rm c})$ is the ratio of the compressive heating rate to the sum of the absolute values of all heating and cooling rates. By construction $-1 \leq \mathcal{C}(\eta) \leq 1$. Terms for which $\mathcal{C} < 0$ represent processes that cool the gas, and vice-versa; the sum of $\mathcal{C}(\eta)$ over all cooling terms is $-1$, and the sum over all heating terms is $+1$, giving $\mathcal{C}(\eta)$ an easy physical interpretation: it is the fraction of the total heating or cooling provided by some particular mechanism, such that mechanisms $\eta$ for which $|\mathcal{C}(\eta)|$ is close to unity are dominant, while those for which is it close to zero do not play a significant role in gas thermodynamics. Note that, in computing $\mathcal{C}$, we group all cooling by metal line emission (i.e., via all lines of \ion{C}{i}, \ion{C}{ii}, \ion{O}{i}, and CO) into a single term $\Lambda_{\rm M}$, and we separate the dust-gas energy exchange ($\Psi_{\rm{gd}}$) into its corresponding heating ($\Gamma_{\rm{gd}}$) and cooling ($\Lambda_{\rm{gd}}$) parts; the latter is critical to ensuring that $\mathcal{C}$ properly captures situations such as the one we have already encountered in starburst conditions, where dust and gas are so tightly coupled that the gas temperature is set almost entirely by the balance between $\Gamma_{\rm gd}$ and $\Lambda_{\rm gd}$, and other mechanisms are unimportant.

\begin{figure*}
\includegraphics[width=1.0\linewidth]{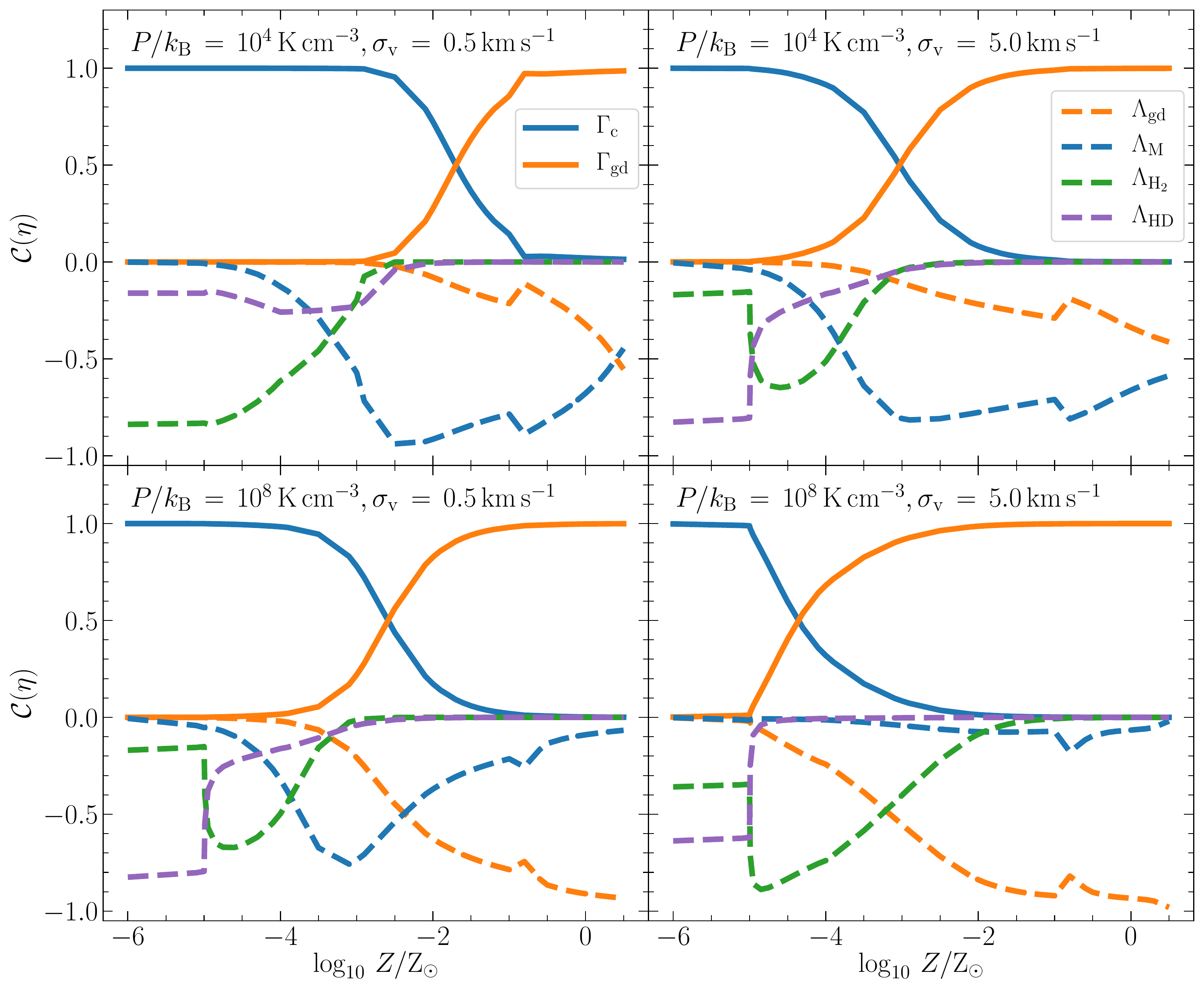}
\caption{\textit{Top panel:} Importance of different heating and cooling processes under thermal balance as a function of metallicity $Z$ (normalised to Solar) for a given $P/k_{\rm{B}}=10^4\,\rm{K\,cm^{-3}}$ and effective velocity dispersion $\sigma_{\rm{v}}=0.5$ and $5\,\rm{km\,s^{-1}}$. $\mathcal{C}(\eta)$ is calculated for each term $\eta$ as shown in the legend following \autoref{eq:term_contribution}. Solid curves show all the heating processes ($\Gamma_{\rm{c}}$ -- adiabatic compression, $\Gamma_{\rm{gd}}$ -- heating due to dust-gas coupling) that have $\mathcal{C}(\eta)>0$ and dashed curves show all the cooling processes ($\Lambda_{\rm{gd}}$ -- cooling due to dust-gas coupling, $\Lambda_{\rm{M}}$ -- cooling due to metals, $\Lambda_{\rm{H_2}}$ -- cooling due to H$_2$, $\Lambda_{\rm{HD}}$ -- cooling due to HD) that have $\mathcal{C}(\eta)<0$. The sum of the magnitudes of all heating and all cooling processes separately is unity. Values of $|\mathcal{C}(\eta)|$ close to unity indicate a dominant process, values close to zero indicate an unimportant process. \textit{Bottom panel:} Same as the top panel but for $P/k_{\rm{B}}=10^8\,\rm{K\,cm^{-3}}$.}
\label{fig:gtb_termcontribution}
\end{figure*}

In \autoref{fig:gtb_termcontribution}, we plot $\mathcal{C}(\eta)$ for all heating and cooling terms $\eta$ as a function of metallicity at values of $P/k_{\rm B} = 10^4$ and $10^8\,\rm{K\,cm^{-3}}$, and $\sigma_{\rm v} = 0.5$ and $5\,\rm{km\,s^{-1}}$ -- values that represent the outermost corners of our parameter space, though we remind the reader that the high $P$, high $\sigma_{\rm{v}}$ case is highly improbable. All these panels reveal a common trend: at very low metallicity, the most important heating process is adiabatic compression. The most important cooling processes are H$_2$ and (at the lowest densities) HD line emission; recall from \autoref{s:gatemp_cooling} that the HD cooling rate we use is an upper limit. At Solar or higher metallicity, by contrast, heating is invariably dominated by dust, which provides a conduit to deposit energy from accretion into the thermal reservoir of the gas. Cooling is also dominated by dust when the density is high (i.e., at low $\sigma_{\rm v}$ or high $P$), with metal line cooling reaching a maximum $\sim 50$ per cent contribution at the lowest-density corner of our grid.

Thus far these findings confirm earlier results on the thermodynamics of present-day (modern) and primordial star formation. However, from \autoref{fig:gtb_termcontribution} we can also see evidence for a distinct, transitionary star formation regime at intermediate metallicity ($-3.5 \lesssim \log_{10}Z/\rm{Z_{\odot}} \lesssim -1.5$), where cooling is dominated by metal line emission. The exact range in metallicity where this regime occurs is somewhat dependent on the physical conditions -- it extends to higher metallicity at low pressure than at high pressure. 

Nonetheless, the overall picture that emerges for gas thermal balance at the critical location that sets $M_{\rm{ch}}$ in the presence of protostellar feedback is that there are three regimes of star formation at all pressures: (1.) primordial-like at $ Z \lesssim 10^{-4}\,\rm{Z_{\odot}}$ where H$_2$ and/or HD cooling and adiabatic compression heating dominate, (2.) transitionary at $10^{-4} \lesssim Z/\rm{Z_{\odot}} \lesssim 10^{-2}$ where metal cooling dominates and dust and compression heating are comparable, and (3.) modern day-like at $Z \gtrsim 10^{-2}\,\rm{Z_{\odot}}$ where the gas temperature is mostly set by dust-gas coupling (with a contribution from metal line cooling in the lowest density regions), and the competition between dust cooling and heating. Since different heating and cooling mechanisms equilibrate at different gas temperatures, we expect the characteristic stellar mass to be different in the three different regimes we identify.

\subsection{Characteristic stellar mass and the IMF}
\label{s:results_IMF}
We now look at the resulting characteristic stellar mass, $M_{\rm{ch}}$, as a function of $Z$ in different star-forming environments. We estimate $M_{\rm{ch}}$ for each $Z$ based on the procedure we outline in \autoref{s:charmass}. As in the previous sections, we study the variations in $M_{\rm{ch}}$ as a function of $Z$ at different $P$ and $\sigma_{\rm{v}}$.

\begin{figure}
\includegraphics[width=1.0\columnwidth]{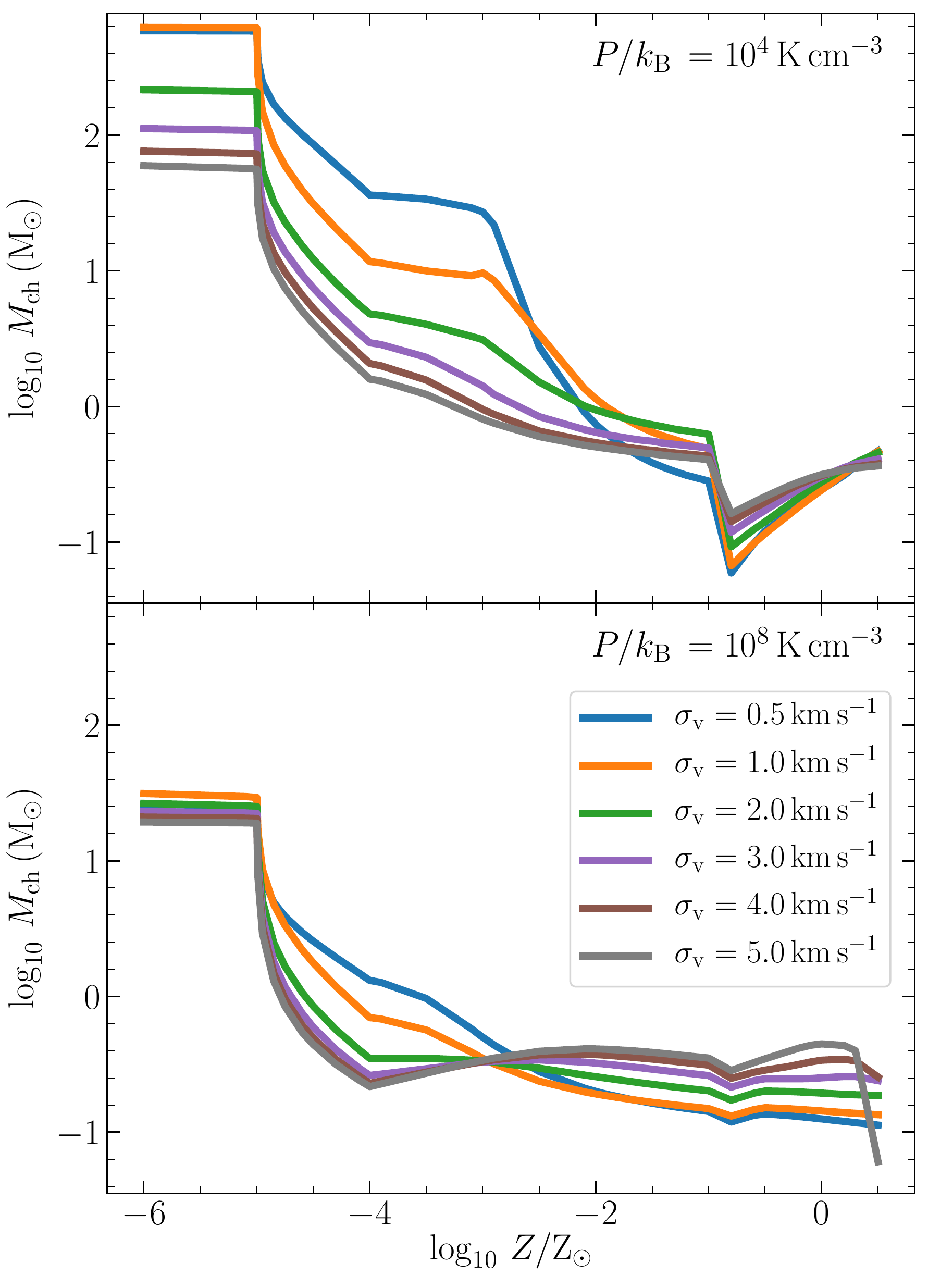}
\caption{\textit{Top panel:} Characteristic stellar mass, $M_{\rm{ch}}$, as a function of metallicity for a fixed cloud pressure $P/k_{\rm{B}}=10^4\,\rm{K\,cm^{-3}}$ at different effective velocity dispersions $\sigma_{\rm{v}}$ as shown in the legend. \textit{Bottom panel:} Same as the top panel but at a high pressure ($P/k_{\rm{B}}=10^8\,\rm{K\,cm^{-3}}$), typical of starburst environments.}
\label{fig:Mch}
\end{figure}

\autoref{fig:Mch} shows $M_{\rm{ch}}$ as a function of $Z$ for a range of $P$ and $\sigma_{\rm{v}}$ values. We see that at all combinations of $P$ and $\sigma_{\rm v}$ the characteristic stellar mass declines steadily from super-solar to sub-solar masses as the metallicity increases from near-zero to $\approx 10^{-2}\,\rm{Z_{\odot}}$. Above this metallicity, the characteristic mass either flattens or rises slightly with $Z$, depending on the choice of $P$ and $\sigma_{\rm v}$. The absolute value of $M_{\rm{ch}}$ is also fairly sensitive to the choice of $P$ and $\sigma_{\rm{v}}$ at low $Z$, but much less so at high $Z$. This is consistent with numerous simulations of Population III star formation that find a much broader mass spectrum than that in Population I star formation \citep{2011ApJ...727..110C,2014ApJ...792...32S,2021arXiv210304997C}. The exact location and depth of the inflection in $M_{\rm{ch}}$ at $10^{-2}\,\rm{Z_{\odot}}$ at low $P$ that delineates the transition from the primordial to the modern regime depends on our assumed model for chemical composition as a function of $Z$ (because CO provides more cooling than atomic lines at low temperature -- \citealt[figure~9]{2014MNRAS.437.1662K}), but the existence of this transition is independent of our chemical assumptions, as we show in \autoref{s:discussions_chemicalcompositions}. It is instead a result of the changeover from an H$_2$-dominated cooling regime to a metal line-dominated regime to a dust-dominated regime, which occurs regardless of the chemical state of C and O.

Another notable, though perhaps not surprising, result is that higher pressure environments favour lower mass stars. This implies that starburst environments at all $Z$ should contain more low mass stars (consistent with earlier results of \citealt{2014ApJ...796...75C} that did not include protostellar feedback), although the peak in very metal-poor environments ($Z \leq 10^{-4}\,\rm{Z_{\odot}}$) still remains top-heavy with $M_{\rm{ch}} \sim 15\,\rm{M_{\odot}}$. At solar metallicity the effects of pressure are much weaker, due to the dominance of dust in the thermodynamics, but we nonetheless find that starburst environments should typically have a slightly more bottom-heavy IMF than the average star-forming environment. We discuss this finding further in the context of massive elliptical galaxies in \autoref{s:discussions_earlytypeSF}.

Thus, we can now answer the question: when does the IMF become bottom-heavy? We find that the transition from a top-heavy to a bottom-heavy IMF occurs between $10^{-4} \leq Z/\rm{Z_{\odot}} \leq 10^{-2}$ at low $P$ (depending on the density), and around $Z \approx 10^{-4}\,\rm{Z_{\odot}}$ at high $P$ irrespective of the density. 

\begin{figure}
\includegraphics[width=1.0\columnwidth]{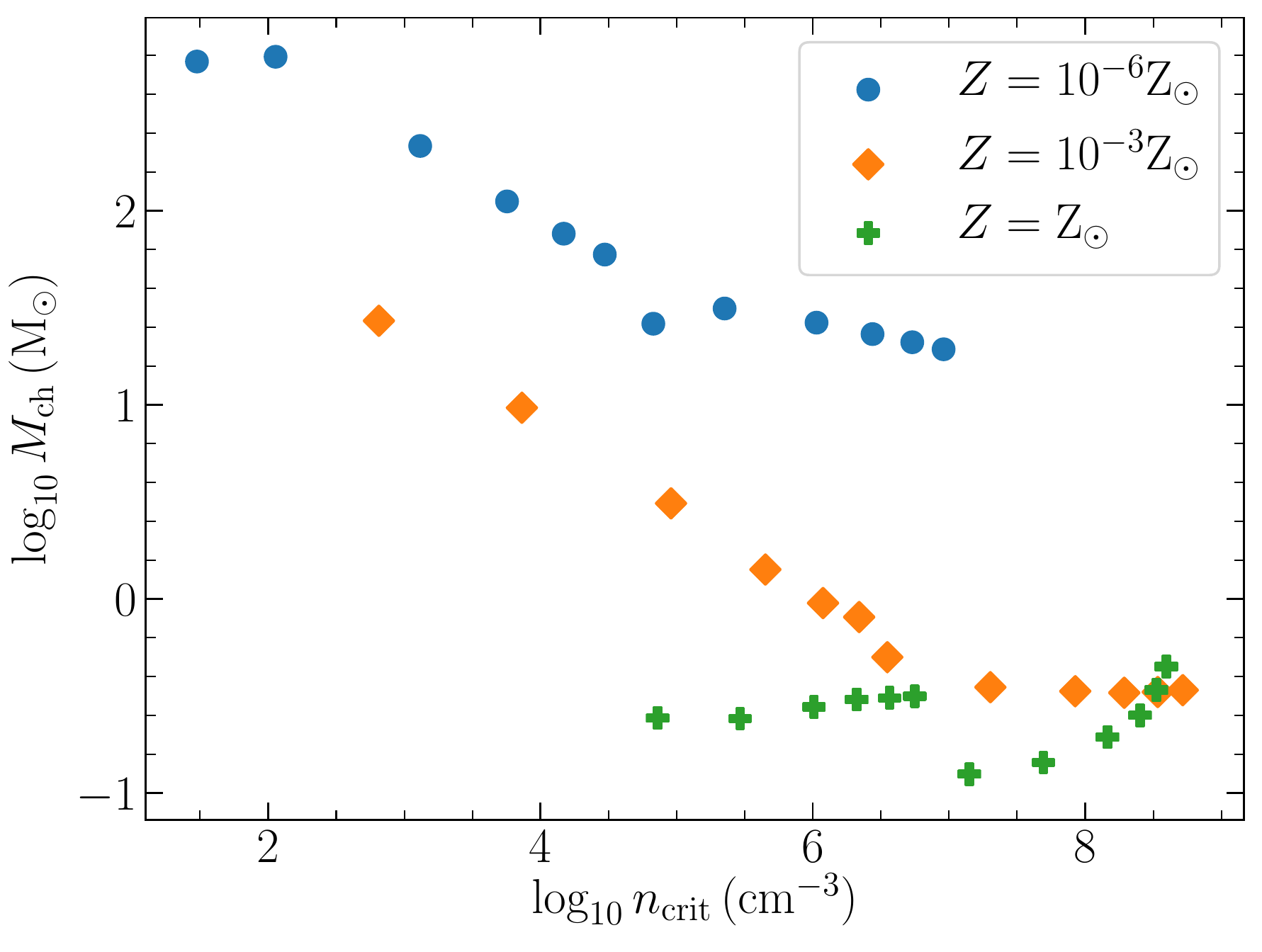}
\caption{Characteristic stellar mass as a function of the critical density (where $M_{\rm{enc}}=M_{\rm{BE}}$) at three different metallicities.}
\label{fig:Mch_ncrit}
\end{figure}

\autoref{fig:Mch_ncrit} plots the characteristic mass as a function of the critical density, \textit{i.e.,} density at the location where $M_{\rm{enc}}=M_{\rm{BE}}$ in the cloud, for three different metallicities. The apparent jitter in the curves arises from plotting all possible values of $P$ and $\sigma_{\rm{v}}$. We find that, at $Z = 10^{-6}\,\rm{Z_{\odot}}$, $M_{\rm{ch}}$ monotonically decreases as a function of $n_{\rm{crit}}$, which can be best fit by a linear function with a slope $\approx -0.3$. There is little evolution in the slope until the metallicity is high enough for dust to take control of gas thermodynamics, after which there is little variation in $M_{\rm{ch}}$ as a function of $n_{\rm{crit}}$.

\begin{figure*}
\includegraphics[width=1.0\linewidth]{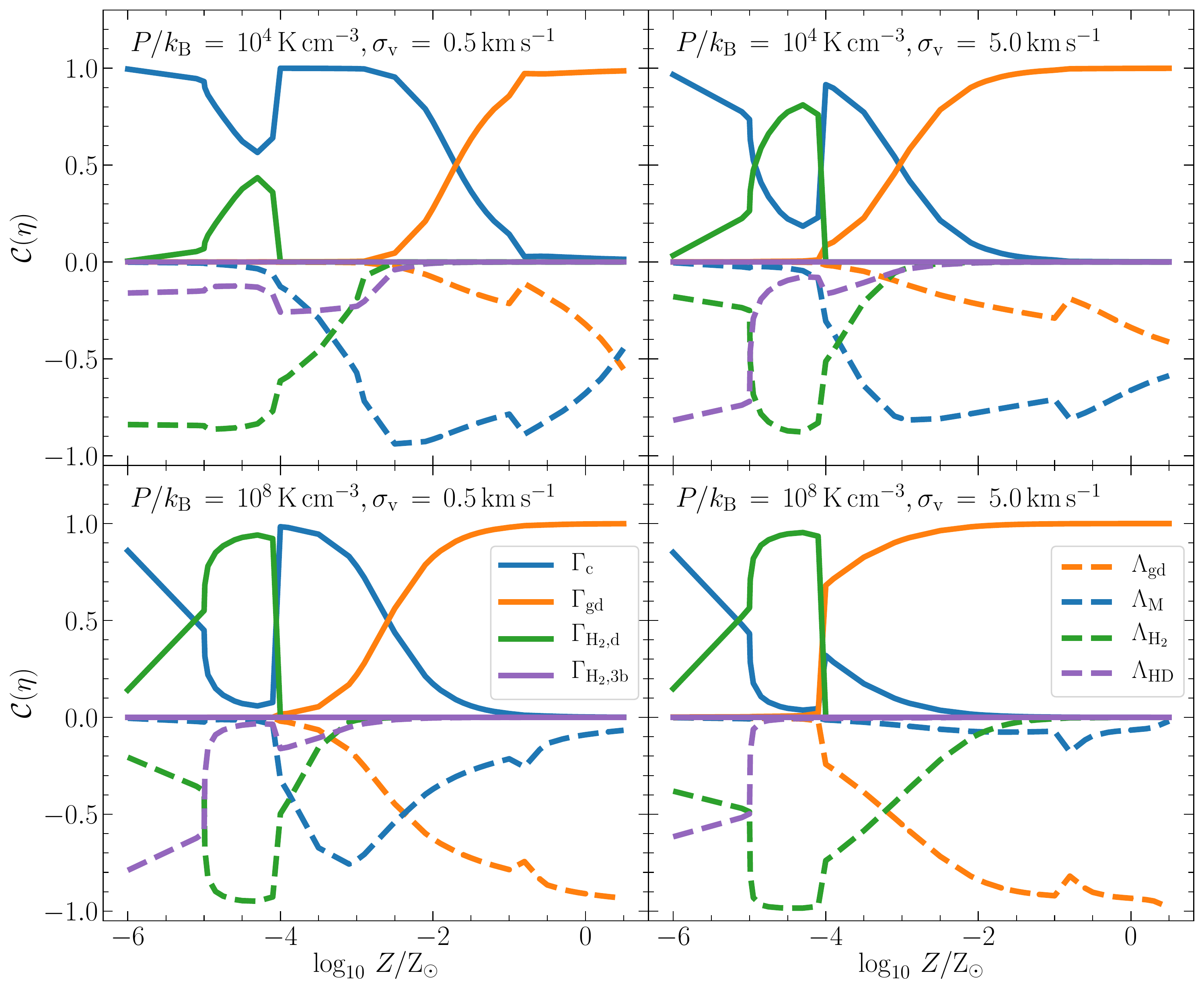}
\caption{Same as \autoref{fig:gtb_termcontribution} but including H$_2$ formation heating due to dust (solid green) and 3-body reactions (solid purple).}
\label{fig:h2formationheating_gtb}
\end{figure*}

\section{Discussion}
\label{s:discussion}

\subsection{Effects of \texorpdfstring{H$_2$}{H2} formation heating}
\label{s:discussions_h2formationheating}
In chemical equilibrium, H$_2$ formation or destruction cannot by itself be a source of heating or cooling, since the chemical energy liberated by formation is balanced by that lost to destruction; at best H$_2$ formation or destruction can act as a conduit by which other processes (e.g., cosmic rays or X-rays -- \citealt{Glassgold12a}) can transfer energy to the gas. Thus, we need to only consider heating from H$_2$ formation when the formation process is non-equilibrium. For this case, we use equation 37 of \citet{2014MNRAS.439.2386G} to write heating due to 3-body formation of H$_2$ as
\begin{equation}
\Gamma_{\rm{H_2,3b}} = E_{\rm{d}} f_{\rm{chem}} \frac{k_{\rm{H_2,3b}}}{\mu_{\rm{H}}m_{\rm{H}}}\left(x_{\ion{H}{i}}n_{\rm{H}}\right)^2\,,
\label{eq:gammaH23b}
\end{equation}
where $E_{\rm{d}} = 4.48\,\rm{eV}$ is the energy released due to formation of H$_2$, $f_{\rm{chem}}$ is the critical density factor that we obtain from equation 33 of \citet{2014MNRAS.439.2386G}, and $k_{\rm{H_2,3b}}$ is the rate coefficient for the reaction that we adopt from table C1 of \citet{2014MNRAS.439.2386G}. Additionally, we follow \cite{2009A&A...496..365C} to include heating due to H$_2$ formation on dust
\begin{equation}
    \Gamma_{\rm{H_2,d}} = R_{\rm{H_2,d}}f_{\rm{H_2,d}}E_{\rm{d}}\frac{1}{{\mu_{\rm{H}} m_{\rm{H}}n_{\rm{H}}}}
\label{eq:dustH2}
\end{equation}
where $R_{\rm{H_2,d}}$ is the formation coefficient (in $\rm{cm^{-3}\,s^{-1}}$) that depends on the metallicity, sticking coefficient (following \citealt{1979ApJS...41..555H}), gas thermal velocity, and H$_2$ formation efficiency on different types of grains \citep[][equation~13]{2009A&A...496..365C}. Here, we only work with C and Si grains. We adopt the formation efficiencies of these grains from \citet[equations 6 and 7]{2009A&A...496..365C}. $f_{\rm{H_2,d}} \approx 0.34$ denotes the fraction of energy released during H$_2$ formation on dust ($E_{\rm{d}}$) that is available to heat the gas \citep{2021arXiv210506843P}.

\begin{figure}
\includegraphics[width=1.0\columnwidth]{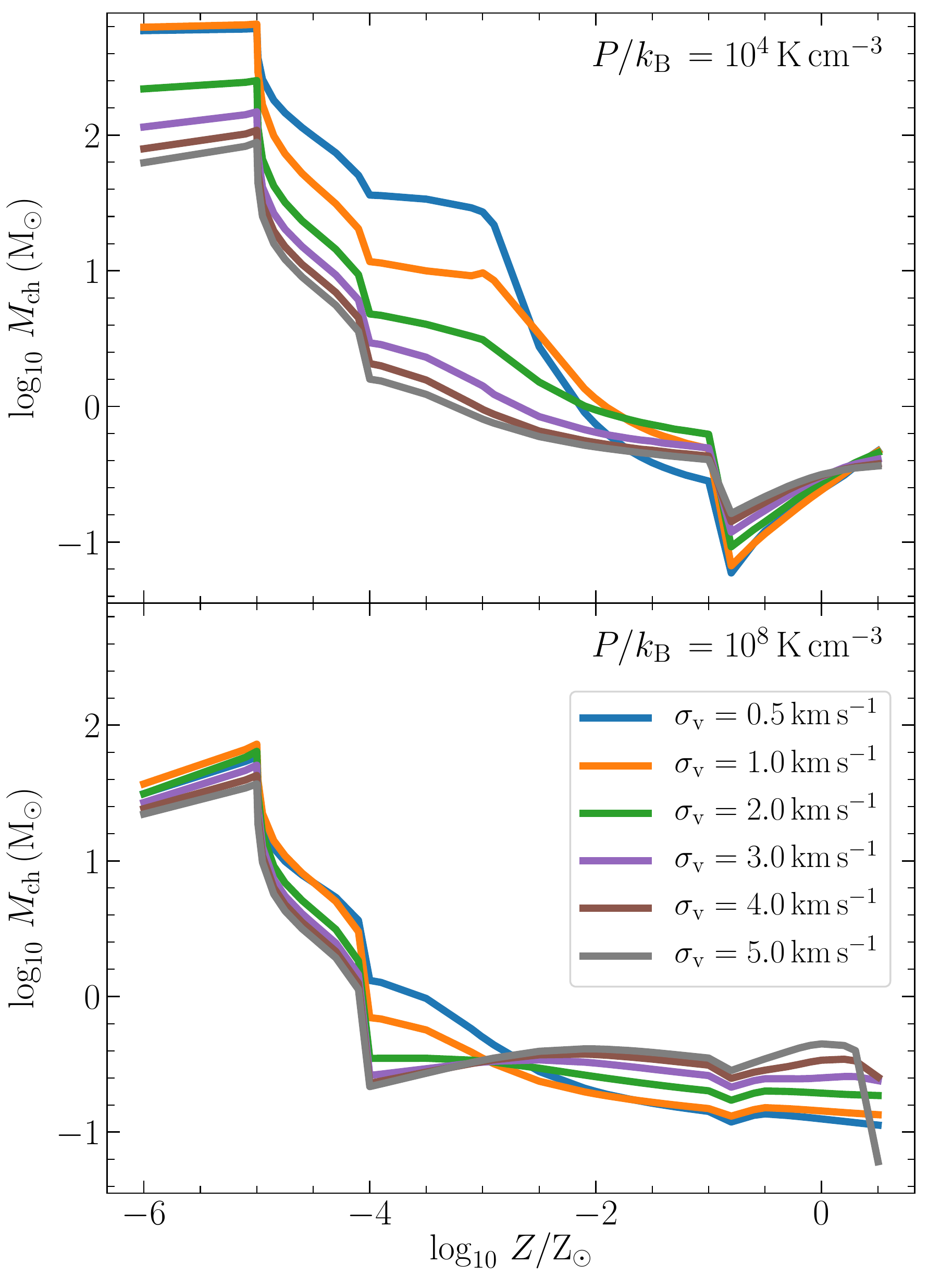}
\caption{Same as \autoref{fig:Mch} but including H$_2$ formation heating as shown in \autoref{fig:h2formationheating_gtb}. Despite being a dominant heating process at very low metallicity, H$_2$ formation heating has no appreciable effect on the characteristic stellar mass.}
\label{fig:h2formationheating_Mch}
\end{figure}

\autoref{fig:h2formationheating_gtb} shows the gas thermal balance at the critical location that sets $M_{\rm{ch}}$, now in the presence of H$_2$ formation heating. We firstly see that heating due to 3-body H$_2$ formation is always negligible. On the other hand, heating due to H$_2$ formation on dust becomes important if the chemical composition is \ion{H}{i}-dominated (typically corresponding to low $Z$), and is zero otherwise. Cooling due to H$_2$ quickly compensates the additional heating provided by H$_2$ formation without a substantial change in the gas temperature since H$_2$ cooling is exponentially sensitive to the gas temperature \citep{1998A&A...335..403G,2008MNRAS.388.1627G}. This yields minimal variation in the characteristic mass due to H$_2$ formation heating, as we illustrate in \autoref{fig:h2formationheating_Mch}.  

We can understand the lack of importance of H$_2$ formation heating as follows. The characteristic timescale to convert a gas that is mostly \ion{H}{i} into one that is mostly H$_2$ is $t_{\rm{H_2}} \sim 1/n_{\rm{H}} R^{\prime}_{\rm{H_2,d}}$, where $R^{\prime}_{\rm{H_2,d}} =  R_{\rm{H_2,d}}/n^2_{\rm{H}}$ is the rate coefficient in $\rm{cm^3\,s^{-1}}$, and we have ignored heating due to 3-body H$_2$ formation. We can compare the timescale for H$_2$ formation to the timescale for collapse, which is $t_{\rm{ff}} \sim 1 / \sqrt{G n_{\rm{H}} m_{\rm{H}}}$, and to the corresponding rate of compressive heating, $\Gamma_{\rm{c}} \sim k_{\rm{B}} T_{\rm{g}} \sqrt{G n_{\rm{H}}/m_{\rm{H}}}$ (see \autoref{eq:compress}). This ratio is $\Gamma_{\rm{H_2,d}} / \Gamma_{\rm{c}} \sim (E_{\rm{d}} / k_{\rm{B}} T_{\rm{g}}) R^{\prime}_{\rm{H_2,d}} \sqrt{n_{\rm{H}} / (G m_{\rm{H}})}$, which is greater than unity, i.e., H$_2$ formation heating is significant compared to compressive heating, only if $n_{\rm{H}} > (G m_{\rm{H}} / R^{\prime 2}_{\rm{H_2,d}}) (k_{\rm{B}} T_{\rm{g}} / E_{\rm{d}})^2$, where we have omitted factors of order unity for simplicity. On the other hand, in order to be out of equilibrium we require $t_{\rm{H_2}} / t_{\rm{ff}} > 1$, which is satisfied only if $n_{\rm{H}} < G m_{\rm{H}} / R^{\prime 2}_{\rm{H_2,d}}$. Adopting the rough scaling $R^{\prime}_{\rm{H_2,d}} \approx 7\times10^{-15} Z/\rm{Z_{\odot}}$ \citep{2009A&A...496..365C}, and for $T_{\rm{g}} \approx 100\,\rm{K}$ (expected if H$_2$ is important), this numerically evaluates to $4\times10^{-6} / (Z/\rm{Z_{\odot}})^{2} < n_{\rm{H}} / \rm{cm^3} < 2 \times10^{-3} / (Z/\rm{Z_{\odot}})^{2}$. We can immediately see that this condition is only satisfied at very low $Z$ for typical values of $n = n_{\rm{crit}}$ we obtain, which is why H$_2$ formation heating does not play a significant role elsewhere.

\subsection{Effects of cosmic rays}
\label{s:discussions_cosmicrays}
So far, we have ignored the effects of cosmic rays. While it is not yet known if cosmic rays threaded primordial/metal-poor star-forming clouds, we can use our models to study if they could have any any effects on the characteristic stellar mass or the IMF \citep[e.g.,][]{2018MNRAS.475.2467F}. Cosmic rays can impact our analysis in two major ways: by providing excess heating to the gas, and by providing free hydrogen atoms needed to form H$_2$. 

Heating due to cosmic rays at different densities, pressures and metallicities is highly uncertain, so we adopt an empirical approach where we express cosmic ray heating normalized to that observed in the Milky Way
\begin{equation}
    \Gamma_{\rm{CR}} = \frac{q_{\rm{CR}}}{\mu_{\rm{H}} m_{\rm{H}}} \zeta\,,
\label{eq:cosmicrays}
\end{equation}
where
\begin{equation}
    \zeta = \zeta_{\rm{MW,CR}} \frac{f_{\rm{CR}} P}{P_{\rm{MW,CR}}}\,.
\label{eq:cosmicrays_zeta}
\end{equation}
Here, $q_{\rm{CR}}=6.5\,\rm{eV}$, $\zeta_{\rm{MW,CR}}=3\times10^{-16}\,\rm{s^{-1}}$ is the cosmic ray primary ionisation rate per H nucleus in the Milky Way \citep[e.g.,][]{2012ApJ...745...91I}, and the product of the two divided by $\mu m_{\rm H}$ is the cosmic ray heating rate in the Milky Way assuming that each primary ionisation yields $6.5$ eV of heating. The remaining factor in \autoref{eq:cosmicrays_zeta}, $f_{\rm CR} P/P_{\rm MW,CR}$, represents our assumed scaling of the cosmic ray heating rate with pressure: $P_{\rm{MW,CR}}/k_{\rm{B}} \approx 3500\,\rm{K\,cm^{-3}}$ is the typical cosmic ray pressure in Milky Way star-forming molecular clouds \citep[e.g.,][]{2002ApJ...568L.121Y,2007ApJ...665L.123Y}, while $f_{\rm{CR}}$ is the ratio of cosmic ray to gas pressure in a given ISM. We compute $f_{\rm{CR}}$ from the semi-analytic models of \citet[Figure~8]{2021MNRAS.502.1312C}, who express it as a function of the gas surface density $\Sigma$, which we can express in terms of the gas pressure; similar calculations are also available in \cite{2018A&A...614A.111P}. 

We also revise our estimate of $x_{\ion{H}{i}}$ in the presence of cosmic rays by solving self-consistently for its equilibrium value in the H$_2$-dominated regime. We do so by equating the rate at which free hydrogen atoms are provided by cosmic ray ionization, $x_{\ion{H}{i},\rm{CR}}$, with the rate of H$_2$ formation on dust, $R_{\rm{H_2,d}}$, that we introduced in \autoref{s:discussions_h2formationheating}:
\begin{equation}
    R_{\rm{H_2,d}} = \zeta n_{\rm{H}} \left(\frac{1-x_{\ion{H}{i},\rm{CR}}}{2}\right)\,.
\label{eq:cosmicrays_xHI}
\end{equation}
Then, we simply take $x_{\ion{H}{i}}$ to be the maximum of $x_{\ion{H}{i},\rm{CR}}$ and our fiducial estimate from \autoref{eq:chemicalcomposition_hydrogen}. We adjust $x_{\rm{H_2}}$ consistently (\autoref{eq:chemicalcomposition_hydrogen}). In practice, since $x_{\ion{H}{i},\rm{CR}}$ is always close to zero, our procedure amounts to changing the minimum \ion{H}{i} fraction that prevails at high metallicity from zero to a value slightly above zero.

\begin{figure*}
\includegraphics[width=1.0\linewidth]{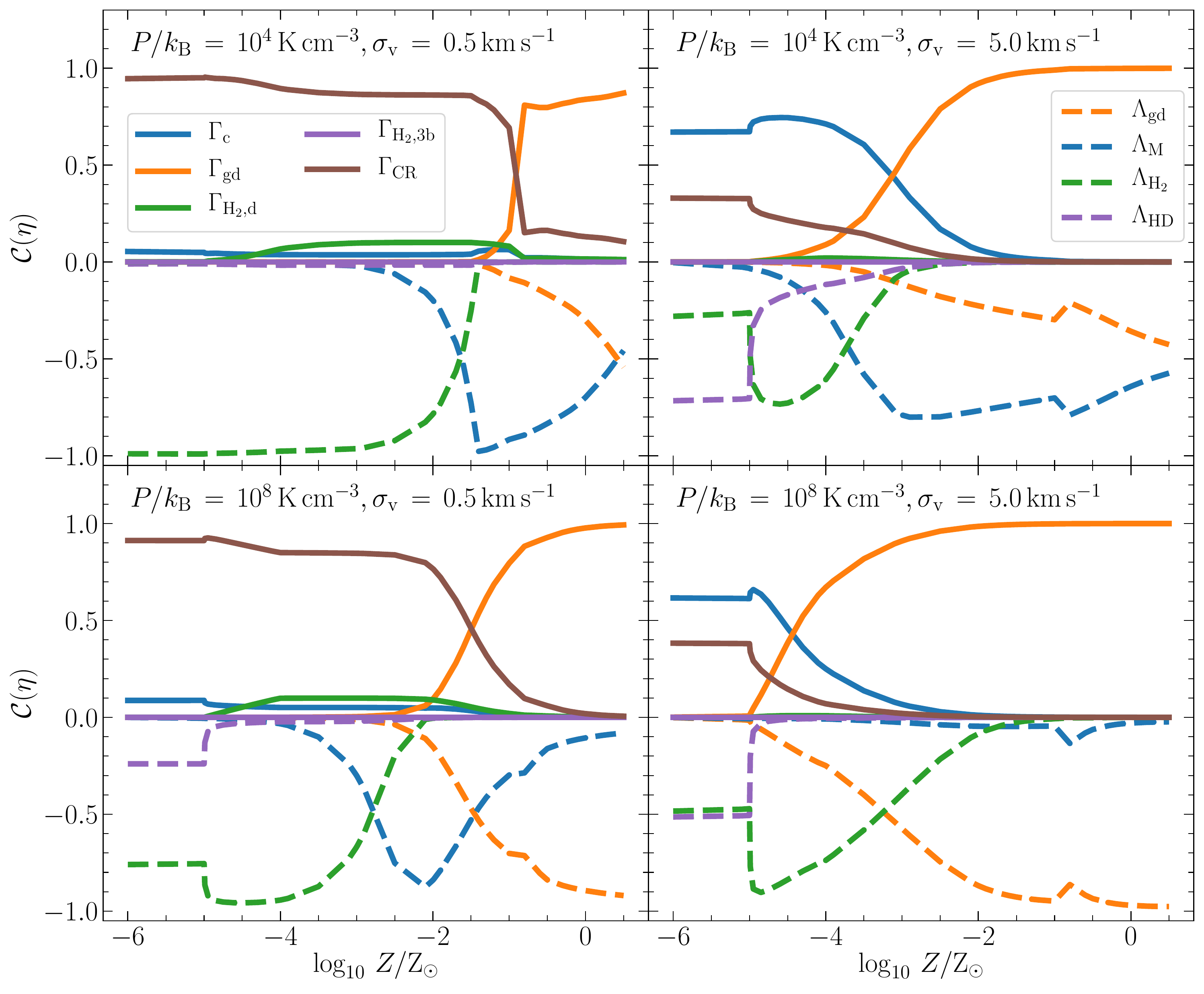}
\caption{Same as \autoref{fig:h2formationheating_gtb} but including cosmic ray heating (solid brown).}
\label{fig:cosmicrays_gtb}
\end{figure*}

\begin{figure}
\includegraphics[width=1.0\columnwidth]{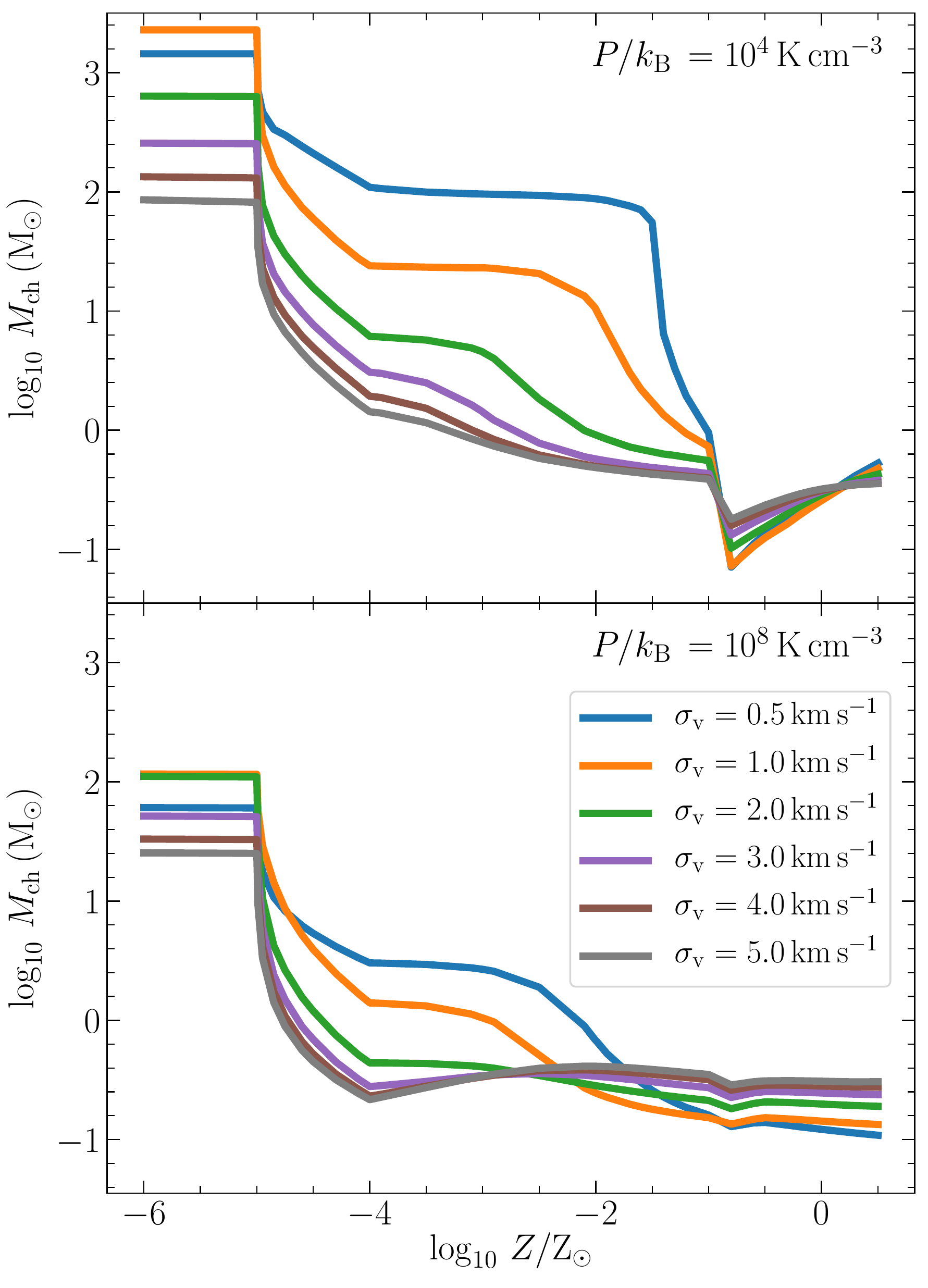}
\caption{Same as \autoref{fig:h2formationheating_Mch} but including cosmic ray heating. Cosmic rays become the dominant heating process at very low metallicity, resulting in a characteristic mass that is a factor of $2-3$ larger than that in the fiducial model (see, however, \autoref{s:discussions_cosmicrays} where we discuss why is this not realistic).}
\label{fig:cosmicrays_Mch}
\end{figure}

We show the effects of cosmic ray heating in \autoref{fig:cosmicrays_gtb} and \autoref{fig:cosmicrays_Mch} at the corner points of our grid; results elsewhere in the grid are qualitatively identical. We observe that heating due to cosmic rays now becomes prominent at low $Z$, leading to changes in the characteristic stellar mass at low $Z$ while maintaining the qualitative trend in $M_{\rm{ch}}$ as a function of $Z$ we know from previous sections. This is because at low $Z$ where cosmic ray heating becomes significant, the gas temperature rises by a factor of a few, which in turn increases the mass needed to collapse ($M_{\rm{BE}}$). The rise in the gas temperature at low $P$ and low $Z$ results in a jump in $M_{\rm{ch}}$ by a factor of $2-3$, yielding $M_{\rm{ch}} > 1000\,\rm{M_{\odot}}$ at $Z = 10^{-6}\,\rm{Z_{\odot}}$.

However, the above result from the model should be treated with caution. This is because we have not adjusted for the increased amount of H$_2$ that would form due to the increased availability of free hydrogen atoms because of cosmic ray ionization. In fact, detailed modeling with time-dependent chemistry from \cite{2007MNRAS.382..229S} and \cite{2016MNRAS.460.2432H} show that the dominant effect of cosmic rays at such low metallicities is that they indirectly provide much more gas cooling due to increased H$_2$ fraction, thus resulting in $10 \lesssim M_{\rm{ch}}/\rm{M_{\odot}} \ll 1000$ even when the cosmic ray strength is changed by five orders of magnitude. So, in reality, we expect cosmic ray \textit{heating} to not play a significant role in setting the characteristic mass at very low metallicities. 

While cosmic rays are of limited significance for our results at high $Z$, we have not accounted for the possibility that cosmic ray pressure in star-forming regions of metal-rich molecular clouds could be substantially higher than the mean pressure across the cloud due to local acceleration of cosmic rays by protostellar outflows \citep{2016A&A...590A...8P, 2020SSRv..216...29P, 2018ApJ...859..136F}. Such an enhancement in cosmic ray pressure can lead to some destruction of CO molecules at high $Z$ \citep{2015ApJ...803...37B,2017ApJ...839...90B}, at the same time leading to $T_{\rm{g}} \gg T_{\rm{d}}$ and hence a higher characteristic mass \citep[because gas closer to the protostar will be hotter and thus unable to collapse; see --][]{2010ApJ...720..226P,2011MNRAS.414.1705P}. This effect is likely not relevant for the moment we are considering, when a protostellar core has first formed, since at this point an outflow will just have been launched, and will not yet have had time to accelerate any significant number of cosmic rays. In a clustered environment where many stars form together over time, it is possible that there could be an enhancement in cosmic rays due to acceleration by neighbouring protostars that formed earlier \citep[e.g.,][]{2018ApJ...861...87G,2019ApJ...878..105G}. This will depend on the degree of clustering and the details of cosmic ray transport, and is beyond the scope of this work. Nevertheless, including cosmic rays does not change the transition metallicity at which we expect the IMF to become bottom-heavy.

\subsection{Effects of different chemical compositions}
\label{s:discussions_chemicalcompositions}
In our fiducial model presented in \autoref{s:results}, we adopt a plausible expression for variation of the chemical composition as a function of $Z$. However, the exact chemical makeup of star-forming clouds, particularly at low metallicities, is poorly known. CO has been detected from galaxies at metallicities down to $\approx 10$ per cent of Solar but not lower \citep{Rubio15a, Shi16a}, so a transition from the majority of the carbon in star-forming regions being in the form of CO to \ion{C}{i} or \ion{C}{ii} likely occurs at around this metallicity, but the lack of a reliable method of detecting ``CO-dark'' H$_2$ means that it is unclear at what metallicity a similar \ion{H}{i} to H$_2$ transition occurs \citep[e.g.,][]{2011ApJ...741...12B,2017MNRAS.470.2890B,2018ApJ...853..111J,2020MNRAS.494.5279C}.

Theoretical models predict a transition from H$_2$ to \ion{H}{i}-dominated star formation at $Z \lesssim 10^{-2} Z_\odot$ \citep{2012ApJ...759....9K, 2021arXiv210501681S}, and simulations of star formation at very low metallicities ($Z \sim 10^{-4}\,\rm{Z_{\odot}}$) find that \ion{O}{i} is the dominant species among all O-bearing species that make up the chemical composition at $n < 10^{8}\,\rm{cm^{-3}}$ \citep[][Figure~4]{2016MNRAS.463.2781C}. Similarly, we see from \citet[Figure~8]{2020arXiv200806107C} that \ion{C}{i} is the dominant C-bearing species at low metallicities for $n \lesssim 10^{5.5}\,\rm{cm^{-3}}$, and CO at high metallicities and densities. Thus, while the general results of these studies motivate the varying chemical composition we adopt for our fiducial models, they are hints only. Moreover, it is likely that, as one proceeds to lower metallicities and equilibration times become longer compared to dynamical ones, chemistry becomes increasingly non-equilibrium, such that a wide range of chemical compositions may coexist in star-forming regions of the same metallicity \citep{Krumholz11a, 2012ApJ...759....9K,Glover12a}.

For this reason, it is important to check to what extent the major qualitative results we have obtained using our fiducial model depend on our uncertain assumptions about chemical composition. To do so, we now fix the chemistry to one of the four main compositions we list in \autoref{s:metallicity_chemicalmodel} for all $Z$, and repeat all our model calculations. We emphasise that many of these cases are not realistic -- we observe that star-forming clouds are dominated by H$_2$ and not \ion{H}{i} at $Z=Z_\odot$, and a CO-dominated composition at $Z \ll 10^{-2} Z_\odot$ is ruled out by observations of local metal-poor dwarf galaxies. We are intentionally exploring a range of variation much wider than plausibly exists in nature.

\begin{figure*}
\includegraphics[width=1.0\linewidth]{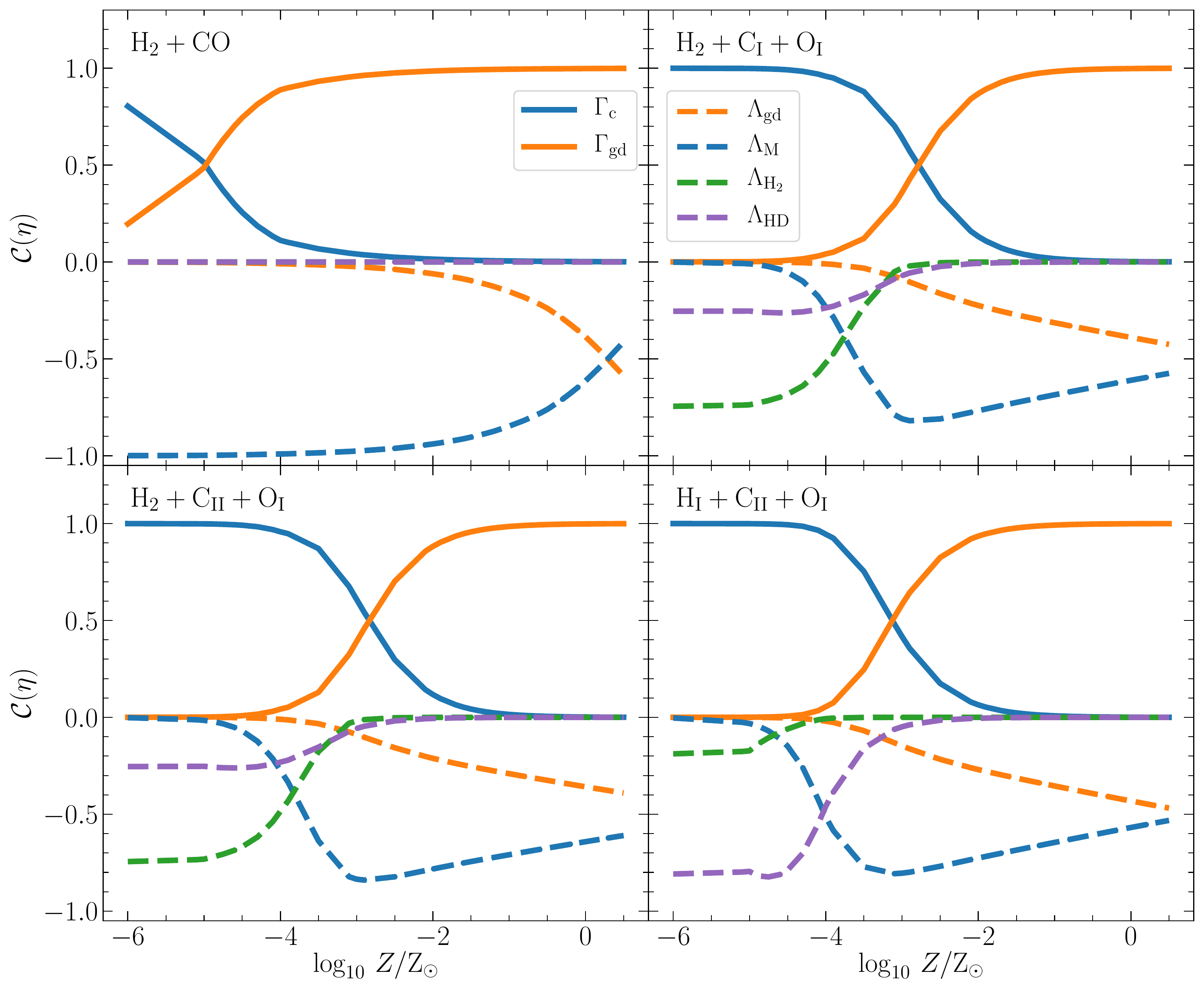}
\caption{Same as \autoref{fig:gtb_termcontribution} but for the four different chemical compositions as noted in the panels, for a fixed $P/k_{\rm{B}}=10^4\,\rm{K\,cm^{-3}}$ and effective velocity dispersion $\sigma_{\rm{v}}=5\,\rm{km\,s^{-1}}$.}
\label{fig:chemicalcomposition_gtb}
\end{figure*}

\autoref{fig:chemicalcomposition_gtb} shows the contribution of different processes to gas thermal balance for the four different chemical compositions at fixed $P/k_{\rm{B}} = 10^{4}\,\rm{K\,cm^{-3}}$ and $\sigma_{\rm{v}} = 5\,\rm{km\,s^{-1}}$. We find that the qualitative behavior of the different heating and cooling processes does not vary within the chemical compositions that use atomic metal line cooling, and is similar to that of the fiducial model we use. However, using a fixed chemical composition consisting of H$_2$ and CO at all $Z$ gives very different results; cooling due to CO becomes dominant at $Z < 0.1\,\rm{Z_{\odot}}$, and heating due to dust-gas coupling becomes dominant at all $Z$, driven by the much cooler gas temperature, and thus much greater dust-gas temperature difference, when CO exists at low $Z$. However, as noted above, a CO-dominated composition is not plausible at low metallicities.

\begin{figure}
\includegraphics[width=1.0\columnwidth]{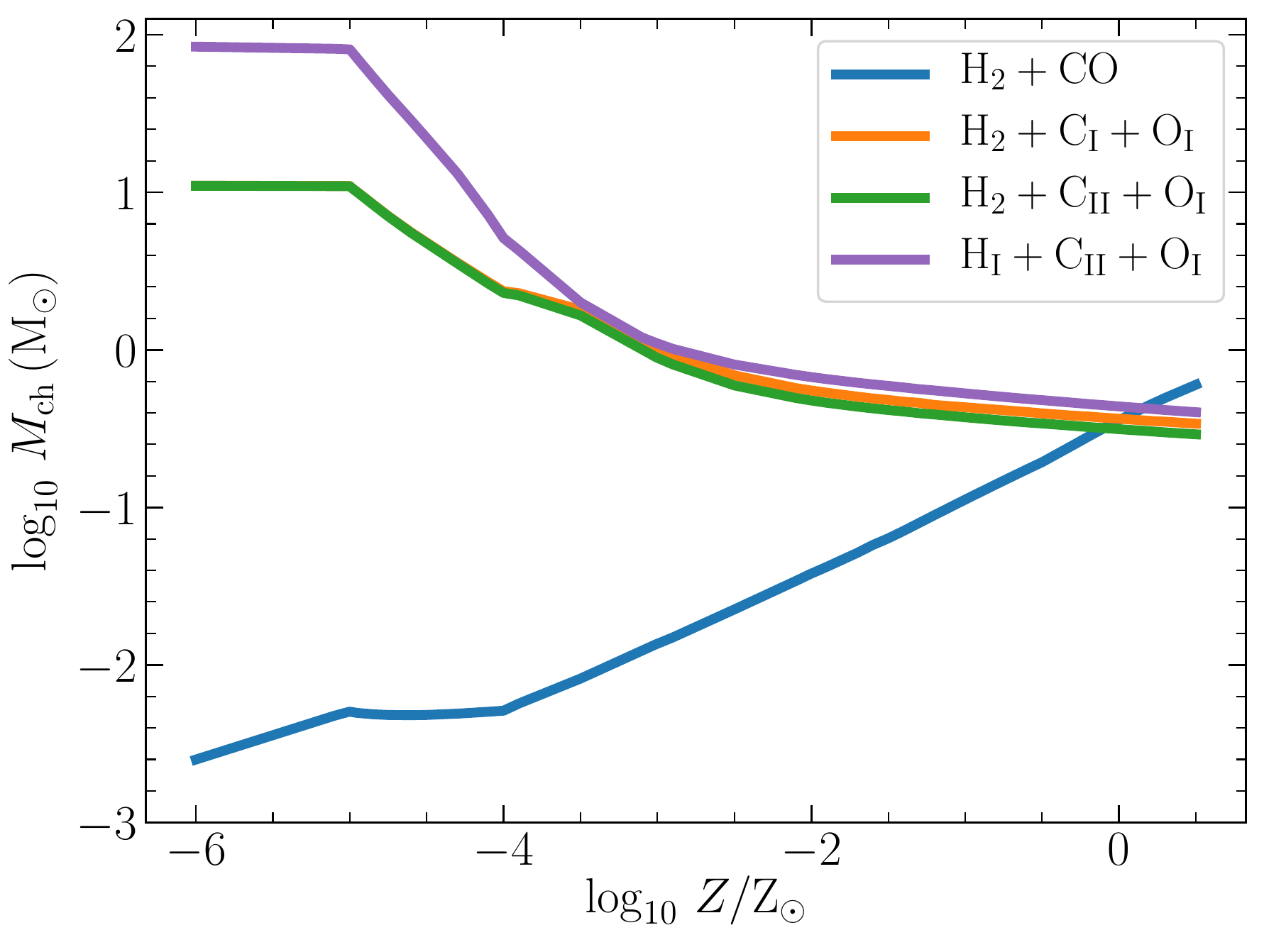}
\caption{Same as the top panel of \autoref{fig:Mch} but for different chemical compositions for a fixed $P/k_{\rm{B}} = 10^4\,\rm{K\,cm^{-3}}$ and $\sigma_{\rm{v}}=5\,\rm{km\,s^{-1}}$.}
\label{fig:chemicalcomposition_Mch}
\end{figure}

\autoref{fig:chemicalcomposition_Mch} extends this conclusion to $M_{\rm ch}$; this figure is identical to the top panel of \autoref{fig:Mch}, but considers fixed chemical composition. Except for the unrealistic case of CO-dominated composition at all $Z$, in all other cases we find the same sharp transition in $M_{\rm{ch}}$ as in our fiducial model. Thus, our results are not sensitive to the choice of the chemical composition we use.

We can understand the apparent lack of sensitivity of $M_{\rm{ch}}$ to the chemical composition as follows. At very low metallicities (primordial star formation), molecular H$_2$ is the dominant coolant, and the metallic composition of the gas (or dust) does not matter, thus yielding $M_{\rm{ch}}$ that is fairly constant across all the four compositions. Similarly, at very high metallicities (modern day star formation), dust takes control of the gas thermodynamics, leaving little room for metals in the gas phase to significantly impact $M_{\rm{ch}}$. Finally, the trends in $M_{\rm{ch}}$ at intermediate metallicities ($10^{-3.5} \leq Z/\rm{Z_{\odot}} \leq 10^{-1.5}$) for different chemical compositions are similar to that we find in the fiducial model because metal cooling dominates in both the cases.

\begin{figure*}
\includegraphics[width=1.0\linewidth]{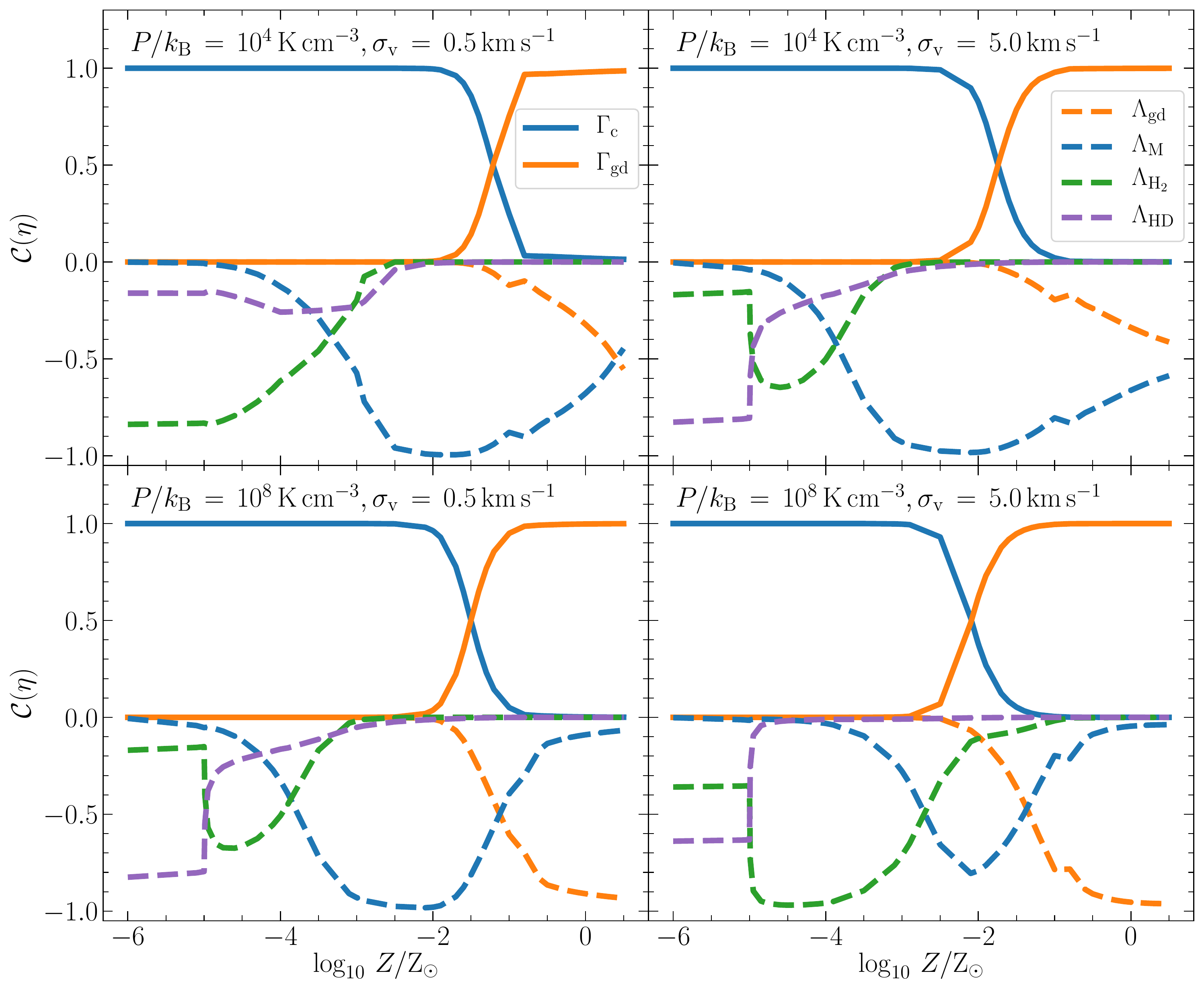}
\caption{Same as \autoref{fig:gtb_termcontribution} but following the dust to gas ratio scaling of metallicity from \citet{2014A&A...563A..31R}. Dust dominates over a narrower range in metallicity at high $P$ in this case as compared to the fiducial model.}
\label{fig:gtb_termcontribution_remyruyer}
\end{figure*}

\subsection{Effects of a varying dust-to-metal ratio}
\label{s:discussions_metalscaling}
The evolution of the dust to gas ratio $\delta$ with metallicity is observed to be linear at $Z \gtrsim 0.01\,\rm{Z_{\odot}}$ \citep{2014A&A...563A..31R,2019MNRAS.490.1425L}, but it is largely unknown at $Z < 0.01\,\rm{Z_{\odot}}$ \citep{2018ARA&A..56..673G} and can even vary within the same cloud for different grain sizes \citep{2017MNRAS.471L..52T,2017MNRAS.465.1089B}. So far, we have assumed a simple linear scaling of $\delta$ with $Z$ (see \autoref{s:metallicity_chemicalmodel}). Now, we try another scaling where we extrapolate the results of \cite{2014A&A...563A..31R} for the variation of $\delta$ with $Z$ down to $10^{-6}\,\rm{Z_{\odot}}$. Specifically, we set $\delta \propto Z$ for $Z \gtrsim 0.2\,\rm{Z_{\odot}}$, and $\delta \propto Z^{3.1}$ for $Z \lesssim 0.2\,\rm{Z_{\odot}}$ \citep[table~1]{2014A&A...563A..31R}, implying that $\delta$ decreases much more steeply with decreasing $Z$ than for our fiducial scaling.

\begin{figure}
\includegraphics[width=1.0\columnwidth]{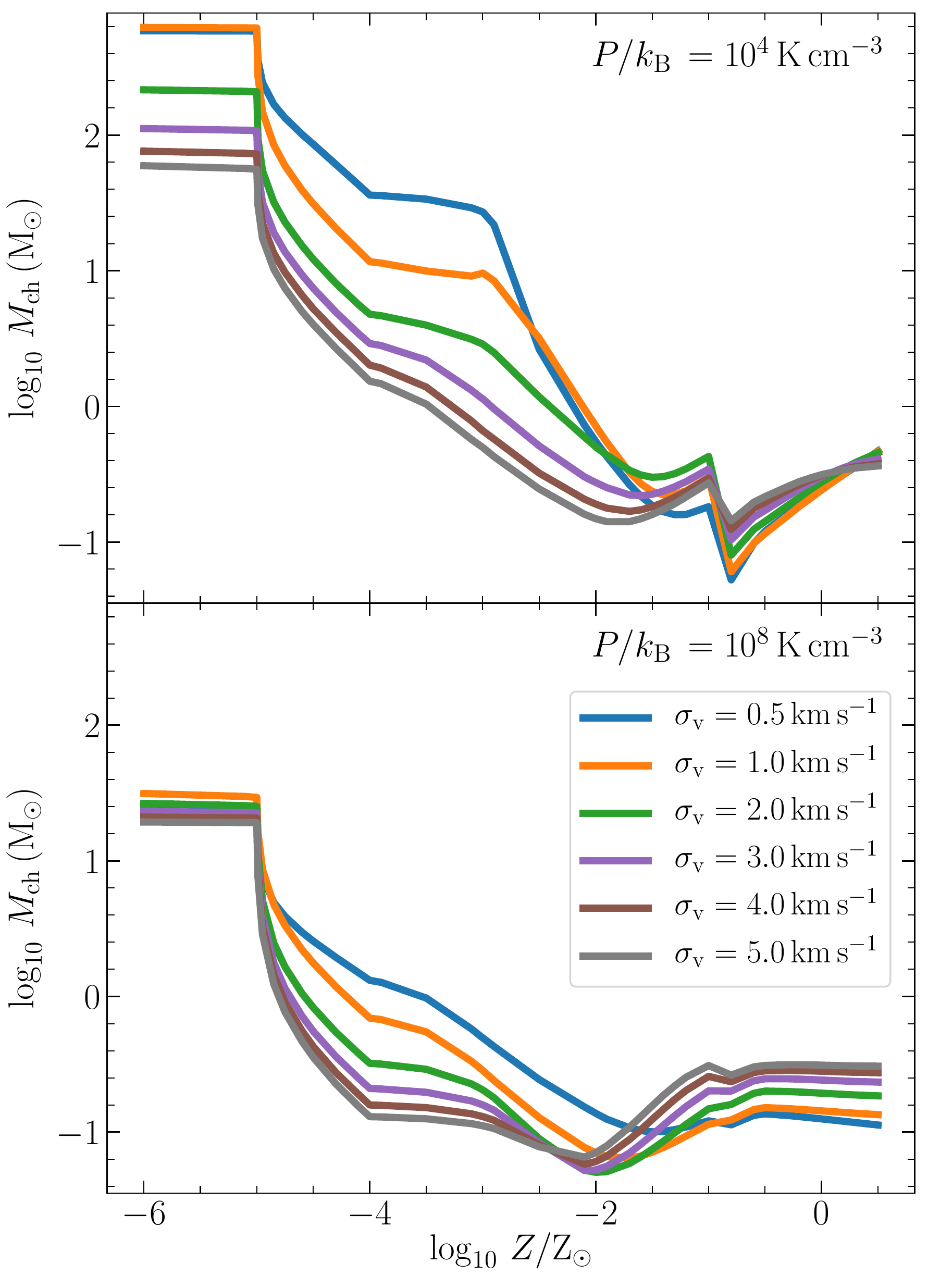}
\caption{Same as \autoref{fig:Mch} but with the dust to gas ratio scaling with metallicity from \citet{2014A&A...563A..31R}. A different scaling only qualitatively impacts the characteristic stellar mass in a starburst environment.}
\label{fig:Mch_remyruyer}
\end{figure}

\autoref{fig:gtb_termcontribution_remyruyer} plots the gas thermal balance at the critical location that sets $M_{\rm{ch}}$ with the alternate $\delta-Z$ scaling. At low $P$, a comparison of \autoref{fig:gtb_termcontribution_remyruyer} with \autoref{fig:gtb_termcontribution} reveals that the dominance of dust and the onset of modern star formation in the case of the alternate $\delta-Z$ scaling is delayed by 0.5 dex as a function of metallicity. The effect is more dramatic at high $P$, and a steeper than linear decline of $\delta$ with $Z$ causes the onset of dust-dominated modern star formation to be delayed by more than 1 dex. Nonetheless, we find from \autoref{fig:Mch_remyruyer} that the effects of a delay in the onset of modern star formation has no impact on the characteristic mass or the transition from top- to bottom-heavy IMF as a function of metallicity. Given the current state of our knowledge of dust-to-metal ratio, we thus do not expect its variations to significantly impact the trends we observe in $M_{\rm{ch}}$ as a function of $Z$.

\subsection{Effects of varying model parameters related to dust}
\label{s:discussions_otherparams}
Since a key driver of our work is studying the role of dust in setting $M_{\rm{ch}}$, we now discuss how uncertainties in various model parameters related to dust can affect our results. We briefly discuss the effects of five such parameters that carry the most uncertainty -- three that characterize the accretion luminosity and by extension, the underlying \citet{2005ApJ...631..792C} model, and two that characterize the dust-gas energy coupling term, $\Psi_{\rm{gd}}$. We limit this discussion to the three categories of star-forming regions we identified in \autoref{s:results_radialprofiles} -- primordial, Galactic and starburst, respectively.

\begin{enumerate}
    \item \textit{$\epsilon_{\rm{L}}$} -- This parameter is defined in \autoref{eq:L}. If we increase its value from 0.75 to 1 (implying that accretion power does not drive any winds or outflows), we find no appreciable impact on $M_{\rm{ch}}$ in any environment. On the other hand, if we decrease it to 0.1 (implying most of the accretion power goes into winds), $M_{\rm{ch}}$ decreases by 20 and 40 per cent for the Galactic and the starburst environments, respectively, while leaving the overall trend of $M_{\rm ch}$ as a function of metallicity the same. The underlying physical reason that the dependence on $\epsilon_L$ is weak is that the gas temperature is relatively insensitive to the luminosity, varying between $T_{\rm d}\propto L^{1/4}$ (for the limit of a completely opaque, optically thick dust atmosphere) to $T_{\rm d} \propto L^{1/6}$ (for optically thin dust with Planck mean opacity that scales as $\kappa\propto T_d^2$).
    
    \item \textit{$\epsilon_{\rm{M}}$} -- So far, we have assumed that approximately half of material infalling onto the protostar is ejected through winds. Not surprisingly, the primary effect of changing this fraction is to produce an almost proportional shift in $M_{\rm ch}$, while leaving the underlying pattern of how $M_{\rm ch}$ varies with metallicity or and all other parameters nearly unchanged. We note that values of $\epsilon_{\rm M}$ outside the range $\approx 1/4-3/4$ are, at least in the Milky Way, ruled out by theory \citep{2000ApJ...545..364M} and observations \citep{2007A&A...462L..17A,2008ApJ...684.1240E}, so the maximum shift in $M_{\rm ch}$ we expect is at the tens of percent level.
    \item \textit{$\mathcal{A}$} -- As we noted in \autoref{s:dusttemperature}, the energy yield per unit mass due to accretion is fairly constant for a wide range of protostellar masses \citep{2011ApJ...743..110K}. We find that variations of an order of magnitude in $\mathcal{A}$ only produce at most a 60 per cent change in $M_{\rm{ch}}$ in dust-dominated environments. Thus, we verify that uncertainties in $\mathcal{A}$ do not significantly alter our results on the evolution of $M_{\rm{ch}}$ as a function of metallicity.
    \item \textit{$\alpha_{\rm{gd}}$} -- The dust-gas energy coupling goes to zero if dust-gas collisions are completely elastic such that dust does not transfer any energy to the gas during collisions. The biggest impact of completely inelastic (implying $\alpha_{\rm{gd}} = 1$) dust-gas collisions is in the Galactic case, where $M_{\rm{ch}}$ increases by 32 per cent. Similarly, if the collisions are nearly mostly elastic ($\alpha_{\rm{gd}} = 0.1$), $M_{\rm{ch}}$ decreases by 50 per cent in the Galactic case. Thus, uncertainties in $\alpha_{\rm{gd}}$ do not significantly change our results. However, the inelasticity of the collisions is not a free parameter as it also depends on the grain size and composition \citep{1972ApJ...174..321W,1983ApJ...265..223B,1985A&A...152..130D}, exploring which is beyond the scope of this work.
    \item \textit{$\mathcal{S}_{\rm{gd}}$} -- As we noted in \autoref{s:dusttemperature}, we have no constraints on the dust cross section per unit dust mass at low metallicities. However, this is not a problem since varying $\mathcal{S}_{\rm{gd}}$ by an order of magnitude only significantly impacts $M_{\rm{ch}}$ in the Galactic case, and has no significant impact at low metallicities due to the diminishing role of dust with decreasing metallicity for the assumed scalings of $\delta$ with metallicity.
\end{enumerate}

\section{Evolution of the IMF with metallicity}
\label{s:discussions_evolutionIMF}

We next seek to put our findings here in the context of other work on the IMF as a function of metallicity, starting with observations (\autoref{s:IMF_observationalevidence}), then considering prior analytic work (\autoref{s:IMF_comparisonmodels}), and finally discussing earlier simulations (\autoref{s:IMF_comparisonsimulations}). We caution that a comprehensive review of the (vast) literature on IMF variations is beyond the scope of this paper, and we refer readers to the large number of reviews that have appeared in the past decade \citep{2010ARA&A..48..339B,2013pss5.book..115K,2014PhR...539...49K,2014prpl.conf...53O,2018PASA...35...39H,2020SSRv..216...70L}. Our focus here will be specifically on variations with metallicity, and to a lesser extent, pressure.

\subsection{Observational evidence}
\label{s:IMF_observationalevidence}
Variations of the IMF have long been suspected \citep{2002Sci...295...82K}. Here we discuss several classes of stellar systems where previous authors have claimed IMF variation, at least potentially due to metallicity effects.

\subsubsection{Metal-poor Milky Way stars}

Based on analysis of metal-poor stars in the Milky Way halo, \cite{2011MNRAS.412..843S,2013MNRAS.432L..46S} favour an IMF that transitions to bottom-heavy around $\rm{[Fe/H]} \sim -2$ based on the abundance of carbon in carbon enhanced metal-poor stars (CEMP, \citealt{2005ARA&A..43..531B}); this is consistent with our finding that the era of modern day star formation begins around $Z \approx 10^{-2}\,\rm{Z_{\odot}}$. These authors also propose an initially top-heavy Galactic IMF with a peak around $10-12\,\rm{M_{\odot}}$. Similar arguments have been made by \cite{2007ApJ...658..367K} who find that the IMF of CEMP stars ($\rm{[Fe/H]} \lesssim -2.5$ in their case) can be well represented by a characteristic mass $\approx 6-10\,\rm{M_{\odot}}$. Factors such as binary star formation and population synthesis also fall short at explaining the observed binary fraction of CEMP stars, requiring other phenomena such as variations in the IMF be considered \citep{2009A&A...508.1359I,2014ApJ...788..131L}. Consistent with this analysis, and with the predictions of our model, \cite{2010A&A...515A..68M} show that the presence of an evolving IMF that was initially top-heavy also provides good fits to the observed C/O versus O/H trends in the Galaxy \citep{2009A&A...500.1143F}.

Observations of metal-poor stars in the Milky Way also reveal an absence of carbon-normal (or CEMP-no) stars below $\rm{[Fe/H]} < -4$ where the slope of the  metallicity distribution function significantly changes (\citealt{2020MNRAS.492.4986Y,2021arXiv210706430Y}, with the exception of the \citealt{2011Natur.477...67C} star), indicating the absence of low mass stars below this metallicity that would otherwise have lived for a Hubble time \citep[see also,][]{2006ApJ...641....1T}. This finding is also consistent with our finding that the transition to a bottom-heavy IMF does not occur at metallicities below $Z \approx 10^{-4}\,\rm{Z_{\odot}}$ anywhere within the parameter space of our model. The presence of more bottom-heavy IMFs at higher metallicities that would result from a lower $M_{\rm{ch}}$ at high $Z$ is also independently confirmed by observations of Wolf Rayet stars in the local Universe \citep{2021arXiv210103217L}. However, we do caution that, while our models are consistent with the available evidence, the data thus far are very limited. Moreover, we have predicted only the characteristic mass, not its spread. For example, simulations of both primordial \citep[e.g.,][]{Clark11a, Greif11a, 2020MNRAS.497..336S} and modern \citep[e.g.,][]{Bate02a, 2010ApJ...717..577T, 2015ApJ...800...72T} star formation show that disc fragmentation can produce a sub-dominant population of stars with mass considerably smaller than the characteristic mass; as the example of \citet{2011Natur.477...67C}'s star shows, the detection of a single star of a particular mass and metallicity cannot be used to deduce the characteristic mass at that metallicity.

\subsubsection{Metal-poor globular clusters and dwarf galaxies}

Our model predicts that star-forming systems with $\log_{10} Z/Z_\odot\lesssim -1.5$ should have increased characteristic masses compared to more metal-rich systems -- slightly higher for systems that formed at high pressure, and more substantially higher for systems that formed at low pressure. The most metal-poor globular clusters and dwarf galaxies reach this metallicity range, so in principle our model is testable by observations of such systems. However, the observational situation is highly-contested. For globular clusters \citet{2000ApJ...530..342D, 2010ApJ...718..105D} find that, once one accounts for preferential evaporation of low-mass stars over $\sim 10$ Gyr timescales, all the globular clusters they survey are consistent with having formed with an IMF with a characteristic mass similar to that found in the Milky Way field \citep[see also,][]{2017MNRAS.464.2174B}. By contrast, \citet{2012MNRAS.422.2246M} argue based on models for the effects of gas ejection by feedback that metal-poor clusters must have had a top-heavy IMF compared to the field. \cite{2014ApJ...796...71Z} report that characteristic masses do vary between globular clusters, but find no systematic variation with metallicity. Thus there is little consensus in the literature, and, in general, searches for IMF variation in globular clusters are challenging due to uncertainties in both the formation channels \citep{2014prpl.conf..291L,2018ARA&A..56...83B,2019ARA&A..57..227K} and the dynamical evolution \citep{1987degc.book.....S,2003MNRAS.340..227B,2012MNRAS.422.1592L,2015MNRAS.453.3278W,2017MNRAS.471.3845W} of these objects. A further challenge is that the globular cluster population only reaches the edge of the ultra-low metallicity region where we expect substantial variations in $M_{\rm ch}$; in the papers discussed above, the median metallicity is close to $\log_{10} Z/Z_\odot \approx -1.5$, and clusters with $\log_{10} Z/Z_\odot < -2$ make up only $\approx 10$ per cent of the sample \citep{1985ApJ...293..424Z,1996AJ....112.1487H}. Thus the expected signal is rather weak.

Compared to globular clusters, metal-poor dwarf galaxies suffer from fewer uncertainties about dynamical evolution, but at the price that, since they are more distant than globular clusters, observations are substantially more difficult. In the most nearby dwarfs, it is possible to measure the IMF from resolved stellar populations. Using this method, \cite{2018ApJ...855...20G} find a strong anti-correlation between the slope of the IMF and metallicity for ultra-faint Milky Way satellites, resulting in a slightly bottom-light IMF (characteristic mass $\approx 0.5-0.6\,\rm{M_{\odot}}$) at lower metallicities ($-2.5 \lesssim \rm{[Fe/H]} \lesssim -1.5$) for some of these galaxies as compared to the Milky Way. This would appear to be consistent with our model. However, the statistical significance of their result is marginal; if they use lognormal rather than powerlaw functional forms to fit their data, the data are consistent with having the same characteristic mass as the Milky Way at the $1\sigma$ level; more generally, \citet{El-Badry17a} show that drawing strong conclusions about the IMF in dwarf galaxies from resolved star counts is exceedingly difficult due to the limited sample sizes available. Similarly, \citet{2021MNRAS.503.6026R} argue for a top-heavy IMF in the ultra-faint dwarf Bo\"otes I based on simulations of its color magnitude diagram, which would be at least qualitatively consistent with our predictions (see, however, \citealt{2020A&A...637A..68Y}). Nonetheless, the same caveat from \citet{El-Badry17a} likely applies.

In more distant dwarfs, only more indirect methods using unresolved stellar populations are available. \citet{2009MNRAS.394.1529D,2010MNRAS.403.1054D,2012ApJ...747...72D} argue for a top-heavy IMF in ultra-compact dwarf galaxies based on the large number of low mass X-ray binaries (LMXBs) found within them. However, this method does not directly probe the characteristic mass, since LMXBs come from substantially more massive stars. Moreover, analysis of a much larger sample of such galaxies has failed to confirm the existence of an LMXB excess (\citealt{2016ApJ...819..162P}; see also, \citealt{2013MNRAS.433.1444P} and \citealt{2017ApJ...841...28P}). Similarly, \citet{2008ApJ...675..163H}, \citet{2009MNRAS.395..394P}, \citet{2009ApJ...706..599L}, \citet{2009ApJ...695..765M}, and \citet{2011MNRAS.415.1647G} all argue for IMF variation based on a variety of photometric indicators that should be sensitive to the slope of the high-mass IMF. Again, these methods do not probe the characteristic mass, and their claims are highly contested. \citet{Fumagalli11a}, \citet{2012MNRAS.422..794E}, and \citet{Weisz12a} all conclude that stochastic fluctuations in IMF sampling, the star formation history, or both are sufficient to explain the observations without any need for IMF variations (however, see \citealt{2013MNRAS.434...84W}). \citet{2014ApJ...793....4A} report direct measurements of the required stochastic fluctuations in low-mass clusters (though see \citealt{2014MNRAS.441.3348W} for a contrary view). In summary, we conclude that at present there is no unambiguous evidence for IMF variation in dwarf galaxies, and that, even if such evidence were found, with present methods it would provide only limited information about the characteristic mass, as opposed to the high-mass slope. JWST observations of high-redshift metal-poor dwarf satellites will help shed some light on IMF variations, if any, in these systems \citep{2021ApJ...913L..25G}.

\subsubsection{Starbursts and young massive clusters}

Our model predicts that starbursts and young massive clusters, which are high-pressure and often metal-rich environments, should have lower characteristic masses, and this too is a testable prediction. However, as with globulars and metal-poor dwarfs, observations of these systems are challenging. The most direct measurement is for the 30 Doradus region in the Large Magellanic Cloud, which is close enough to permit resolution of individual stars. \citet{2018Sci...359...69S} report that 30 Dor has a flatter high-mass IMF slope than is found in the Milky Way (see also, \citealt{2012A&A...547A..23B}). While this might at first blush seem to be inconsistent with our findings, it is important to note that this study only looked at stars more massive than $15\,\rm{M_{\odot}}$ without investigating the low mass part of the IMF, and previous work on the central cluster R136 in 30 Doradus by \cite{2009ApJ...707.1347A} that covered stars with masses as low as $0.5\,\rm{M_{\odot}}$ did not find any IMF variations. Thus, it is quite plausible that the characteristic stellar mass in 30 Doradus is not very different from that in the Milky Way, or is even lower, even if the slope at the high mass end is flatter. Moreover, it is worth noting that, while 30 Doradus is a starburst, its metallicity is only about half of Solar, so the effects of its high pressure may be offset by its somewhat low metallicity.

All other starburst systems for which claims of IMF variation exist in the literature are more distant, and thus the evidence is more indirect. \citet{2018Natur.558..260Z} and \citet{2019ApJ...879...17B} argue that the observed ratios of $^{13}$CO/C$^{18}$O in some local and high-redshift starburst galaxies provide evidence for a top-heavy IMF. Taken at face value, this again seem inconsistent with our results. However, as with 30 Dor, these observations only constrain the high mass slope of the IMF, not the characteristic mass. Moreover, \citet{2019A&A...624A.125M} show that isotopic ratios derived from unresolved observations may be confused by optical depth effects, leading to systematic errors. 

A final caveat is that chemical evolution models that call for IMF variation implicitly assume that the ejecta of supernovae (the primary source of $^{18}$O) and AGB stars (the primary source of $^{13}$C) are ejected in galactic winds in the same proportion. There is no reason to believe this to be the case, and excellent reason to believe the opposite, given that models \citep{2021MNRAS.502.5935S,2021MNRAS.504...53S,2021MNRAS.506.1295S} as well as observations \citep{Lopez20a, Cameron21a} suggest galactic winds in starbursts are preferentially enriched in supernova ejecta.

\subsubsection{Massive early type galaxies}

\label{s:discussions_earlytypeSF}
The centres of massive early type galaxies are very metal-rich, and, given their extremely high present-day surface densities, must have formed at very high gas pressure. There have been a number of observational results that focus on the nature of the IMF in massive early type galaxies (see the review by \citealt{2020ARA&A..58..577S}). These observations can be broadly divided into categories - one where spectroscopic measurements are taken \citep[e.g.,][]{2010Natur.468..940V,2012ApJ...760...71C,2013MNRAS.433.3017L,2013MNRAS.429L..15F,2015MNRAS.448L..82F,2017ApJ...837..166C}, and other where dynamical measurements often aided by gravitational lens modeling are used \citep[e.g.,][]{2010ApJ...709.1195T,2012Natur.484..485C,2014ApJ...792L..37M,2015MNRAS.449.3441S,2017ApJ...845..157N,2018MNRAS.474.4169O}. Some studies report a tight correlation between local metallicity and the IMF in these galaxies \citep{2015ApJ...806L..31M}. While both approaches find the presence of a more bottom-heavy IMF than the Milky Way in centres of massive ellipticals (see, however, \citealt{2015MNRAS.449.3441S}), the systematic differences between the two methodological approaches are still not well understood \citep{2014MNRAS.443L..69S,2020ARA&A..58..577S}. Based on our results in \autoref{s:results_IMF}, we expect a more bottom-heavy IMF in such an environment compared to that in the Milky Way, with a characteristic stellar mass $\lesssim 0.1\,\rm{M_{\odot}}$ (see also, \citealt{2021arXiv210703388Y}). This result from the model seems rather robust given that the variations in $M_{\rm{ch}}$ are tiny even when we use different chemical compositions, include additional processes, or change the dust-to-metal ratio. Thus, dust-dominated star formation at high pressure that naturally leads to a more bottom-heavy stellar population is a compelling candidate to explain the observations of the centres of massive elliptical galaxies. 

Recent studies have also discovered the presence of an IMF gradient in elliptical galaxies, where the central regions show a more bottom-heavy IMF than the Milky Way but the outskirts show an IMF compatible with the Milky Way \citep[e.g.,][]{2015MNRAS.447.1033M,2016MNRAS.457.1468L,2018MNRAS.474.4169O,2018MNRAS.478.4084S,2018Sci...360.1342C,2018MNRAS.477.3954P}. In the context of our model, such an IMF gradient seems viable if the pressure in galaxy centres is systemically larger than that in the disc or the outskirts in elliptical galaxies \citep[see also,][]{2015MNRAS.447.1033M}. Moreover, the presence of a negative metallicity gradient in early type galaxies that implies lower metallicities in the outskirts \citep{2018MNRAS.477.3954P} further strengthens the agreement between our model and observations. Most recently, \citet{2021arXiv210714243M} extended the spectroscopic IMF determination method to a sample of less massive quenched galaxies in the Fornax cluster. They find a strong positive correlation of metallicity with the slope of the IMF in the mass range $0.2-1\,\rm{M_{\odot}}$, indicating the presence of more high mass stars at lower metallicities, again qualitatively consistent with our model.

\subsubsection{Cosmological observations}

Claims of IMF variation based on cosmological observations represent a final observational category. \cite{2007MNRAS.379..985F} and \cite{2019MNRAS.487.3082C} study the total extragalactic background radiation observed today to constrain the global star formation history. These authors also find that a universal IMF all the way to very early times is in tension with their work, and a top-heavy IMF in the past is needed to explain the observed background radiation and stellar density at the present day. In another study, \cite{2011ApJ...727L..34W} find that an evolving IMF that becomes increasingly top-heavy at higher redshifts \citep{2008MNRAS.385..147D} can also easily reproduce the observed redshift distribution of gamma ray bursts. However, we emphasise that none of these results are unchallenged, and some are contradictory. Evidence that the IMF varies at all, let alone for the nature of that variation, remains hard to come by.

\subsection{Comparison with theoretical models}
\label{s:IMF_comparisonmodels}
It is also helpful to put our work in the context of previous theoretical studies, though we offer only a short summary of an extensive field; see \citet{2008IAUS..255..285S} for a review of earlier work. Some of the earliest models that studied the transition in the ISM as a function of metallicity are those of \cite{1997ApJ...480..145N} and \cite{1997ApJ...483...87S}, where the authors found a phase transition occurs in the ISM between $0.03 - 0.1\,\rm{Z_{\odot}}$. These models were however developed to study star formation in dwarf galaxies, did not include protostellar feedback, and did not explore the very metal-poor regime. Later models that followed, giving particular attention to the role of dust \citep{2006MNRAS.369.1437S,2010MNRAS.402..429S}, found a qualitatively similar transition in the characteristic mass (defined in these studies as the mass of a typical fragment) from highly super-solar to sub-solar as a function of metallicity \citep[figure~5][]{2006MNRAS.369.1437S}. However, the transition to a sub-solar characteristic mass occurs in these studies at very low metallicities ($Z \approx 10^{-6}\,\rm{Z_{\odot}}$), owing to efficient dust-induced fragmentation. The difference between our results and theirs is probably due to heating of dust grains post protostar formation due to feedback.

Based on theoretical modeling of the Jeans mass in collapsing clouds including dust-gas coupling, \cite{2008ApJ...681..365E} proposed that the reason for a universal IMF at high $Z$ is because there is little variation in the gas temperature as a function of metallicity for $ Z \gtrsim 0.2\,\rm{Z_{\odot}}$. This is consistent with our conclusions (see \autoref{fig:gdt_Zprofile}), though the underlying physical picture of gas thermodynamics presented in \citeauthor{2008ApJ...681..365E} is quite different from ours.

Finally, number of authors have also studied how an IMF that varies in response to metallicity, pressure, star formation rate, or other large-scale galactic properties would influence a range of other galactic properties, for example present-day mass functions, photometric correlations, and rates of compact object mergers \citep[e.g.][]{2018A&A...620A..39J, 2019MNRAS.485.4852G, 2020A&A...636A..10C}. Since these authors generally seek to explore the implications of IMF variation rather than develop theoretical models for why such variation should exist, their focus is somewhat different from ours, and we will therefore not discuss these works further.

\subsection{Comparison with simulations}
\label{s:IMF_comparisonsimulations}
Several simulations have investigated the properties of the IMF to search for variations with metallicity. These simulations can be divided in two categories -- one that includes a sub-grid model for a varying IMF in cosmological simulations \citep[e.g.,][]{2013MNRAS.436.2254B,2014MNRAS.444.3845F,2019MNRAS.482.2515B,2019MNRAS.482..118G,2020MNRAS.492....8A,2021arXiv210700663P}, and another focused on star formation where the IMF is self-consistently constructed based on fragmentation and distribution of stellar masses. We only discuss the latter approach because our work is directly comparable to those studies.

\cite{2009ApJ...696.1065J} find that the change in the dominant coolant from molecular H$_2$ to metals at $Z \sim 10^{-3.5}\,\rm{Z_{\odot}}$ metallicity can give rise to an apparent transition from primordial to modern day star formation based on the fragmentation characteristics \citep{2001MNRAS.328..969B}. While these authors did not include dust-gas coupling or protostellar feedback, they proposed that the transition in star formation is not caused by metals but by the formation of dust. Thus, our work confirms their hypothesis that it is indeed dust taking control of gas thermodynamics that sets the beginning of modern day star formation. Earlier simulations by \cite{2005A&A...435..611J} based on idealised stiff equations of state (i.e., not directly including dust-gas coupling or protostellar feedback) found that the transition in the regime of star formation at $Z \approx 10^{-3.5}\,\rm{Z_{\odot}}$ proposed by \cite{2001MNRAS.328..969B} is not real, and is simply a case of metals taking over from molecular H$_2$ to act as the dominant cooling agents at this metallicity. By contrast, our results suggest that the transition around $10^{-3.5}\,\rm{Z_{\odot}}$ is indeed real. This is because the effect of metals taking over from molecular H$_2$ to cool the gas is directly reflected in the characteristic stellar mass, which turns out to be different for the case of metal cooling as compared to that of molecular H$_2$ cooling.

Particularly noteworthy in the context of an evolving IMF are the results from numerical simulations run by \cite{2011ApJ...735...49M}, and \cite{2021arXiv210304997C}. The simulation setup used by \cite{2011ApJ...735...49M} is similar in essence to our theoretical framework, since protostellar feedback and dust-gas coupling are the key ingredients that are common in both studies. Additionally, \cite{2011ApJ...735...49M} also use the \cite{2005ApJ...631..792C} model to look at the evolution of the IMF as a function of metallicity. The key conclusion of \cite{2011ApJ...735...49M} is that varying the dust opacity by a large factor does not make any difference in fragmentation in a molecular cloud affected by protostellar radiation feedback, thus leading to a characteristic stellar mass independent of metallicity. Similar results were also obtained by \cite{2014MNRAS.442..285B} from their simulations of star cluster formation. However, both these works did not consider metallicities less than $0.01\,\rm{Z_{\odot}}$, and thus they could not study the transition in star formation from primordial to modern day, nor the transition in the characteristic stellar mass as a function of metallicity that only occurs below $0.01\,\rm{Z_{\odot}}$.

\cite{2021arXiv210304997C} use hydrodynamic simulations to study the transition in the IMF as a function of $Z$, finding that dust already starts to control fragmentation at $Z \gtrsim 10^{-5}\,\rm{Z_{\odot}}$ and that the IMF remains top-heavy for $Z \lesssim 10^{-2}\,\rm{Z_{\odot}}$. They use sink particles to follow the the evolution of star clusters up to $10^{4-5}\,\rm{yr}$ past the onset of star formation, and include cooling due to H$_2$, HD, and fine structure atomic lines (\ion{C}{ii} and \ion{O}{i}). However, they do not take into account the effects of protostellar feedback, which is the key ingredient in our model. While our results are broadly consistent with theirs, there are small differences. For example, we find that dust only contributes at the $\approx 10-30$ per cent level in setting the characteristic mass at $10^{-5} \leq Z/\rm{Z_{\odot}} \leq 10^{-3.5}$ at all pressures, in contrast to their findings (see their figure~3). A possible explanation of this discrepancy could be that we only look at the importance of dust at the critical location where $M_{\rm{enc}}=M_{\rm{BE}}$, which occurs at relatively low $n$ (and thus in a region of weak dust-gas coupling) in primordial-like environments, whereas they study the importance of dust at much higher densities. Another reason could be that protostellar feedback leads to much higher dust temperatures, which reduce the efficiency of dust in cooling the gas; in this context, the results of \cite{2021arXiv210304997C} (as well as similar earlier results by \citealt{2006MNRAS.369.1437S,2010MNRAS.402..429S,2011ApJ...729L...3D,2013ApJ...766..103D,2012ApJ...756L..35N}) can be separated from ours based on the inclusion of protostellar feedback. Further progress in this area requires more simulations like those of \cite{2021arXiv210304997C} where one can follow a time-dependent chemical evolution in 3D at different metallicities, combined with protostellar feedback models such as ours to yield realistic dust temperatures, rather than ignoring feedback and treating dust solely as a coolant.   

\section{Implications for cosmic star formation history}
\label{s:implications}

Our result that the IMF is top-heavy in extremely-metal poor environments, and our finding of a characteristic mass set to value in the range $0.1-1$ M$_\odot$ in metal-rich environments, are consistent with previous work. However, the IMF for the intermediate metal-poor regime ($10^{-3.5}\lesssim Z/\rm{Z_{\odot}} \lesssim 10^{-1.5}$) has received very limited investigation \citep{2018MNRAS.475.4378C,2021arXiv210304997C}, despite the growing number of observed metal-poor stars with $-3.5 \lesssim \rm{[Fe/H]} \lesssim -1.5$ \citep{2013ApJ...762...26Y,2015ARA&A..53..631F,2017MNRAS.471.2587S,2018ApJ...852...49H,2020MNRAS.496.4964A}. Such low metallicities can exist (even if only briefly) in ISM that has been recently enriched by Population III supernovae or AGB stars \citep{2007ApJ...665.1361T,2019MNRAS.488L.109N,2019ApJ...876...97E,2021Natur.595..223Y,2021arXiv210611592S,2021ApJ...912L..32P}, or in very low mass dwarf galaxies (e.g., Leo P -- \citealt{2015ApJ...812..158M}; J0811+4730 -- \citealt{2018MNRAS.473.1956I}; simulations -- \citealt{2018ApJ...865L..22E,2018ApJ...869...94E}). In fact, it is also proposed that a large fraction of the Universe can still exist in a metal-free regime even at redshifts as low as 5, and continue to form metal-free or extremely metal-poor stars \citep{2013ApJ...773...19M,2020MNRAS.497.2839L}. These results indicate the existence of an extended metal-poor phase of star formation in cosmic history. 

In our models, the existence of a distinct mode of star formation at $10^{-3.5}\leq Z/\rm{Z_{\odot}} \leq 10^{-1.5}$ is most clearly visible when we examine heating and cooling processes, which at all $P$ and $\sigma_{\rm v}$ exhibit a metal-cooling dominated regime that separates primordial star formation from modern day star formation. However, the signal is less clear simply from examining $M_{\rm ch}$. We see from \autoref{fig:Mch} and analogous figures that the characteristic stellar mass for this regime in $Z$ can be either sub-solar or super-solar, depending on the cloud pressure, velocity dispersion and metallicity. The exact value of $M_{\rm{ch}}$ in this metallicity regime depends rather sensitively on the chemical composition, as well as the effects of cosmic ray heating. As the first supernovae explode and produce copious amounts of dust within a short time ($\approx\,0.1-10\,\rm{Myr}$, \citealt{2018SSRv..214...63S}), or binary-rich massive stellar populations and low mass AGB stars expel dust forming elements after they form, subsequent star formation can already be heavily dust-dominated \citep{2001MNRAS.325..726T,2003ApJ...598..785N,2010ApJ...713....1C,2020MNRAS.491..440W,2020MNRAS.498.5652K,2018ApJ...857..132K,2021A&A...645A..10G,2021arXiv210804471V}. Such a scenario would lead to a bottom-heavy characteristic stellar mass within a short period of time, considerably shortening the period of transitionary, metal-poor star formation.

Given these findings, we speculate that the scatter in the duration of such a transitionary star formation phase can be fairly large. In fact, the rise in metallicity in certain environments can be very sharp, thereby completely skipping the transitionary phase. Another possibility is that star formation remains fairly quiescent for a long period of time, thus delaying the onset of the modern phase. Such star formation histories (and the corresponding metallicity histories) have been retrieved through detailed SED modeling for several galaxies \citep{2020MNRAS.498.5581B,2021MNRAS.tmp.1310T}, although without invoking IMF variations even at very low metallicities. This implies that while we can predict when the IMF becomes bottom-heavy, we cannot place strong constraints on the time it would take for it to do so. Convolution of models like ours with studies like those of \cite{2017MNRAS.472.2107G} where the authors test different IMF models in cosmological simulations at early times will be able to quantify the scatter present in the characteristic mass at low metallicities.

\section{Summary}
\label{s:conclusions}
In this work, we focus on understanding the evolution of the characteristic mass that sets the peak of the stellar initial mass function (IMF) as a function of metallicity. We consider collapsing dusty gas clouds that have just begun to fragment, at metallicities from $10^{-6} - 2\,\rm{Z_{\odot}}$, and pay careful attention to how radiation feedback from the first objects to form influences subsequent fragmentation. Our work thus compliments studies that focus on fragmentation in star-forming clouds at different metallicities prior to protostar formation \citep[e.g.,][]{2005ApJ...626..627O,2010ApJ...722.1793O}; such an extension is necessary to correctly capture the transition to modern-day star formation, where such feedback plays a decisive role in shaping the IMF. In addition to covering a wide range in metallicity, our models also span a large range in pressure and velocity dispersion of star-forming clouds, as observed in Milky Way and dwarf galaxies (low pressure environments) to super star clusters and the centres of massive early type galaxies (high pressure environments).

We demonstrate the existence of three phases of star formation that can be separated based on the ISM metallicity: (1.) the primordial phase below $Z \lesssim 10^{-4}\,\rm{Z_{\odot}}$ where gas cooling is dominated by molecular H$_2$ and HD, (2.) the transitionary phase between $10^{-4} \lesssim Z/\rm{Z_{\odot}} \lesssim 10^{-2}$ where metal cooling dominates gas thermodynamics, and (3.) the modern phase above $Z \gtrsim 10^{-2}\,\rm{Z_{\odot}}$ where dust governs gas thermodynamics. The effects of the changes between these thermodynamic regimes is reflected in the characteristic stellar mass that sets the peak of the IMF. We find that at low pressures ($P/k_{\rm{B}} \sim 10^{4}\,\rm{K\,cm^{-3}}$), the characteristic stellar mass is of the order of $50-100\,\rm{M_{\odot}}$ at extremely-low metallicites ($Z = 10^{-6}\,\rm{Z_{\odot}}$), and drops down to $0.3\,\rm{M_{\odot}}$ at $Z=\rm{Z_{\odot}}$ (see \autoref{fig:Mch}). At very high pressures ($P/k_{\rm{B}} \sim 10^8\,\rm{K\,cm^{-3}}$) and high metallicity, the characteristic stellar mass drops to $\sim 0.1\,\rm{M_{\odot}}$ (see \autoref{fig:Mch}), which is $3\times$ smaller than that we find above for a typical Milky Way molecular cloud. This provides a natural explanation for the even more bottom-heavy IMF found in early-type galaxy centres.

Our results thus suggest that the IMF became bottom-heavy around $Z \sim 10^{-2}\,\rm{Z_{\odot}}$ in the progenitors of late type galaxies like the Milky Way, whereas it became bottom-heavy around $Z \sim 10^{-4}\,\rm{Z_{\odot}}$ in the progenitors of massive, early type galaxies like NGC 1407. The earlier transition to a bottom-heavy IMF in early type galaxies is a result of the high-pressure ISM that existed in these galaxies. However, our models remain simplistic in the sense that they do not have the capability to predict the full distribution of the IMF. It is also possible that the trends in the characteristic mass with metallicity, velocity dispersion and pressure that we find are not noticeable in certain environments where other factors have a stronger influence on the IMF. Further advancements in our understanding of the evolution of the IMF as a function of metallicity will require chemodynamical simulations covering a wide range in metallicities coupled with models that properly account for protostellar feedback so that we can study fragmentation both pre- and post-collapse, and over metallicities ranging from the primordial to the modern.

\section*{Acknowledgements}
We thank Christoph Federrath, Patrick Hennebelle, Tereza Je{\v{r}}{\'a}bkov{\'a}, and Boyuan Liu for going through a preprint of this paper and providing feedback. We are indebted to an anonymous referee for critically reading and commenting on the manuscript. We also thank Thomas Nordlander for sharing a compilation of the most metal-poor stars discovered so far. We acknowledge insightful discussions with Roland Crocker on cosmic ray heating, with Gary Da Costa on very metal-poor stars, and with Stephanie Monty and Madeleine McKenzie on globular clusters. PS is supported by the Australian Government Research Training Program (RTP) Scholarship, and also acknowledges support by the Australian Research Council Centre of Excellence for All Sky Astrophysics in 3 Dimensions (ASTRO 3D) through project number CE170100013. MRK acknowledges funding provided by the Australian Research Council (ARC) through Discovery Project DP190101258 and Future Fellowship FT180100375. MRK is also the recipient of an Alexander von Humboldt award. Analysis was performed using \texttt{NUMPY} \citep{oliphant2006guide,2020arXiv200610256H} and \texttt{SCIPY} \citep{2020NatMe..17..261V}; plots were created using \texttt{MATPLOTLIB} \citep{Hunter:2007}. This research has made extensive use of NASA's Astrophysics Data System Bibliographic Services, image to data software \texttt{WEBPLOT} \texttt{DIGITIZER}, and the Leiden Atomic and Molecular Database \citep[LAMDA,][]{2005A&A...432..369S,2020Atoms...8...15V}. This research has also made extensive use of \texttt{MATHEMATICA} for numerical analyses.

\section*{Data Availability}
No data were generated for this work.
 


\bibliographystyle{mnras}
\bibliography{references}

\begin{thebibliography}{}
\makeatletter
\relax
\def\mn@urlcharsother{\let\do\@makeother \do\$\do\&\do\#\do\^\do\_\do\%\do\~}
\def\mn@doi{\begingroup\mn@urlcharsother \@ifnextchar [ {\mn@doi@}
  {\mn@doi@[]}}
\def\mn@doi@[#1]#2{\def\@tempa{#1}\ifx\@tempa\@empty \href
  {http://dx.doi.org/#2} {doi:#2}\else \href {http://dx.doi.org/#2} {#1}\fi
  \endgroup}
\def\mn@eprint#1#2{\mn@eprint@#1:#2::\@nil}
\def\mn@eprint@arXiv#1{\href {http://arxiv.org/abs/#1} {{\tt arXiv:#1}}}
\def\mn@eprint@dblp#1{\href {http://dblp.uni-trier.de/rec/bibtex/#1.xml}
  {dblp:#1}}
\def\mn@eprint@#1:#2:#3:#4\@nil{\def\@tempa {#1}\def\@tempb {#2}\def\@tempc
  {#3}\ifx \@tempc \@empty \let \@tempc \@tempb \let \@tempb \@tempa \fi \ifx
  \@tempb \@empty \def\@tempb {arXiv}\fi \@ifundefined
  {mn@eprint@\@tempb}{\@tempb:\@tempc}{\expandafter \expandafter \csname
  mn@eprint@\@tempb\endcsname \expandafter{\@tempc}}}

\bibitem[\protect\citeauthoryear{{Adams} \& {Fatuzzo}}{{Adams} \&
  {Fatuzzo}}{1996}]{1996ApJ...464..256A}
{Adams} F.~C.,  {Fatuzzo} M.,  1996, \mn@doi [\apj] {10.1086/177318}, \href
  {https://ui.adsabs.harvard.edu/abs/1996ApJ...464..256A} {464, 256}

\bibitem[\protect\citeauthoryear{{Adams}, {Ruden}  \& {Shu}}{{Adams}
  et~al.}{1989}]{1989ApJ...347..959A}
{Adams} F.~C.,  {Ruden} S.~P.,   {Shu} F.~H.,  1989, \mn@doi [\apj]
  {10.1086/168187}, \href
  {https://ui.adsabs.harvard.edu/abs/1989ApJ...347..959A} {347, 959}

\bibitem[\protect\citeauthoryear{{Aguado}, {Allende Prieto}, {Gonz{\'a}lez
  Hern{\'a}ndez}, {Rebolo}  \& {Caffau}}{{Aguado}
  et~al.}{2017}]{2017A&A...604A...9A}
{Aguado} D.~S.,  {Allende Prieto} C.,  {Gonz{\'a}lez Hern{\'a}ndez} J.~I.,
  {Rebolo} R.,   {Caffau} E.,  2017, \mn@doi [\aap]
  {10.1051/0004-6361/201731320}, \href
  {https://ui.adsabs.harvard.edu/abs/2017A&A...604A...9A} {604, A9}

\bibitem[\protect\citeauthoryear{{Aguado}, {Allende Prieto}, {Gonz{\'a}lez
  Hern{\'a}ndez}  \& {Rebolo}}{{Aguado} et~al.}{2018}]{2018ApJ...854L..34A}
{Aguado} D.~S.,  {Allende Prieto} C.,  {Gonz{\'a}lez Hern{\'a}ndez} J.~I.,
  {Rebolo} R.,  2018, \mn@doi [\apjl] {10.3847/2041-8213/aaadb8}, \href
  {https://ui.adsabs.harvard.edu/abs/2018ApJ...854L..34A} {854, L34}

\bibitem[\protect\citeauthoryear{{Alves}, {Lombardi}  \& {Lada}}{{Alves}
  et~al.}{2007}]{2007A&A...462L..17A}
{Alves} J.,  {Lombardi} M.,   {Lada} C.~J.,  2007, \mn@doi [\aap]
  {10.1051/0004-6361:20066389}, \href
  {https://ui.adsabs.harvard.edu/abs/2007A&A...462L..17A} {462, L17}

\bibitem[\protect\citeauthoryear{{Andersen}, {Zinnecker}, {Moneti},
  {McCaughrean}, {Brandl}, {Brandner}, {Meylan}  \& {Hunter}}{{Andersen}
  et~al.}{2009}]{2009ApJ...707.1347A}
{Andersen} M.,  {Zinnecker} H.,  {Moneti} A.,  {McCaughrean} M.~J.,  {Brandl}
  B.,  {Brandner} W.,  {Meylan} G.,   {Hunter} D.,  2009, \mn@doi [\apj]
  {10.1088/0004-637X/707/2/1347}, \href
  {https://ui.adsabs.harvard.edu/abs/2009ApJ...707.1347A} {707, 1347}

\bibitem[\protect\citeauthoryear{{Andrews} et~al.,}{{Andrews}
  et~al.}{2014}]{2014ApJ...793....4A}
{Andrews} J.~E.,  et~al., 2014, \mn@doi [\apj] {10.1088/0004-637X/793/1/4},
  \href {https://ui.adsabs.harvard.edu/abs/2014ApJ...793....4A} {793, 4}

\bibitem[\protect\citeauthoryear{{Applebaum}, {Brooks}, {Quinn}  \&
  {Christensen}}{{Applebaum} et~al.}{2020}]{2020MNRAS.492....8A}
{Applebaum} E.,  {Brooks} A.~M.,  {Quinn} T.~R.,   {Christensen} C.~R.,  2020,
  \mn@doi [\mnras] {10.1093/mnras/stz3331}, \href
  {https://ui.adsabs.harvard.edu/abs/2020MNRAS.492....8A} {492, 8}

\bibitem[\protect\citeauthoryear{{Arentsen} et~al.,}{{Arentsen}
  et~al.}{2020}]{2020MNRAS.496.4964A}
{Arentsen} A.,  et~al., 2020, \mn@doi [\mnras] {10.1093/mnras/staa1661}, \href
  {https://ui.adsabs.harvard.edu/abs/2020MNRAS.496.4964A} {496, 4964}

\bibitem[\protect\citeauthoryear{{Balashev} et~al.,}{{Balashev}
  et~al.}{2017}]{2017MNRAS.470.2890B}
{Balashev} S.~A.,  et~al., 2017, \mn@doi [\mnras] {10.1093/mnras/stx1339},
  \href {https://ui.adsabs.harvard.edu/abs/2017MNRAS.470.2890B} {470, 2890}

\bibitem[\protect\citeauthoryear{{Banerjee} \& {Kroupa}}{{Banerjee} \&
  {Kroupa}}{2012}]{2012A&A...547A..23B}
{Banerjee} S.,  {Kroupa} P.,  2012, \mn@doi [\aap]
  {10.1051/0004-6361/201218972}, \href
  {https://ui.adsabs.harvard.edu/abs/2012A&A...547A..23B} {547, A23}

\bibitem[\protect\citeauthoryear{{Barber}, {Schaye}  \& {Crain}}{{Barber}
  et~al.}{2019}]{2019MNRAS.482.2515B}
{Barber} C.,  {Schaye} J.,   {Crain} R.~A.,  2019, \mn@doi [\mnras]
  {10.1093/mnras/sty2825}, \href
  {https://ui.adsabs.harvard.edu/abs/2019MNRAS.482.2515B} {482, 2515}

\bibitem[\protect\citeauthoryear{{Barinovs}, {van Hemert}, {Krems}  \&
  {Dalgarno}}{{Barinovs} et~al.}{2005}]{2005ApJ...620..537B}
{Barinovs} {\u{G}}.,  {van Hemert} M.~C.,  {Krems} R.,   {Dalgarno} A.,  2005,
  \mn@doi [\apj] {10.1086/426860}, \href
  {https://ui.adsabs.harvard.edu/abs/2005ApJ...620..537B} {620, 537}

\bibitem[\protect\citeauthoryear{{Bastian} \& {Lardo}}{{Bastian} \&
  {Lardo}}{2018}]{2018ARA&A..56...83B}
{Bastian} N.,  {Lardo} C.,  2018, \mn@doi [\araa]
  {10.1146/annurev-astro-081817-051839}, \href
  {https://ui.adsabs.harvard.edu/abs/2018ARA&A..56...83B} {56, 83}

\bibitem[\protect\citeauthoryear{{Bastian}, {Saglia}, {Goudfrooij},
  {Kissler-Patig}, {Maraston}, {Schweizer}  \& {Zoccali}}{{Bastian}
  et~al.}{2006}]{2006A&A...448..881B}
{Bastian} N.,  {Saglia} R.~P.,  {Goudfrooij} P.,  {Kissler-Patig} M.,
  {Maraston} C.,  {Schweizer} F.,   {Zoccali} M.,  2006, \mn@doi [\aap]
  {10.1051/0004-6361:20054177}, \href
  {https://ui.adsabs.harvard.edu/abs/2006A&A...448..881B} {448, 881}

\bibitem[\protect\citeauthoryear{{Bastian}, {Covey}  \& {Meyer}}{{Bastian}
  et~al.}{2010}]{2010ARA&A..48..339B}
{Bastian} N.,  {Covey} K.~R.,   {Meyer} M.~R.,  2010, \mn@doi [\araa]
  {10.1146/annurev-astro-082708-101642}, \href
  {https://ui.adsabs.harvard.edu/abs/2010ARA&A..48..339B} {48, 339}

\bibitem[\protect\citeauthoryear{{Bate}}{{Bate}}{2005}]{2005MNRAS.363..363B}
{Bate} M.~R.,  2005, \mn@doi [\mnras] {10.1111/j.1365-2966.2005.09476.x}, \href
  {https://ui.adsabs.harvard.edu/abs/2005MNRAS.363..363B} {363, 363}

\bibitem[\protect\citeauthoryear{{Bate}}{{Bate}}{2009}]{Bate09a}
{Bate} M.~R.,  2009, \mn@doi [\mnras] {10.1111/j.1365-2966.2008.14165.x}, \href
  {http://adsabs.harvard.edu/abs/2009MNRAS.392.1363B} {392, 1363}

\bibitem[\protect\citeauthoryear{{Bate}}{{Bate}}{2012}]{Bate12a}
{Bate} M.~R.,  2012, \mn@doi [\mnras] {10.1111/j.1365-2966.2011.19955.x}, \href
  {http://adsabs.harvard.edu/abs/2012MNRAS.419.3115B} {419, 3115}

\bibitem[\protect\citeauthoryear{{Bate}}{{Bate}}{2014}]{2014MNRAS.442..285B}
{Bate} M.~R.,  2014, \mn@doi [\mnras] {10.1093/mnras/stu795}, \href
  {https://ui.adsabs.harvard.edu/abs/2014MNRAS.442..285B} {442, 285}

\bibitem[\protect\citeauthoryear{{Bate}}{{Bate}}{2019}]{2019MNRAS.484.2341B}
{Bate} M.~R.,  2019, \mn@doi [\mnras] {10.1093/mnras/stz103}, \href
  {https://ui.adsabs.harvard.edu/abs/2019MNRAS.484.2341B} {484, 2341}

\bibitem[\protect\citeauthoryear{{Bate} \& {Keto}}{{Bate} \&
  {Keto}}{2015}]{Bate15a}
{Bate} M.~R.,  {Keto} E.~R.,  2015, \mn@doi [\mnras] {10.1093/mnras/stv451},
  \href {http://adsabs.harvard.edu/abs/2015MNRAS.449.2643B} {449, 2643}

\bibitem[\protect\citeauthoryear{{Bate} \& {Lor{\'e}n-Aguilar}}{{Bate} \&
  {Lor{\'e}n-Aguilar}}{2017}]{2017MNRAS.465.1089B}
{Bate} M.~R.,  {Lor{\'e}n-Aguilar} P.,  2017, \mn@doi [\mnras]
  {10.1093/mnras/stw2853}, \href
  {https://ui.adsabs.harvard.edu/abs/2017MNRAS.465.1089B} {465, 1089}

\bibitem[\protect\citeauthoryear{{Bate}, {Bonnell}  \& {Bromm}}{{Bate}
  et~al.}{2002}]{Bate02a}
{Bate} M.~R.,  {Bonnell} I.~A.,   {Bromm} V.,  2002, \mn@doi [\mnras]
  {10.1046/j.1365-8711.2002.05539.x}, \href
  {https://ui.adsabs.harvard.edu/abs/2002MNRAS.332L..65B} {332, L65}

\bibitem[\protect\citeauthoryear{{Bate}, {Bonnell}  \& {Bromm}}{{Bate}
  et~al.}{2003}]{Bate03a}
{Bate} M.~R.,  {Bonnell} I.~A.,   {Bromm} V.,  2003, \mn@doi [\mnras]
  {10.1046/j.1365-8711.2003.06210.x}, \href
  {http://adsabs.harvard.edu/abs/2003MNRAS.339..577B} {339, 577}

\bibitem[\protect\citeauthoryear{{Baumgardt}}{{Baumgardt}}{2017}]{2017MNRAS.464.2174B}
{Baumgardt} H.,  2017, \mn@doi [\mnras] {10.1093/mnras/stw2488}, \href
  {https://ui.adsabs.harvard.edu/abs/2017MNRAS.464.2174B} {464, 2174}

\bibitem[\protect\citeauthoryear{{Baumgardt} \& {Makino}}{{Baumgardt} \&
  {Makino}}{2003}]{2003MNRAS.340..227B}
{Baumgardt} H.,  {Makino} J.,  2003, \mn@doi [\mnras]
  {10.1046/j.1365-8711.2003.06286.x}, \href
  {https://ui.adsabs.harvard.edu/abs/2003MNRAS.340..227B} {340, 227}

\bibitem[\protect\citeauthoryear{{Beers} \& {Christlieb}}{{Beers} \&
  {Christlieb}}{2005}]{2005ARA&A..43..531B}
{Beers} T.~C.,  {Christlieb} N.,  2005, \mn@doi [\araa]
  {10.1146/annurev.astro.42.053102.134057}, \href
  {https://ui.adsabs.harvard.edu/abs/2005ARA&A..43..531B} {43, 531}

\bibitem[\protect\citeauthoryear{{Bekki}}{{Bekki}}{2013}]{2013MNRAS.436.2254B}
{Bekki} K.,  2013, \mn@doi [\mnras] {10.1093/mnras/stt1735}, \href
  {https://ui.adsabs.harvard.edu/abs/2013MNRAS.436.2254B} {436, 2254}

\bibitem[\protect\citeauthoryear{{Bellstedt} et~al.,}{{Bellstedt}
  et~al.}{2020}]{2020MNRAS.498.5581B}
{Bellstedt} S.,  et~al., 2020, \mn@doi [\mnras] {10.1093/mnras/staa2620}, \href
  {https://ui.adsabs.harvard.edu/abs/2020MNRAS.498.5581B} {498, 5581}

\bibitem[\protect\citeauthoryear{{Bialy} \& {Sternberg}}{{Bialy} \&
  {Sternberg}}{2015}]{2015MNRAS.450.4424B}
{Bialy} S.,  {Sternberg} A.,  2015, \mn@doi [\mnras] {10.1093/mnras/stv851},
  \href {https://ui.adsabs.harvard.edu/abs/2015MNRAS.450.4424B} {450, 4424}

\bibitem[\protect\citeauthoryear{{Bisbas}, {Papadopoulos}  \& {Viti}}{{Bisbas}
  et~al.}{2015}]{2015ApJ...803...37B}
{Bisbas} T.~G.,  {Papadopoulos} P.~P.,   {Viti} S.,  2015, \mn@doi [\apj]
  {10.1088/0004-637X/803/1/37}, \href
  {https://ui.adsabs.harvard.edu/abs/2015ApJ...803...37B} {803, 37}

\bibitem[\protect\citeauthoryear{{Bisbas}, {van Dishoeck}, {Papadopoulos},
  {Sz{\H{u}}cs}, {Bialy}  \& {Zhang}}{{Bisbas}
  et~al.}{2017}]{2017ApJ...839...90B}
{Bisbas} T.~G.,  {van Dishoeck} E.~F.,  {Papadopoulos} P.~P.,  {Sz{\H{u}}cs}
  L.,  {Bialy} S.,   {Zhang} Z.-Y.,  2017, \mn@doi [\apj]
  {10.3847/1538-4357/aa696d}, \href
  {https://ui.adsabs.harvard.edu/abs/2017ApJ...839...90B} {839, 90}

\bibitem[\protect\citeauthoryear{{Bolatto}, {Leroy}, {Rosolowsky}, {Walter}  \&
  {Blitz}}{{Bolatto} et~al.}{2008}]{2008ApJ...686..948B}
{Bolatto} A.~D.,  {Leroy} A.~K.,  {Rosolowsky} E.,  {Walter} F.,   {Blitz} L.,
  2008, \mn@doi [\apj] {10.1086/591513}, \href
  {https://ui.adsabs.harvard.edu/abs/2008ApJ...686..948B} {686, 948}

\bibitem[\protect\citeauthoryear{{Bolatto} et~al.,}{{Bolatto}
  et~al.}{2011}]{2011ApJ...741...12B}
{Bolatto} A.~D.,  et~al., 2011, \mn@doi [\apj] {10.1088/0004-637X/741/1/12},
  \href {https://ui.adsabs.harvard.edu/abs/2011ApJ...741...12B} {741, 12}

\bibitem[\protect\citeauthoryear{{Bonnell} \& {Bate}}{{Bonnell} \&
  {Bate}}{1994}]{1994MNRAS.269L..45B}
{Bonnell} I.~A.,  {Bate} M.~R.,  1994, \mn@doi [\mnras]
  {10.1093/mnras/269.1.L45}, \href
  {https://ui.adsabs.harvard.edu/abs/1994MNRAS.269L..45B} {269, L45}

\bibitem[\protect\citeauthoryear{{Bonnor}}{{Bonnor}}{1956}]{1956MNRAS.116..351B}
{Bonnor} W.~B.,  1956, \mn@doi [\mnras] {10.1093/mnras/116.3.351}, \href
  {https://ui.adsabs.harvard.edu/abs/1956MNRAS.116..351B} {116, 351}

\bibitem[\protect\citeauthoryear{{Bonnor}}{{Bonnor}}{1957}]{1957MNRAS.117..104B}
{Bonnor} W.~B.,  1957, \mn@doi [\mnras] {10.1093/mnras/117.1.104}, \href
  {https://ui.adsabs.harvard.edu/abs/1957MNRAS.117..104B} {117, 104}

\bibitem[\protect\citeauthoryear{{Bromm}}{{Bromm}}{2013}]{2013RPPh...76k2901B}
{Bromm} V.,  2013, \mn@doi [Reports on Progress in Physics]
  {10.1088/0034-4885/76/11/112901}, \href
  {https://ui.adsabs.harvard.edu/abs/2013RPPh...76k2901B} {76, 112901}

\bibitem[\protect\citeauthoryear{{Bromm}, {Ferrara}, {Coppi}  \&
  {Larson}}{{Bromm} et~al.}{2001}]{2001MNRAS.328..969B}
{Bromm} V.,  {Ferrara} A.,  {Coppi} P.~S.,   {Larson} R.~B.,  2001, \mn@doi
  [\mnras] {10.1046/j.1365-8711.2001.04915.x}, \href
  {https://ui.adsabs.harvard.edu/abs/2001MNRAS.328..969B} {328, 969}

\bibitem[\protect\citeauthoryear{{Bromm}, {Yoshida}, {Hernquist}  \&
  {McKee}}{{Bromm} et~al.}{2009}]{2009Natur.459...49B}
{Bromm} V.,  {Yoshida} N.,  {Hernquist} L.,   {McKee} C.~F.,  2009, \mn@doi
  [\nat] {10.1038/nature07990}, \href
  {https://ui.adsabs.harvard.edu/abs/2009Natur.459...49B} {459, 49}

\bibitem[\protect\citeauthoryear{{Brown} \& {Wilson}}{{Brown} \&
  {Wilson}}{2019}]{2019ApJ...879...17B}
{Brown} T.,  {Wilson} C.~D.,  2019, \mn@doi [\apj] {10.3847/1538-4357/ab2246},
  \href {https://ui.adsabs.harvard.edu/abs/2019ApJ...879...17B} {879, 17}

\bibitem[\protect\citeauthoryear{{Burke} \& {Hollenbach}}{{Burke} \&
  {Hollenbach}}{1983}]{1983ApJ...265..223B}
{Burke} J.~R.,  {Hollenbach} D.~J.,  1983, \mn@doi [\apj] {10.1086/160667},
  \href {https://ui.adsabs.harvard.edu/abs/1983ApJ...265..223B} {265, 223}

\bibitem[\protect\citeauthoryear{{Caffau} et~al.,}{{Caffau}
  et~al.}{2011}]{2011Natur.477...67C}
{Caffau} E.,  et~al., 2011, \mn@doi [\nat] {10.1038/nature10377}, \href
  {https://ui.adsabs.harvard.edu/abs/2011Natur.477...67C} {477, 67}

\bibitem[\protect\citeauthoryear{{Cameron} et~al.,}{{Cameron}
  et~al.}{2021}]{Cameron21a}
{Cameron} A.~J.,  et~al., 2021, \apjl, \href
  {https://ui.adsabs.harvard.edu/abs/2021arXiv210813211C} {918, L16}

\bibitem[\protect\citeauthoryear{{Cappellari} et~al.,}{{Cappellari}
  et~al.}{2012}]{2012Natur.484..485C}
{Cappellari} M.,  et~al., 2012, \mn@doi [\nat] {10.1038/nature10972}, \href
  {https://ui.adsabs.harvard.edu/abs/2012Natur.484..485C} {484, 485}

\bibitem[\protect\citeauthoryear{{Caselli} \& {Myers}}{{Caselli} \&
  {Myers}}{1995}]{1995ApJ...446..665C}
{Caselli} P.,  {Myers} P.~C.,  1995, \mn@doi [\apj] {10.1086/175825}, \href
  {https://ui.adsabs.harvard.edu/abs/1995ApJ...446..665C} {446, 665}

\bibitem[\protect\citeauthoryear{{Cazaux} \& {Spaans}}{{Cazaux} \&
  {Spaans}}{2009}]{2009A&A...496..365C}
{Cazaux} S.,  {Spaans} M.,  2009, \mn@doi [\aap] {10.1051/0004-6361:200811302},
  \href {https://ui.adsabs.harvard.edu/abs/2009A&A...496..365C} {496, 365}

\bibitem[\protect\citeauthoryear{{Chabrier}}{{Chabrier}}{2003}]{2003PASP..115..763C}
{Chabrier} G.,  2003, \mn@doi [\pasp] {10.1086/376392}, \href
  {https://ui.adsabs.harvard.edu/abs/2003PASP..115..763C} {115, 763}

\bibitem[\protect\citeauthoryear{{Chabrier}}{{Chabrier}}{2005}]{2005ASSL..327...41C}
{Chabrier} G.,  2005, in {Corbelli} E.,  {Palla} F.,   {Zinnecker} H.,  eds, ~
  Vol. 327, The Initial Mass Function 50 Years Later. Springer, p.~41,
  \mn@doi{10.1007/978-1-4020-3407-7\_5}

\bibitem[\protect\citeauthoryear{{Chabrier}, {Hennebelle}  \&
  {Charlot}}{{Chabrier} et~al.}{2014}]{2014ApJ...796...75C}
{Chabrier} G.,  {Hennebelle} P.,   {Charlot} S.,  2014, \mn@doi [\apj]
  {10.1088/0004-637X/796/2/75}, \href
  {https://ui.adsabs.harvard.edu/abs/2014ApJ...796...75C} {796, 75}

\bibitem[\protect\citeauthoryear{{Chakrabarti} \& {McKee}}{{Chakrabarti} \&
  {McKee}}{2005}]{2005ApJ...631..792C}
{Chakrabarti} S.,  {McKee} C.~F.,  2005, \mn@doi [\apj] {10.1086/432659}, \href
  {https://ui.adsabs.harvard.edu/abs/2005ApJ...631..792C} {631, 792}

\bibitem[\protect\citeauthoryear{{Chakrabarti} \& {McKee}}{{Chakrabarti} \&
  {McKee}}{2008}]{2008ApJ...683..693C}
{Chakrabarti} S.,  {McKee} C.~F.,  2008, \mn@doi [\apj] {10.1086/589637}, \href
  {https://ui.adsabs.harvard.edu/abs/2008ApJ...683..693C} {683, 693}

\bibitem[\protect\citeauthoryear{{Chakrabarti}, {Magnelli}, {McKee}, {Lutz},
  {Berta}, {Popesso}  \& {Pozzi}}{{Chakrabarti}
  et~al.}{2013}]{2013ApJ...773..113C}
{Chakrabarti} S.,  {Magnelli} B.,  {McKee} C.~F.,  {Lutz} D.,  {Berta} S.,
  {Popesso} P.,   {Pozzi} F.,  2013, \mn@doi [\apj]
  {10.1088/0004-637X/773/2/113}, \href
  {https://ui.adsabs.harvard.edu/abs/2013ApJ...773..113C} {773, 113}

\bibitem[\protect\citeauthoryear{{Cherchneff} \& {Dwek}}{{Cherchneff} \&
  {Dwek}}{2010}]{2010ApJ...713....1C}
{Cherchneff} I.,  {Dwek} E.,  2010, \mn@doi [\apj] {10.1088/0004-637X/713/1/1},
  \href {https://ui.adsabs.harvard.edu/abs/2010ApJ...713....1C} {713, 1}

\bibitem[\protect\citeauthoryear{{Chevance} et~al.,}{{Chevance}
  et~al.}{2020}]{2020MNRAS.494.5279C}
{Chevance} M.,  et~al., 2020, \mn@doi [\mnras] {10.1093/mnras/staa1106}, \href
  {https://ui.adsabs.harvard.edu/abs/2020MNRAS.494.5279C} {494, 5279}

\bibitem[\protect\citeauthoryear{{Chiaki} \& {Yoshida}}{{Chiaki} \&
  {Yoshida}}{2020}]{2020arXiv200806107C}
{Chiaki} G.,  {Yoshida} N.,  2020, arXiv e-prints, \href
  {https://ui.adsabs.harvard.edu/abs/2020arXiv200806107C} {p. arXiv:2008.06107}

\bibitem[\protect\citeauthoryear{{Chiaki}, {Marassi}, {Nozawa}, {Yoshida},
  {Schneider}, {Omukai}, {Limongi}  \& {Chieffi}}{{Chiaki}
  et~al.}{2015}]{2015MNRAS.446.2659C}
{Chiaki} G.,  {Marassi} S.,  {Nozawa} T.,  {Yoshida} N.,  {Schneider} R.,
  {Omukai} K.,  {Limongi} M.,   {Chieffi} A.,  2015, \mn@doi [\mnras]
  {10.1093/mnras/stu2298}, \href
  {https://ui.adsabs.harvard.edu/abs/2015MNRAS.446.2659C} {446, 2659}

\bibitem[\protect\citeauthoryear{{Chiaki}, {Yoshida}  \& {Hirano}}{{Chiaki}
  et~al.}{2016}]{2016MNRAS.463.2781C}
{Chiaki} G.,  {Yoshida} N.,   {Hirano} S.,  2016, \mn@doi [\mnras]
  {10.1093/mnras/stw2120}, \href
  {https://ui.adsabs.harvard.edu/abs/2016MNRAS.463.2781C} {463, 2781}

\bibitem[\protect\citeauthoryear{{Chiaki}, {Susa}  \& {Hirano}}{{Chiaki}
  et~al.}{2018}]{2018MNRAS.475.4378C}
{Chiaki} G.,  {Susa} H.,   {Hirano} S.,  2018, \mn@doi [\mnras]
  {10.1093/mnras/sty040}, \href
  {https://ui.adsabs.harvard.edu/abs/2018MNRAS.475.4378C} {475, 4378}

\bibitem[\protect\citeauthoryear{{Chon} \& {Hosokawa}}{{Chon} \&
  {Hosokawa}}{2019}]{2019MNRAS.488.2658C}
{Chon} S.,  {Hosokawa} T.,  2019, \mn@doi [\mnras] {10.1093/mnras/stz1824},
  \href {https://ui.adsabs.harvard.edu/abs/2019MNRAS.488.2658C} {488, 2658}

\bibitem[\protect\citeauthoryear{{Chon}, {Omukai}  \& {Schneider}}{{Chon}
  et~al.}{2021}]{2021arXiv210304997C}
{Chon} S.,  {Omukai} K.,   {Schneider} R.,  2021, \mn@doi [\mnras]
  {10.1093/mnras/stab2497}, \href
  {https://ui.adsabs.harvard.edu/abs/2021MNRAS.tmp.2268C} {}

\bibitem[\protect\citeauthoryear{{Christlieb}, {Gustafsson}, {Korn}, {Barklem},
  {Beers}, {Bessell}, {Karlsson}  \& {Mizuno-Wiedner}}{{Christlieb}
  et~al.}{2004}]{2004ApJ...603..708C}
{Christlieb} N.,  {Gustafsson} B.,  {Korn} A.~J.,  {Barklem} P.~S.,  {Beers}
  T.~C.,  {Bessell} M.~S.,  {Karlsson} T.,   {Mizuno-Wiedner} M.,  2004,
  \mn@doi [\apj] {10.1086/381237}, \href
  {https://ui.adsabs.harvard.edu/abs/2004ApJ...603..708C} {603, 708}

\bibitem[\protect\citeauthoryear{{Chru{\'s}li{\'n}ska},
  {Je{\v{r}}{\'a}bkov{\'a}}, {Nelemans}  \& {Yan}}{{Chru{\'s}li{\'n}ska}
  et~al.}{2020}]{2020A&A...636A..10C}
{Chru{\'s}li{\'n}ska} M.,  {Je{\v{r}}{\'a}bkov{\'a}} T.,  {Nelemans} G.,
  {Yan} Z.,  2020, \mn@doi [\aap] {10.1051/0004-6361/202037688}, \href
  {https://ui.adsabs.harvard.edu/abs/2020A&A...636A..10C} {636, A10}

\bibitem[\protect\citeauthoryear{{Clark}, {Glover}, {Smith}, {Greif}, {Klessen}
   \& {Bromm}}{{Clark} et~al.}{2011a}]{Clark11a}
{Clark} P.~C.,  {Glover} S.~C.~O.,  {Smith} R.~J.,  {Greif} T.~H.,  {Klessen}
  R.~S.,   {Bromm} V.,  2011a, \mn@doi [Science] {10.1126/science.1198027},
  \href {http://adsabs.harvard.edu/abs/2011Sci...331.1040C} {331, 1040}

\bibitem[\protect\citeauthoryear{{Clark}, {Glover}, {Klessen}  \&
  {Bromm}}{{Clark} et~al.}{2011b}]{2011ApJ...727..110C}
{Clark} P.~C.,  {Glover} S. C.~O.,  {Klessen} R.~S.,   {Bromm} V.,  2011b,
  \mn@doi [\apj] {10.1088/0004-637X/727/2/110}, \href
  {https://ui.adsabs.harvard.edu/abs/2011ApJ...727..110C} {727, 110}

\bibitem[\protect\citeauthoryear{{Clarke} \& {Pringle}}{{Clarke} \&
  {Pringle}}{1991}]{1991MNRAS.249..588C}
{Clarke} C.~J.,  {Pringle} J.~E.,  1991, \mn@doi [\mnras]
  {10.1093/mnras/249.4.588}, \href
  {https://ui.adsabs.harvard.edu/abs/1991MNRAS.249..588C} {249, 588}

\bibitem[\protect\citeauthoryear{{Collett} et~al.,}{{Collett}
  et~al.}{2018}]{2018Sci...360.1342C}
{Collett} T.~E.,  et~al., 2018, \mn@doi [Science] {10.1126/science.aao2469},
  \href {https://ui.adsabs.harvard.edu/abs/2018Sci...360.1342C} {360, 1342}

\bibitem[\protect\citeauthoryear{{Colman} \& {Teyssier}}{{Colman} \&
  {Teyssier}}{2020}]{2020MNRAS.492.4727C}
{Colman} T.,  {Teyssier} R.,  2020, \mn@doi [\mnras] {10.1093/mnras/staa075},
  \href {https://ui.adsabs.harvard.edu/abs/2020MNRAS.492.4727C} {492, 4727}

\bibitem[\protect\citeauthoryear{{Conroy} \& {van Dokkum}}{{Conroy} \& {van
  Dokkum}}{2012}]{2012ApJ...760...71C}
{Conroy} C.,  {van Dokkum} P.~G.,  2012, \mn@doi [\apj]
  {10.1088/0004-637X/760/1/71}, \href
  {https://ui.adsabs.harvard.edu/abs/2012ApJ...760...71C} {760, 71}

\bibitem[\protect\citeauthoryear{{Conroy}, {van Dokkum}  \&
  {Villaume}}{{Conroy} et~al.}{2017}]{2017ApJ...837..166C}
{Conroy} C.,  {van Dokkum} P.~G.,   {Villaume} A.,  2017, \mn@doi [\apj]
  {10.3847/1538-4357/aa6190}, \href
  {https://ui.adsabs.harvard.edu/abs/2017ApJ...837..166C} {837, 166}

\bibitem[\protect\citeauthoryear{{Cooke}, {Pettini}, {Jorgenson}, {Murphy}  \&
  {Steidel}}{{Cooke} et~al.}{2014}]{2014ApJ...781...31C}
{Cooke} R.~J.,  {Pettini} M.,  {Jorgenson} R.~A.,  {Murphy} M.~T.,   {Steidel}
  C.~C.,  2014, \mn@doi [\apj] {10.1088/0004-637X/781/1/31}, \href
  {https://ui.adsabs.harvard.edu/abs/2014ApJ...781...31C} {781, 31}

\bibitem[\protect\citeauthoryear{{Cowley}, {Lacey}, {Baugh}, {Cole}, {Frenk}
  \& {Lagos}}{{Cowley} et~al.}{2019}]{2019MNRAS.487.3082C}
{Cowley} W.~I.,  {Lacey} C.~G.,  {Baugh} C.~M.,  {Cole} S.,  {Frenk} C.~S.,
  {Lagos} C. d.~P.,  2019, \mn@doi [\mnras] {10.1093/mnras/stz1398}, \href
  {https://ui.adsabs.harvard.edu/abs/2019MNRAS.487.3082C} {487, 3082}

\bibitem[\protect\citeauthoryear{{Crocker}, {Krumholz}  \&
  {Thompson}}{{Crocker} et~al.}{2021}]{2021MNRAS.502.1312C}
{Crocker} R.~M.,  {Krumholz} M.~R.,   {Thompson} T.~A.,  2021, \mn@doi [\mnras]
  {10.1093/mnras/stab148}, \href
  {https://ui.adsabs.harvard.edu/abs/2021MNRAS.502.1312C} {502, 1312}

\bibitem[\protect\citeauthoryear{{Cunningham}, {Krumholz}, {McKee}  \&
  {Klein}}{{Cunningham} et~al.}{2018}]{2018MNRAS.476..771C}
{Cunningham} A.~J.,  {Krumholz} M.~R.,  {McKee} C.~F.,   {Klein} R.~I.,  2018,
  \mn@doi [\mnras] {10.1093/mnras/sty154}, \href
  {https://ui.adsabs.harvard.edu/abs/2018MNRAS.476..771C} {476, 771}

\bibitem[\protect\citeauthoryear{{D'Hendecourt}, {Allamandola}  \&
  {Greenberg}}{{D'Hendecourt} et~al.}{1985}]{1985A&A...152..130D}
{D'Hendecourt} L.~B.,  {Allamandola} L.~J.,   {Greenberg} J.~M.,  1985, \aap,
  \href {https://ui.adsabs.harvard.edu/abs/1985A&A...152..130D} {152, 130}

\bibitem[\protect\citeauthoryear{{Dabringhausen}, {Kroupa}  \&
  {Baumgardt}}{{Dabringhausen} et~al.}{2009}]{2009MNRAS.394.1529D}
{Dabringhausen} J.,  {Kroupa} P.,   {Baumgardt} H.,  2009, \mn@doi [\mnras]
  {10.1111/j.1365-2966.2009.14425.x}, \href
  {https://ui.adsabs.harvard.edu/abs/2009MNRAS.394.1529D} {394, 1529}

\bibitem[\protect\citeauthoryear{{Dabringhausen}, {Fellhauer}  \&
  {Kroupa}}{{Dabringhausen} et~al.}{2010}]{2010MNRAS.403.1054D}
{Dabringhausen} J.,  {Fellhauer} M.,   {Kroupa} P.,  2010, \mn@doi [\mnras]
  {10.1111/j.1365-2966.2009.16189.x}, \href
  {https://ui.adsabs.harvard.edu/abs/2010MNRAS.403.1054D} {403, 1054}

\bibitem[\protect\citeauthoryear{{Dabringhausen}, {Kroupa}, {Pflamm-Altenburg}
  \& {Mieske}}{{Dabringhausen} et~al.}{2012}]{2012ApJ...747...72D}
{Dabringhausen} J.,  {Kroupa} P.,  {Pflamm-Altenburg} J.,   {Mieske} S.,  2012,
  \mn@doi [\apj] {10.1088/0004-637X/747/1/72}, \href
  {https://ui.adsabs.harvard.edu/abs/2012ApJ...747...72D} {747, 72}

\bibitem[\protect\citeauthoryear{{Dame}, {Hartmann}  \& {Thaddeus}}{{Dame}
  et~al.}{2001}]{2001ApJ...547..792D}
{Dame} T.~M.,  {Hartmann} D.,   {Thaddeus} P.,  2001, \mn@doi [\apj]
  {10.1086/318388}, \href
  {https://ui.adsabs.harvard.edu/abs/2001ApJ...547..792D} {547, 792}

\bibitem[\protect\citeauthoryear{{Dav{\'e}}}{{Dav{\'e}}}{2008}]{2008MNRAS.385..147D}
{Dav{\'e}} R.,  2008, \mn@doi [\mnras] {10.1111/j.1365-2966.2008.12866.x},
  \href {https://ui.adsabs.harvard.edu/abs/2008MNRAS.385..147D} {385, 147}

\bibitem[\protect\citeauthoryear{{De Marchi}, {Paresce}  \& {Pulone}}{{De
  Marchi} et~al.}{2000}]{2000ApJ...530..342D}
{De Marchi} G.,  {Paresce} F.,   {Pulone} L.,  2000, \mn@doi [\apj]
  {10.1086/308334}, \href
  {https://ui.adsabs.harvard.edu/abs/2000ApJ...530..342D} {530, 342}

\bibitem[\protect\citeauthoryear{{De Marchi}, {Paresce}  \& {Portegies
  Zwart}}{{De Marchi} et~al.}{2010}]{2010ApJ...718..105D}
{De Marchi} G.,  {Paresce} F.,   {Portegies Zwart} S.,  2010, \mn@doi [\apj]
  {10.1088/0004-637X/718/1/105}, \href
  {https://ui.adsabs.harvard.edu/abs/2010ApJ...718..105D} {718, 105}

\bibitem[\protect\citeauthoryear{{Dopcke}, {Glover}, {Clark}  \&
  {Klessen}}{{Dopcke} et~al.}{2011}]{2011ApJ...729L...3D}
{Dopcke} G.,  {Glover} S. C.~O.,  {Clark} P.~C.,   {Klessen} R.~S.,  2011,
  \mn@doi [\apjl] {10.1088/2041-8205/729/1/L3}, \href
  {https://ui.adsabs.harvard.edu/abs/2011ApJ...729L...3D} {729, L3}

\bibitem[\protect\citeauthoryear{{Dopcke}, {Glover}, {Clark}  \&
  {Klessen}}{{Dopcke} et~al.}{2013}]{2013ApJ...766..103D}
{Dopcke} G.,  {Glover} S. C.~O.,  {Clark} P.~C.,   {Klessen} R.~S.,  2013,
  \mn@doi [\apj] {10.1088/0004-637X/766/2/103}, \href
  {https://ui.adsabs.harvard.edu/abs/2013ApJ...766..103D} {766, 103}

\bibitem[\protect\citeauthoryear{{Draine}}{{Draine}}{2011}]{2011piim.book.....D}
{Draine} B.~T.,  2011, {Physics of the Interstellar and Intergalactic Medium}.
Princeton University Press

\bibitem[\protect\citeauthoryear{{Ebert}}{{Ebert}}{1955}]{1955ZA.....37..217E}
{Ebert} R.,  1955, \zap, \href
  {https://ui.adsabs.harvard.edu/abs/1955ZA.....37..217E} {37, 217}

\bibitem[\protect\citeauthoryear{{El-Badry}, {Weisz}  \& {Quataert}}{{El-Badry}
  et~al.}{2017}]{El-Badry17a}
{El-Badry} K.,  {Weisz} D.~R.,   {Quataert} E.,  2017, \mn@doi [\mnras]
  {10.1093/mnras/stx436}, \href
  {https://ui.adsabs.harvard.edu/abs/2017MNRAS.468..319E} {468, 319}

\bibitem[\protect\citeauthoryear{{Eldridge}}{{Eldridge}}{2012}]{2012MNRAS.422..794E}
{Eldridge} J.~J.,  2012, \mn@doi [\mnras] {10.1111/j.1365-2966.2012.20662.x},
  \href {https://ui.adsabs.harvard.edu/abs/2012MNRAS.422..794E} {422, 794}

\bibitem[\protect\citeauthoryear{{Elmegreen}, {Klessen}  \&
  {Wilson}}{{Elmegreen} et~al.}{2008}]{2008ApJ...681..365E}
{Elmegreen} B.~G.,  {Klessen} R.~S.,   {Wilson} C.~D.,  2008, \mn@doi [\apj]
  {10.1086/588725}, \href
  {https://ui.adsabs.harvard.edu/abs/2008ApJ...681..365E} {681, 365}

\bibitem[\protect\citeauthoryear{{Emerick}, {Bryan}  \& {Mac Low}}{{Emerick}
  et~al.}{2018a}]{2018ApJ...865L..22E}
{Emerick} A.,  {Bryan} G.~L.,   {Mac Low} M.-M.,  2018a, \mn@doi [\apjl]
  {10.3847/2041-8213/aae315}, \href
  {https://ui.adsabs.harvard.edu/abs/2018ApJ...865L..22E} {865, L22}

\bibitem[\protect\citeauthoryear{{Emerick}, {Bryan}, {Mac Low}, {C{\^o}t{\'e}},
  {Johnston}  \& {O'Shea}}{{Emerick} et~al.}{2018b}]{2018ApJ...869...94E}
{Emerick} A.,  {Bryan} G.~L.,  {Mac Low} M.-M.,  {C{\^o}t{\'e}} B.,  {Johnston}
  K.~V.,   {O'Shea} B.~W.,  2018b, \mn@doi [\apj] {10.3847/1538-4357/aaec7d},
  \href {https://ui.adsabs.harvard.edu/abs/2018ApJ...869...94E} {869, 94}

\bibitem[\protect\citeauthoryear{{Enoch}, {Evans}, {Sargent}, {Glenn},
  {Rosolowsky}  \& {Myers}}{{Enoch} et~al.}{2008}]{2008ApJ...684.1240E}
{Enoch} M.~L.,  {Evans} Neal~J. I.,  {Sargent} A.~I.,  {Glenn} J.,
  {Rosolowsky} E.,   {Myers} P.,  2008, \mn@doi [\apj] {10.1086/589963}, \href
  {https://ui.adsabs.harvard.edu/abs/2008ApJ...684.1240E} {684, 1240}

\bibitem[\protect\citeauthoryear{{Ezzeddine} et~al.,}{{Ezzeddine}
  et~al.}{2019}]{2019ApJ...876...97E}
{Ezzeddine} R.,  et~al., 2019, \mn@doi [\apj] {10.3847/1538-4357/ab14e7}, \href
  {https://ui.adsabs.harvard.edu/abs/2019ApJ...876...97E} {876, 97}

\bibitem[\protect\citeauthoryear{{Fabbian}, {Nissen}, {Asplund}, {Pettini}  \&
  {Akerman}}{{Fabbian} et~al.}{2009}]{2009A&A...500.1143F}
{Fabbian} D.,  {Nissen} P.~E.,  {Asplund} M.,  {Pettini} M.,   {Akerman} C.,
  2009, \mn@doi [\aap] {10.1051/0004-6361/200810095}, \href
  {https://ui.adsabs.harvard.edu/abs/2009A&A...500.1143F} {500, 1143}

\bibitem[\protect\citeauthoryear{{Fardal}, {Katz}, {Weinberg}  \&
  {Dav{\'e}}}{{Fardal} et~al.}{2007}]{2007MNRAS.379..985F}
{Fardal} M.~A.,  {Katz} N.,  {Weinberg} D.~H.,   {Dav{\'e}} R.,  2007, \mn@doi
  [\mnras] {10.1111/j.1365-2966.2007.11522.x}, \href
  {https://ui.adsabs.harvard.edu/abs/2007MNRAS.379..985F} {379, 985}

\bibitem[\protect\citeauthoryear{{Favre} et~al.,}{{Favre}
  et~al.}{2018}]{2018ApJ...859..136F}
{Favre} C.,  et~al., 2018, \mn@doi [\apj] {10.3847/1538-4357/aabfd4}, \href
  {https://ui.adsabs.harvard.edu/abs/2018ApJ...859..136F} {859, 136}

\bibitem[\protect\citeauthoryear{{Federrath} \& {Klessen}}{{Federrath} \&
  {Klessen}}{2012}]{2012ApJ...761..156F}
{Federrath} C.,  {Klessen} R.~S.,  2012, \mn@doi [\apj]
  {10.1088/0004-637X/761/2/156}, \href
  {https://ui.adsabs.harvard.edu/abs/2012ApJ...761..156F} {761, 156}

\bibitem[\protect\citeauthoryear{{Federrath} \& {Klessen}}{{Federrath} \&
  {Klessen}}{2013}]{2013ApJ...763...51F}
{Federrath} C.,  {Klessen} R.~S.,  2013, \mn@doi [\apj]
  {10.1088/0004-637X/763/1/51}, \href
  {https://ui.adsabs.harvard.edu/abs/2013ApJ...763...51F} {763, 51}

\bibitem[\protect\citeauthoryear{{Federrath}, {Schr{\"o}n}, {Banerjee}  \&
  {Klessen}}{{Federrath} et~al.}{2014}]{2014ApJ...790..128F}
{Federrath} C.,  {Schr{\"o}n} M.,  {Banerjee} R.,   {Klessen} R.~S.,  2014,
  \mn@doi [\apj] {10.1088/0004-637X/790/2/128}, \href
  {https://ui.adsabs.harvard.edu/abs/2014ApJ...790..128F} {790, 128}

\bibitem[\protect\citeauthoryear{{Federrath}, {Krumholz}  \&
  {Hopkins}}{{Federrath} et~al.}{2017}]{2017JPhCS.837a2007F}
{Federrath} C.,  {Krumholz} M.,   {Hopkins} P.~F.,  2017, in Journal of Physics
  Conference Series. p. 012007, \mn@doi{10.1088/1742-6596/837/1/012007}

\bibitem[\protect\citeauthoryear{{Ferreras}, {La Barbera}, {de La Rosa},
  {Vazdekis}, {de Carvalho}, {Falcon-Barroso}  \& {Ricciardelli}}{{Ferreras}
  et~al.}{2013}]{2013MNRAS.429L..15F}
{Ferreras} I.,  {La Barbera} F.,  {de La Rosa} I.~G.,  {Vazdekis} A.,  {de
  Carvalho} R.~R.,  {Falcon-Barroso} J.,   {Ricciardelli} E.,  2013, \mn@doi
  [\mnras] {10.1093/mnrasl/sls014}, \href
  {https://ui.adsabs.harvard.edu/abs/2013MNRAS.429L..15F} {429, L15}

\bibitem[\protect\citeauthoryear{{Ferreras}, {Weidner}, {Vazdekis}  \& {La
  Barbera}}{{Ferreras} et~al.}{2015}]{2015MNRAS.448L..82F}
{Ferreras} I.,  {Weidner} C.,  {Vazdekis} A.,   {La Barbera} F.,  2015, \mn@doi
  [\mnras] {10.1093/mnrasl/slv003}, \href
  {https://ui.adsabs.harvard.edu/abs/2015MNRAS.448L..82F} {448, L82}

\bibitem[\protect\citeauthoryear{{Few}, {Courty}, {Gibson}, {Michel-Dansac}  \&
  {Calura}}{{Few} et~al.}{2014}]{2014MNRAS.444.3845F}
{Few} C.~G.,  {Courty} S.,  {Gibson} B.~K.,  {Michel-Dansac} L.,   {Calura} F.,
   2014, \mn@doi [\mnras] {10.1093/mnras/stu1709}, \href
  {https://ui.adsabs.harvard.edu/abs/2014MNRAS.444.3845F} {444, 3845}

\bibitem[\protect\citeauthoryear{{Fisher}}{{Fisher}}{2004}]{2004ApJ...600..769F}
{Fisher} R.~T.,  2004, \mn@doi [\apj] {10.1086/380111}, \href
  {https://ui.adsabs.harvard.edu/abs/2004ApJ...600..769F} {600, 769}

\bibitem[\protect\citeauthoryear{{Flower} \& {Pineau des For{\^e}ts}}{{Flower}
  \& {Pineau des For{\^e}ts}}{2000}]{2000MNRAS.316..901F}
{Flower} D.~R.,  {Pineau des For{\^e}ts} G.,  2000, \mn@doi [\mnras]
  {10.1046/j.1365-8711.2000.03611.x}, \href
  {https://ui.adsabs.harvard.edu/abs/2000MNRAS.316..901F} {316, 901}

\bibitem[\protect\citeauthoryear{{Fontanot}, {De Lucia}, {Xie}, {Hirschmann},
  {Bruzual}  \& {Charlot}}{{Fontanot} et~al.}{2018}]{2018MNRAS.475.2467F}
{Fontanot} F.,  {De Lucia} G.,  {Xie} L.,  {Hirschmann} M.,  {Bruzual} G.,
  {Charlot} S.,  2018, \mn@doi [\mnras] {10.1093/mnras/stx3323}, \href
  {https://ui.adsabs.harvard.edu/abs/2018MNRAS.475.2467F} {475, 2467}

\bibitem[\protect\citeauthoryear{{Frebel} \& {Norris}}{{Frebel} \&
  {Norris}}{2015}]{2015ARA&A..53..631F}
{Frebel} A.,  {Norris} J.~E.,  2015, \mn@doi [\araa]
  {10.1146/annurev-astro-082214-122423}, \href
  {https://ui.adsabs.harvard.edu/abs/2015ARA&A..53..631F} {53, 631}

\bibitem[\protect\citeauthoryear{{Frebel}, {Chiti}, {Ji}, {Jacobson}  \&
  {Placco}}{{Frebel} et~al.}{2015}]{2015ApJ...810L..27F}
{Frebel} A.,  {Chiti} A.,  {Ji} A.~P.,  {Jacobson} H.~R.,   {Placco} V.~M.,
  2015, \mn@doi [\apjl] {10.1088/2041-8205/810/2/L27}, \href
  {https://ui.adsabs.harvard.edu/abs/2015ApJ...810L..27F} {810, L27}

\bibitem[\protect\citeauthoryear{{Fumagalli}, {da Silva}  \&
  {Krumholz}}{{Fumagalli} et~al.}{2011}]{Fumagalli11a}
{Fumagalli} M.,  {da Silva} R.~L.,   {Krumholz} M.~R.,  2011, \mn@doi [\apjl]
  {10.1088/2041-8205/741/2/L26}, \href
  {http://adsabs.harvard.edu/abs/2011ApJ...741L..26F} {741, L26}

\bibitem[\protect\citeauthoryear{{Gaches} \& {Offner}}{{Gaches} \&
  {Offner}}{2018}]{2018ApJ...861...87G}
{Gaches} B. A.~L.,  {Offner} S. S.~R.,  2018, \mn@doi [\apj]
  {10.3847/1538-4357/aac94d}, \href
  {https://ui.adsabs.harvard.edu/abs/2018ApJ...861...87G} {861, 87}

\bibitem[\protect\citeauthoryear{{Gaches}, {Offner}  \& {Bisbas}}{{Gaches}
  et~al.}{2019}]{2019ApJ...878..105G}
{Gaches} B. A.~L.,  {Offner} S. S.~R.,   {Bisbas} T.~G.,  2019, \mn@doi [\apj]
  {10.3847/1538-4357/ab20c7}, \href
  {https://ui.adsabs.harvard.edu/abs/2019ApJ...878..105G} {878, 105}

\bibitem[\protect\citeauthoryear{{Galli} \& {Palla}}{{Galli} \&
  {Palla}}{1998}]{1998A&A...335..403G}
{Galli} D.,  {Palla} F.,  1998, \aap, \href
  {https://ui.adsabs.harvard.edu/abs/1998A&A...335..403G} {335, 403}

\bibitem[\protect\citeauthoryear{{Galliano}, {Galametz}  \& {Jones}}{{Galliano}
  et~al.}{2018}]{2018ARA&A..56..673G}
{Galliano} F.,  {Galametz} M.,   {Jones} A.~P.,  2018, \mn@doi [\araa]
  {10.1146/annurev-astro-081817-051900}, \href
  {https://ui.adsabs.harvard.edu/abs/2018ARA&A..56..673G} {56, 673}

\bibitem[\protect\citeauthoryear{{Gelli}, {Salvadori}, {Ferrara}, {Pallottini}
  \& {Carniani}}{{Gelli} et~al.}{2021}]{2021ApJ...913L..25G}
{Gelli} V.,  {Salvadori} S.,  {Ferrara} A.,  {Pallottini} A.,   {Carniani} S.,
  2021, \mn@doi [\apjl] {10.3847/2041-8213/abfe6c}, \href
  {https://ui.adsabs.harvard.edu/abs/2021ApJ...913L..25G} {913, L25}

\bibitem[\protect\citeauthoryear{{Gennaro} et~al.,}{{Gennaro}
  et~al.}{2018}]{2018ApJ...855...20G}
{Gennaro} M.,  et~al., 2018, \mn@doi [\apj] {10.3847/1538-4357/aaa973}, \href
  {https://ui.adsabs.harvard.edu/abs/2018ApJ...855...20G} {855, 20}

\bibitem[\protect\citeauthoryear{{Gieser} et~al.,}{{Gieser}
  et~al.}{2021}]{2021A&A...648A..66G}
{Gieser} C.,  et~al., 2021, \mn@doi [\aap] {10.1051/0004-6361/202039670}, \href
  {https://ui.adsabs.harvard.edu/abs/2021A&A...648A..66G} {648, A66}

\bibitem[\protect\citeauthoryear{{Gil-Pons}, {Doherty}, {Guti{\'e}rrez},
  {Campbell}, {Siess}  \& {Lattanzio}}{{Gil-Pons}
  et~al.}{2021}]{2021A&A...645A..10G}
{Gil-Pons} P.,  {Doherty} C.~L.,  {Guti{\'e}rrez} J.,  {Campbell} S.~W.,
  {Siess} L.,   {Lattanzio} J.~C.,  2021, \mn@doi [\aap]
  {10.1051/0004-6361/201937264}, \href
  {https://ui.adsabs.harvard.edu/abs/2021A&A...645A..10G} {645, A10}

\bibitem[\protect\citeauthoryear{{Glassgold}, {Galli}  \&
  {Padovani}}{{Glassgold} et~al.}{2012}]{Glassgold12a}
{Glassgold} A.~E.,  {Galli} D.,   {Padovani} M.,  2012, \mn@doi [\apj]
  {10.1088/0004-637X/756/2/157}, \href
  {http://adsabs.harvard.edu/abs/2012ApJ...756..157G} {756, 157}

\bibitem[\protect\citeauthoryear{{Glover} \& {Abel}}{{Glover} \&
  {Abel}}{2008}]{2008MNRAS.388.1627G}
{Glover} S.~C.~O.,  {Abel} T.,  2008, \mn@doi [\mnras]
  {10.1111/j.1365-2966.2008.13224.x}, \href
  {https://ui.adsabs.harvard.edu/abs/2008MNRAS.388.1627G} {388, 1627}

\bibitem[\protect\citeauthoryear{{Glover} \& {Clark}}{{Glover} \&
  {Clark}}{2012a}]{Glover12a}
{Glover} S.~C.~O.,  {Clark} P.~C.,  2012a, \mn@doi [\mnras]
  {10.1111/j.1365-2966.2011.19648.x}, \href
  {http://adsabs.harvard.edu/abs/2012MNRAS.421....9G} {421, 9}

\bibitem[\protect\citeauthoryear{{Glover} \& {Clark}}{{Glover} \&
  {Clark}}{2012b}]{2012MNRAS.426..377G}
{Glover} S. C.~O.,  {Clark} P.~C.,  2012b, \mn@doi [\mnras]
  {10.1111/j.1365-2966.2012.21737.x}, \href
  {https://ui.adsabs.harvard.edu/abs/2012MNRAS.426..377G} {426, 377}

\bibitem[\protect\citeauthoryear{{Glover} \& {Clark}}{{Glover} \&
  {Clark}}{2016}]{2016MNRAS.456.3596G}
{Glover} S. C.~O.,  {Clark} P.~C.,  2016, \mn@doi [\mnras]
  {10.1093/mnras/stv2863}, \href
  {https://ui.adsabs.harvard.edu/abs/2016MNRAS.456.3596G} {456, 3596}

\bibitem[\protect\citeauthoryear{{Goldsmith}}{{Goldsmith}}{2001}]{Goldsmith01a}
{Goldsmith} P.~F.,  2001, \mn@doi [\apj] {10.1086/322255}, \href
  {http://adsabs.harvard.edu/abs/2001ApJ...557..736G} {557, 736}

\bibitem[\protect\citeauthoryear{{Goodman}}{{Goodman}}{1974}]{1974PrSS....5..261G}
{Goodman} F.~O.,  1974, \mn@doi [Progress In Surface Science]
  {10.1016/0079-6816(74)90005-7}, \href
  {https://ui.adsabs.harvard.edu/abs/1974PrSS....5..261G} {5, 261}

\bibitem[\protect\citeauthoryear{{Goodman} \& {Wachman}}{{Goodman} \&
  {Wachman}}{1967}]{1967JChPh..46.2376G}
{Goodman} F.~O.,  {Wachman} H.~Y.,  1967, \mn@doi [\jcp] {10.1063/1.1841046},
  \href {https://ui.adsabs.harvard.edu/abs/1967JChPh..46.2376G} {46, 2376}

\bibitem[\protect\citeauthoryear{{Goodwin}, {Whitworth}  \&
  {Ward-Thompson}}{{Goodwin} et~al.}{2004}]{2004A&A...414..633G}
{Goodwin} S.~P.,  {Whitworth} A.~P.,   {Ward-Thompson} D.,  2004, \mn@doi
  [\aap] {10.1051/0004-6361:20031594}, \href
  {https://ui.adsabs.harvard.edu/abs/2004A&A...414..633G} {414, 633}

\bibitem[\protect\citeauthoryear{{Grassi}, {Bovino}, {Schleicher}, {Prieto},
  {Seifried}, {Simoncini}  \& {Gianturco}}{{Grassi}
  et~al.}{2014}]{2014MNRAS.439.2386G}
{Grassi} T.,  {Bovino} S.,  {Schleicher} D.~R.~G.,  {Prieto} J.,  {Seifried}
  D.,  {Simoncini} E.,   {Gianturco} F.~A.,  2014, \mn@doi [\mnras]
  {10.1093/mnras/stu114}, \href
  {https://ui.adsabs.harvard.edu/abs/2014MNRAS.439.2386G} {439, 2386}

\bibitem[\protect\citeauthoryear{{Greif}, {Springel}, {White}, {Glover},
  {Clark}, {Smith}, {Klessen}  \& {Bromm}}{{Greif} et~al.}{2011}]{Greif11a}
{Greif} T.~H.,  {Springel} V.,  {White} S.~D.~M.,  {Glover} S.~C.~O.,  {Clark}
  P.~C.,  {Smith} R.~J.,  {Klessen} R.~S.,   {Bromm} V.,  2011, \mn@doi [\apj]
  {10.1088/0004-637X/737/2/75}, \href
  {http://adsabs.harvard.edu/abs/2011ApJ...737...75G} {737, 75}

\bibitem[\protect\citeauthoryear{{Gunawardhana} et~al.,}{{Gunawardhana}
  et~al.}{2011}]{2011MNRAS.415.1647G}
{Gunawardhana} M.~L.~P.,  et~al., 2011, \mn@doi [\mnras]
  {10.1111/j.1365-2966.2011.18800.x}, \href
  {https://ui.adsabs.harvard.edu/abs/2011MNRAS.415.1647G} {415, 1647}

\bibitem[\protect\citeauthoryear{{Guszejnov}, {Krumholz}  \&
  {Hopkins}}{{Guszejnov} et~al.}{2016}]{2016MNRAS.458..673G}
{Guszejnov} D.,  {Krumholz} M.~R.,   {Hopkins} P.~F.,  2016, \mn@doi [\mnras]
  {10.1093/mnras/stw315}, \href
  {https://ui.adsabs.harvard.edu/abs/2016MNRAS.458..673G} {458, 673}

\bibitem[\protect\citeauthoryear{{Guszejnov}, {Hopkins}  \&
  {Krumholz}}{{Guszejnov} et~al.}{2017a}]{2017MNRAS.468.4093G}
{Guszejnov} D.,  {Hopkins} P.~F.,   {Krumholz} M.~R.,  2017a, \mn@doi [\mnras]
  {10.1093/mnras/stx725}, \href
  {https://ui.adsabs.harvard.edu/abs/2017MNRAS.468.4093G} {468, 4093}

\bibitem[\protect\citeauthoryear{{Guszejnov}, {Hopkins}  \& {Ma}}{{Guszejnov}
  et~al.}{2017b}]{2017MNRAS.472.2107G}
{Guszejnov} D.,  {Hopkins} P.~F.,   {Ma} X.,  2017b, \mn@doi [\mnras]
  {10.1093/mnras/stx2067}, \href
  {https://ui.adsabs.harvard.edu/abs/2017MNRAS.472.2107G} {472, 2107}

\bibitem[\protect\citeauthoryear{{Guszejnov}, {Hopkins}, {Grudi{\'c}},
  {Krumholz}  \& {Federrath}}{{Guszejnov} et~al.}{2018}]{Guszejnov18a}
{Guszejnov} D.,  {Hopkins} P.~F.,  {Grudi{\'c}} M.~Y.,  {Krumholz} M.~R.,
  {Federrath} C.,  2018, \mn@doi [\mnras] {10.1093/mnras/sty1847}, \href
  {https://ui.adsabs.harvard.edu/#abs/2018MNRAS.480..182G} {480, 182}

\bibitem[\protect\citeauthoryear{{Guszejnov}, {Hopkins}  \&
  {Graus}}{{Guszejnov} et~al.}{2019}]{2019MNRAS.485.4852G}
{Guszejnov} D.,  {Hopkins} P.~F.,   {Graus} A.~S.,  2019, \mn@doi [\mnras]
  {10.1093/mnras/stz736}, \href
  {https://ui.adsabs.harvard.edu/abs/2019MNRAS.485.4852G} {485, 4852}

\bibitem[\protect\citeauthoryear{{Gutcke} \& {Springel}}{{Gutcke} \&
  {Springel}}{2019}]{2019MNRAS.482..118G}
{Gutcke} T.~A.,  {Springel} V.,  2019, \mn@doi [\mnras]
  {10.1093/mnras/sty2688}, \href
  {https://ui.adsabs.harvard.edu/abs/2019MNRAS.482..118G} {482, 118}

\bibitem[\protect\citeauthoryear{{Harris}}{{Harris}}{1996}]{1996AJ....112.1487H}
{Harris} W.~E.,  1996, \mn@doi [\aj] {10.1086/118116}, \href
  {https://ui.adsabs.harvard.edu/abs/1996AJ....112.1487H} {112, 1487}

\bibitem[\protect\citeauthoryear{{Harris} et~al.,}{{Harris}
  et~al.}{2020}]{2020arXiv200610256H}
{Harris} C.~R.,  et~al., 2020, \mn@doi [\nat] {10.1038/s41586-020-2649-2},
  \href {https://ui.adsabs.harvard.edu/abs/2020arXiv200610256H} {585, 357}

\bibitem[\protect\citeauthoryear{{Hartwig}, {Clark}, {Glover}, {Klessen}  \&
  {Sasaki}}{{Hartwig} et~al.}{2015}]{2015ApJ...799..114H}
{Hartwig} T.,  {Clark} P.~C.,  {Glover} S. C.~O.,  {Klessen} R.~S.,   {Sasaki}
  M.,  2015, \mn@doi [\apj] {10.1088/0004-637X/799/2/114}, \href
  {https://ui.adsabs.harvard.edu/abs/2015ApJ...799..114H} {799, 114}

\bibitem[\protect\citeauthoryear{{Hayes} et~al.,}{{Hayes}
  et~al.}{2018}]{2018ApJ...852...49H}
{Hayes} C.~R.,  et~al., 2018, \mn@doi [\apj] {10.3847/1538-4357/aa9cec}, \href
  {https://ui.adsabs.harvard.edu/abs/2018ApJ...852...49H} {852, 49}

\bibitem[\protect\citeauthoryear{{Hennebelle}, {Lee}  \&
  {Chabrier}}{{Hennebelle} et~al.}{2019}]{2019ApJ...883..140H}
{Hennebelle} P.,  {Lee} Y.-N.,   {Chabrier} G.,  2019, \mn@doi [\apj]
  {10.3847/1538-4357/ab3d46}, \href
  {https://ui.adsabs.harvard.edu/abs/2019ApJ...883..140H} {883, 140}

\bibitem[\protect\citeauthoryear{{Hennebelle}, {Commer{\c{c}}on}, {Lee}  \&
  {Chabrier}}{{Hennebelle} et~al.}{2020}]{Hennebelle20a}
{Hennebelle} P.,  {Commer{\c{c}}on} B.,  {Lee} Y.-N.,   {Chabrier} G.,  2020,
  \mn@doi [\apj] {10.3847/1538-4357/abbfab}, \href
  {https://ui.adsabs.harvard.edu/abs/2020ApJ...904..194H} {904, 194}

\bibitem[\protect\citeauthoryear{{Hirano}, {Hosokawa}, {Yoshida}, {Umeda},
  {Omukai}, {Chiaki}  \& {Yorke}}{{Hirano} et~al.}{2014}]{2014ApJ...781...60H}
{Hirano} S.,  {Hosokawa} T.,  {Yoshida} N.,  {Umeda} H.,  {Omukai} K.,
  {Chiaki} G.,   {Yorke} H.~W.,  2014, \mn@doi [\apj]
  {10.1088/0004-637X/781/2/60}, \href
  {https://ui.adsabs.harvard.edu/abs/2014ApJ...781...60H} {781, 60}

\bibitem[\protect\citeauthoryear{{Hollenbach} \& {McKee}}{{Hollenbach} \&
  {McKee}}{1979}]{1979ApJS...41..555H}
{Hollenbach} D.,  {McKee} C.~F.,  1979, \mn@doi [\apjs] {10.1086/190631}, \href
  {https://ui.adsabs.harvard.edu/abs/1979ApJS...41..555H} {41, 555}

\bibitem[\protect\citeauthoryear{{Hopkins}}{{Hopkins}}{2018}]{2018PASA...35...39H}
{Hopkins} A.~M.,  2018, \mn@doi [\pasa] {10.1017/pasa.2018.29}, \href
  {https://ui.adsabs.harvard.edu/abs/2018PASA...35...39H} {35, e039}

\bibitem[\protect\citeauthoryear{{Hosokawa}, {Offner}  \&
  {Krumholz}}{{Hosokawa} et~al.}{2011}]{2011ApJ...738..140H}
{Hosokawa} T.,  {Offner} S. S.~R.,   {Krumholz} M.~R.,  2011, \mn@doi [\apj]
  {10.1088/0004-637X/738/2/140}, \href
  {https://ui.adsabs.harvard.edu/abs/2011ApJ...738..140H} {738, 140}

\bibitem[\protect\citeauthoryear{{Hoversten} \& {Glazebrook}}{{Hoversten} \&
  {Glazebrook}}{2008}]{2008ApJ...675..163H}
{Hoversten} E.~A.,  {Glazebrook} K.,  2008, \mn@doi [\apj] {10.1086/524095},
  \href {https://ui.adsabs.harvard.edu/abs/2008ApJ...675..163H} {675, 163}

\bibitem[\protect\citeauthoryear{{Hu}, {Sternberg}  \& {van Dishoeck}}{{Hu}
  et~al.}{2021}]{2021arXiv210303889H}
{Hu} C.-Y.,  {Sternberg} A.,   {van Dishoeck} E.~F.,  2021, arXiv e-prints,
  \href {https://ui.adsabs.harvard.edu/abs/2021arXiv210303889H} {p.
  arXiv:2103.03889}

\bibitem[\protect\citeauthoryear{{Hummel}, {Stacy}  \& {Bromm}}{{Hummel}
  et~al.}{2016}]{2016MNRAS.460.2432H}
{Hummel} J.~A.,  {Stacy} A.,   {Bromm} V.,  2016, \mn@doi [\mnras]
  {10.1093/mnras/stw1127}, \href
  {https://ui.adsabs.harvard.edu/abs/2016MNRAS.460.2432H} {460, 2432}

\bibitem[\protect\citeauthoryear{Hunter}{Hunter}{2007}]{Hunter:2007}
Hunter J.~D.,  2007, \mn@doi [Computing in Science \& Engineering]
  {10.1109/MCSE.2007.55}, 9, 90

\bibitem[\protect\citeauthoryear{{Indriolo} \& {McCall}}{{Indriolo} \&
  {McCall}}{2012}]{2012ApJ...745...91I}
{Indriolo} N.,  {McCall} B.~J.,  2012, \mn@doi [\apj]
  {10.1088/0004-637X/745/1/91}, \href
  {https://ui.adsabs.harvard.edu/abs/2012ApJ...745...91I} {745, 91}

\bibitem[\protect\citeauthoryear{{Izotov}, {Thuan}, {Guseva}  \&
  {Liss}}{{Izotov} et~al.}{2018}]{2018MNRAS.473.1956I}
{Izotov} Y.~I.,  {Thuan} T.~X.,  {Guseva} N.~G.,   {Liss} S.~E.,  2018, \mn@doi
  [\mnras] {10.1093/mnras/stx2478}, \href
  {https://ui.adsabs.harvard.edu/abs/2018MNRAS.473.1956I} {473, 1956}

\bibitem[\protect\citeauthoryear{{Izzard}, {Glebbeek}, {Stancliffe}  \&
  {Pols}}{{Izzard} et~al.}{2009}]{2009A&A...508.1359I}
{Izzard} R.~G.,  {Glebbeek} E.,  {Stancliffe} R.~J.,   {Pols} O.~R.,  2009,
  \mn@doi [\aap] {10.1051/0004-6361/200912827}, \href
  {https://ui.adsabs.harvard.edu/abs/2009A&A...508.1359I} {508, 1359}

\bibitem[\protect\citeauthoryear{{Jameson} et~al.,}{{Jameson}
  et~al.}{2018}]{2018ApJ...853..111J}
{Jameson} K.~E.,  et~al., 2018, \mn@doi [\apj] {10.3847/1538-4357/aaa4bb},
  \href {https://ui.adsabs.harvard.edu/abs/2018ApJ...853..111J} {853, 111}

\bibitem[\protect\citeauthoryear{{Jappsen}, {Klessen}, {Larson}, {Li}  \& {Mac
  Low}}{{Jappsen} et~al.}{2005}]{2005A&A...435..611J}
{Jappsen} A.~K.,  {Klessen} R.~S.,  {Larson} R.~B.,  {Li} Y.,   {Mac Low}
  M.~M.,  2005, \mn@doi [\aap] {10.1051/0004-6361:20042178}, \href
  {https://ui.adsabs.harvard.edu/abs/2005A&A...435..611J} {435, 611}

\bibitem[\protect\citeauthoryear{{Jappsen}, {Klessen}, {Glover}  \& {Mac
  Low}}{{Jappsen} et~al.}{2009}]{2009ApJ...696.1065J}
{Jappsen} A.-K.,  {Klessen} R.~S.,  {Glover} S. C.~O.,   {Mac Low} M.-M.,
  2009, \mn@doi [\apj] {10.1088/0004-637X/696/2/1065}, \href
  {https://ui.adsabs.harvard.edu/abs/2009ApJ...696.1065J} {696, 1065}

\bibitem[\protect\citeauthoryear{{Je{\v{r}}{\'a}bkov{\'a}}, {Hasani Zonoozi},
  {Kroupa}, {Beccari}, {Yan}, {Vazdekis}  \& {Zhang}}{{Je{\v{r}}{\'a}bkov{\'a}}
  et~al.}{2018}]{2018A&A...620A..39J}
{Je{\v{r}}{\'a}bkov{\'a}} T.,  {Hasani Zonoozi} A.,  {Kroupa} P.,  {Beccari}
  G.,  {Yan} Z.,  {Vazdekis} A.,   {Zhang} Z.~Y.,  2018, \mn@doi [\aap]
  {10.1051/0004-6361/201833055}, \href
  {https://ui.adsabs.harvard.edu/abs/2018A&A...620A..39J} {620, A39}

\bibitem[\protect\citeauthoryear{{J{\o}rgensen}, {Sch{\"o}ier}  \& {van
  Dishoeck}}{{J{\o}rgensen} et~al.}{2002}]{2002A&A...389..908J}
{J{\o}rgensen} J.~K.,  {Sch{\"o}ier} F.~L.,   {van Dishoeck} E.~F.,  2002,
  \mn@doi [\aap] {10.1051/0004-6361:20020681}, \href
  {https://ui.adsabs.harvard.edu/abs/2002A&A...389..908J} {389, 908}

\bibitem[\protect\citeauthoryear{{Kalari}, {Carraro}, {Evans}  \&
  {Rubio}}{{Kalari} et~al.}{2018}]{2018ApJ...857..132K}
{Kalari} V.~M.,  {Carraro} G.,  {Evans} C.~J.,   {Rubio} M.,  2018, \mn@doi
  [\apj] {10.3847/1538-4357/aab609}, \href
  {https://ui.adsabs.harvard.edu/abs/2018ApJ...857..132K} {857, 132}

\bibitem[\protect\citeauthoryear{{Kauffmann}, {Pillai}  \&
  {Goldsmith}}{{Kauffmann} et~al.}{2013}]{2013ApJ...779..185K}
{Kauffmann} J.,  {Pillai} T.,   {Goldsmith} P.~F.,  2013, \mn@doi [\apj]
  {10.1088/0004-637X/779/2/185}, \href
  {https://ui.adsabs.harvard.edu/abs/2013ApJ...779..185K} {779, 185}

\bibitem[\protect\citeauthoryear{{Keller} et~al.,}{{Keller}
  et~al.}{2014}]{2014Natur.506..463K}
{Keller} S.~C.,  et~al., 2014, \mn@doi [\nat] {10.1038/nature12990}, \href
  {https://ui.adsabs.harvard.edu/abs/2014Natur.506..463K} {506, 463}

\bibitem[\protect\citeauthoryear{{Klessen}}{{Klessen}}{2019}]{2019ffbh.book...67K}
{Klessen} R.,  2019, {Formation of the first stars}.
World Scientific Publishing Company, pp 67--97,
  \mn@doi{10.1142/9789813227958\_0004}

\bibitem[\protect\citeauthoryear{{Komiya}, {Suda}, {Minaguchi}, {Shigeyama},
  {Aoki}  \& {Fujimoto}}{{Komiya} et~al.}{2007}]{2007ApJ...658..367K}
{Komiya} Y.,  {Suda} T.,  {Minaguchi} H.,  {Shigeyama} T.,  {Aoki} W.,
  {Fujimoto} M.~Y.,  2007, \mn@doi [\apj] {10.1086/510826}, \href
  {https://ui.adsabs.harvard.edu/abs/2007ApJ...658..367K} {658, 367}

\bibitem[\protect\citeauthoryear{{Kratter} \& {Matzner}}{{Kratter} \&
  {Matzner}}{2006}]{2006MNRAS.373.1563K}
{Kratter} K.~M.,  {Matzner} C.~D.,  2006, \mn@doi [\mnras]
  {10.1111/j.1365-2966.2006.11103.x}, \href
  {https://ui.adsabs.harvard.edu/abs/2006MNRAS.373.1563K} {373, 1563}

\bibitem[\protect\citeauthoryear{{Kraus}, {Ireland}, {Martinache}  \&
  {Hillenbrand}}{{Kraus} et~al.}{2011}]{2011ApJ...731....8K}
{Kraus} A.~L.,  {Ireland} M.~J.,  {Martinache} F.,   {Hillenbrand} L.~A.,
  2011, \mn@doi [\apj] {10.1088/0004-637X/731/1/8}, \href
  {https://ui.adsabs.harvard.edu/abs/2011ApJ...731....8K} {731, 8}

\bibitem[\protect\citeauthoryear{{Kroupa}}{{Kroupa}}{2001}]{2001MNRAS.322..231K}
{Kroupa} P.,  2001, \mn@doi [\mnras] {10.1046/j.1365-8711.2001.04022.x}, \href
  {https://ui.adsabs.harvard.edu/abs/2001MNRAS.322..231K} {322, 231}

\bibitem[\protect\citeauthoryear{{Kroupa}}{{Kroupa}}{2002}]{2002Sci...295...82K}
{Kroupa} P.,  2002, \mn@doi [Science] {10.1126/science.1067524}, \href
  {https://ui.adsabs.harvard.edu/abs/2002Sci...295...82K} {295, 82}

\bibitem[\protect\citeauthoryear{{Kroupa}, {Weidner}, {Pflamm-Altenburg},
  {Thies}, {Dabringhausen}, {Marks}  \& {Maschberger}}{{Kroupa}
  et~al.}{2013}]{2013pss5.book..115K}
{Kroupa} P.,  {Weidner} C.,  {Pflamm-Altenburg} J.,  {Thies} I.,
  {Dabringhausen} J.,  {Marks} M.,   {Maschberger} T.,  2013, in {Oswalt}
  T.~D.,  {Gilmore} G.,  eds, ~ Vol. 5, Planets, Stars and Stellar Systems.
  Volume 5: Galactic Structure and Stellar Populations. Springer, p.~115,
  \mn@doi{10.1007/978-94-007-5612-0\_4}

\bibitem[\protect\citeauthoryear{{Kroupa}, {Subr}, {Je{\v{r}}{\'a}bkov{\'a}}
  \& {Wang}}{{Kroupa} et~al.}{2020}]{2020MNRAS.498.5652K}
{Kroupa} P.,  {Subr} L.,  {Je{\v{r}}{\'a}bkov{\'a}} T.,   {Wang} L.,  2020,
  \mn@doi [\mnras] {10.1093/mnras/staa2276}, \href
  {https://ui.adsabs.harvard.edu/abs/2020MNRAS.498.5652K} {498, 5652}

\bibitem[\protect\citeauthoryear{{Krumholz}}{{Krumholz}}{2011}]{2011ApJ...743..110K}
{Krumholz} M.~R.,  2011, \mn@doi [\apj] {10.1088/0004-637X/743/2/110}, \href
  {https://ui.adsabs.harvard.edu/abs/2011ApJ...743..110K} {743, 110}

\bibitem[\protect\citeauthoryear{{Krumholz}}{{Krumholz}}{2012}]{2012ApJ...759....9K}
{Krumholz} M.~R.,  2012, \mn@doi [\apj] {10.1088/0004-637X/759/1/9}, \href
  {https://ui.adsabs.harvard.edu/abs/2012ApJ...759....9K} {759, 9}

\bibitem[\protect\citeauthoryear{{Krumholz}}{{Krumholz}}{2014a}]{2014MNRAS.437.1662K}
{Krumholz} M.~R.,  2014a, \mn@doi [\mnras] {10.1093/mnras/stt2000}, \href
  {https://ui.adsabs.harvard.edu/abs/2014MNRAS.437.1662K} {437, 1662}

\bibitem[\protect\citeauthoryear{{Krumholz}}{{Krumholz}}{2014b}]{2014PhR...539...49K}
{Krumholz} M.~R.,  2014b, \mn@doi [\physrep] {10.1016/j.physrep.2014.02.001},
  \href {https://ui.adsabs.harvard.edu/abs/2014PhR...539...49K} {539, 49}

\bibitem[\protect\citeauthoryear{{Krumholz} \& {Gnedin}}{{Krumholz} \&
  {Gnedin}}{2011}]{Krumholz11a}
{Krumholz} M.~R.,  {Gnedin} N.~Y.,  2011, \mn@doi [\apj]
  {10.1088/0004-637X/729/1/36}, \href
  {http://adsabs.harvard.edu/abs/2011ApJ...729...36K} {729, 36}

\bibitem[\protect\citeauthoryear{{Krumholz}, {Klein}, {McKee}, {Offner}  \&
  {Cunningham}}{{Krumholz} et~al.}{2009a}]{2009Sci...323..754K}
{Krumholz} M.~R.,  {Klein} R.~I.,  {McKee} C.~F.,  {Offner} S. S.~R.,
  {Cunningham} A.~J.,  2009a, \mn@doi [Science] {10.1126/science.1165857},
  \href {https://ui.adsabs.harvard.edu/abs/2009Sci...323..754K} {323, 754}

\bibitem[\protect\citeauthoryear{{Krumholz}, {McKee}  \&
  {Tumlinson}}{{Krumholz} et~al.}{2009b}]{2009ApJ...693..216K}
{Krumholz} M.~R.,  {McKee} C.~F.,   {Tumlinson} J.,  2009b, \mn@doi [\apj]
  {10.1088/0004-637X/693/1/216}, \href
  {https://ui.adsabs.harvard.edu/abs/2009ApJ...693..216K} {693, 216}

\bibitem[\protect\citeauthoryear{{Krumholz}, {Cunningham}, {Klein}  \&
  {McKee}}{{Krumholz} et~al.}{2010}]{2010ApJ...713.1120K}
{Krumholz} M.~R.,  {Cunningham} A.~J.,  {Klein} R.~I.,   {McKee} C.~F.,  2010,
  \mn@doi [\apj] {10.1088/0004-637X/713/2/1120}, \href
  {https://ui.adsabs.harvard.edu/abs/2010ApJ...713.1120K} {713, 1120}

\bibitem[\protect\citeauthoryear{{Krumholz}, {Klein}  \& {McKee}}{{Krumholz}
  et~al.}{2011}]{2011ApJ...740...74K}
{Krumholz} M.~R.,  {Klein} R.~I.,   {McKee} C.~F.,  2011, \mn@doi [\apj]
  {10.1088/0004-637X/740/2/74}, \href
  {https://ui.adsabs.harvard.edu/abs/2011ApJ...740...74K} {740, 74}

\bibitem[\protect\citeauthoryear{{Krumholz}, {Klein}  \& {McKee}}{{Krumholz}
  et~al.}{2012}]{Krumholz12b}
{Krumholz} M.~R.,  {Klein} R.~I.,   {McKee} C.~F.,  2012, \mn@doi [\apj]
  {10.1088/0004-637X/754/1/71}, \href
  {http://adsabs.harvard.edu/abs/2012ApJ...754...71K} {754, 71}

\bibitem[\protect\citeauthoryear{{Krumholz}, {Myers}, {Klein}  \&
  {McKee}}{{Krumholz} et~al.}{2016}]{2016MNRAS.460.3272K}
{Krumholz} M.~R.,  {Myers} A.~T.,  {Klein} R.~I.,   {McKee} C.~F.,  2016,
  \mn@doi [\mnras] {10.1093/mnras/stw1236}, \href
  {https://ui.adsabs.harvard.edu/abs/2016MNRAS.460.3272K} {460, 3272}

\bibitem[\protect\citeauthoryear{{Krumholz}, {McKee}  \&
  {Bland-Hawthorn}}{{Krumholz} et~al.}{2019}]{2019ARA&A..57..227K}
{Krumholz} M.~R.,  {McKee} C.~F.,   {Bland-Hawthorn} J.,  2019, \mn@doi [\araa]
  {10.1146/annurev-astro-091918-104430}, \href
  {https://ui.adsabs.harvard.edu/abs/2019ARA&A..57..227K} {57, 227}

\bibitem[\protect\citeauthoryear{{La Barbera}, {Ferreras}, {Vazdekis}, {de la
  Rosa}, {de Carvalho}, {Trevisan}, {Falc{\'o}n-Barroso}  \&
  {Ricciardelli}}{{La Barbera} et~al.}{2013}]{2013MNRAS.433.3017L}
{La Barbera} F.,  {Ferreras} I.,  {Vazdekis} A.,  {de la Rosa} I.~G.,  {de
  Carvalho} R.~R.,  {Trevisan} M.,  {Falc{\'o}n-Barroso} J.,   {Ricciardelli}
  E.,  2013, \mn@doi [\mnras] {10.1093/mnras/stt943}, \href
  {https://ui.adsabs.harvard.edu/abs/2013MNRAS.433.3017L} {433, 3017}

\bibitem[\protect\citeauthoryear{{La Barbera}, {Vazdekis}, {Ferreras},
  {Pasquali}, {Cappellari}, {Mart{\'\i}n-Navarro}, {Sch{\"o}nebeck}  \&
  {Falc{\'o}n-Barroso}}{{La Barbera} et~al.}{2016}]{2016MNRAS.457.1468L}
{La Barbera} F.,  {Vazdekis} A.,  {Ferreras} I.,  {Pasquali} A.,  {Cappellari}
  M.,  {Mart{\'\i}n-Navarro} I.,  {Sch{\"o}nebeck} F.,   {Falc{\'o}n-Barroso}
  J.,  2016, \mn@doi [\mnras] {10.1093/mnras/stv2996}, \href
  {https://ui.adsabs.harvard.edu/abs/2016MNRAS.457.1468L} {457, 1468}

\bibitem[\protect\citeauthoryear{{Langer}}{{Langer}}{2009}]{2009ASPC..417...71L}
{Langer} W.~D.,  2009, in {Lis} D.~C.,  {Vaillancourt} J.~E.,  {Goldsmith}
  P.~F.,  {Bell} T.~A.,  {Scoville} N.~Z.,   {Zmuidzinas} J.,  eds,
  Astronomical Society of the Pacific Conference Series Vol. 417, Submillimeter
  Astrophysics and Technology: a Symposium Honoring Thomas G. Phillips. p.~71

\bibitem[\protect\citeauthoryear{{Larson}}{{Larson}}{1969}]{1969MNRAS.145..271L}
{Larson} R.~B.,  1969, \mn@doi [\mnras] {10.1093/mnras/145.3.271}, \href
  {https://ui.adsabs.harvard.edu/abs/1969MNRAS.145..271L} {145, 271}

\bibitem[\protect\citeauthoryear{{Latif} \& {Schleicher}}{{Latif} \&
  {Schleicher}}{2015}]{2015MNRAS.449...77L}
{Latif} M.~A.,  {Schleicher} D.~R.~G.,  2015, \mn@doi [\mnras]
  {10.1093/mnras/stu2573}, \href
  {https://ui.adsabs.harvard.edu/abs/2015MNRAS.449...77L} {449, 77}

\bibitem[\protect\citeauthoryear{{Launay} \& {Roueff}}{{Launay} \&
  {Roueff}}{1977}]{1977A&A....56..289L}
{Launay} J.~M.,  {Roueff} E.,  1977, \aap, \href
  {https://ui.adsabs.harvard.edu/abs/1977A&A....56..289L} {56, 289}

\bibitem[\protect\citeauthoryear{{Lee} \& {Hennebelle}}{{Lee} \&
  {Hennebelle}}{2018}]{2018A&A...611A..89L}
{Lee} Y.-N.,  {Hennebelle} P.,  2018, \mn@doi [\aap]
  {10.1051/0004-6361/201731523}, \href
  {https://ui.adsabs.harvard.edu/abs/2018A&A...611A..89L} {611, A89}

\bibitem[\protect\citeauthoryear{{Lee} et~al.,}{{Lee}
  et~al.}{2009}]{2009ApJ...706..599L}
{Lee} J.~C.,  et~al., 2009, \mn@doi [\apj] {10.1088/0004-637X/706/1/599}, \href
  {https://ui.adsabs.harvard.edu/abs/2009ApJ...706..599L} {706, 599}

\bibitem[\protect\citeauthoryear{{Lee}, {Suda}, {Beers}  \& {Stancliffe}}{{Lee}
  et~al.}{2014}]{2014ApJ...788..131L}
{Lee} Y.~S.,  {Suda} T.,  {Beers} T.~C.,   {Stancliffe} R.~J.,  2014, \mn@doi
  [\apj] {10.1088/0004-637X/788/2/131}, \href
  {https://ui.adsabs.harvard.edu/abs/2014ApJ...788..131L} {788, 131}

\bibitem[\protect\citeauthoryear{{Lee}, {Lee}, {Dunham}, {Tatematsu}, {Choi},
  {Bergin}  \& {Evans}}{{Lee} et~al.}{2017}]{2017NatAs...1E.172L}
{Lee} J.-E.,  {Lee} S.,  {Dunham} M.~M.,  {Tatematsu} K.,  {Choi} M.,  {Bergin}
  E.~A.,   {Evans} N.~J.,  2017, \mn@doi [Nature Astronomy]
  {10.1038/s41550-017-0172}, \href
  {https://ui.adsabs.harvard.edu/abs/2017NatAs...1E.172L} {1, 0172}

\bibitem[\protect\citeauthoryear{{Lee}, {Offner}, {Hennebelle}, {Andr{\'e}},
  {Zinnecker}, {Ballesteros-Paredes}, {Inutsuka}  \& {Kruijssen}}{{Lee}
  et~al.}{2020}]{2020SSRv..216...70L}
{Lee} Y.-N.,  {Offner} S. S.~R.,  {Hennebelle} P.,  {Andr{\'e}} P.,
  {Zinnecker} H.,  {Ballesteros-Paredes} J.,  {Inutsuka} S.-i.,   {Kruijssen}
  J.~M.~D.,  2020, \mn@doi [\ssr] {10.1007/s11214-020-00699-2}, \href
  {https://ui.adsabs.harvard.edu/abs/2020SSRv..216...70L} {216, 70}

\bibitem[\protect\citeauthoryear{{Leigh}, {Umbreit}, {Sills}, {Knigge}, {de
  Marchi}, {Glebbeek}  \& {Sarajedini}}{{Leigh}
  et~al.}{2012}]{2012MNRAS.422.1592L}
{Leigh} N.,  {Umbreit} S.,  {Sills} A.,  {Knigge} C.,  {de Marchi} G.,
  {Glebbeek} E.,   {Sarajedini} A.,  2012, \mn@doi [\mnras]
  {10.1111/j.1365-2966.2012.20735.x}, \href
  {https://ui.adsabs.harvard.edu/abs/2012MNRAS.422.1592L} {422, 1592}

\bibitem[\protect\citeauthoryear{{Lepp} \& {Shull}}{{Lepp} \&
  {Shull}}{1984}]{1984ApJ...280..465L}
{Lepp} S.,  {Shull} J.~M.,  1984, \mn@doi [\apj] {10.1086/162013}, \href
  {https://ui.adsabs.harvard.edu/abs/1984ApJ...280..465L} {280, 465}

\bibitem[\protect\citeauthoryear{{Li}, {Narayanan}  \& {Dav{\'e}}}{{Li}
  et~al.}{2019}]{2019MNRAS.490.1425L}
{Li} Q.,  {Narayanan} D.,   {Dav{\'e}} R.,  2019, \mn@doi [\mnras]
  {10.1093/mnras/stz2684}, \href
  {https://ui.adsabs.harvard.edu/abs/2019MNRAS.490.1425L} {490, 1425}

\bibitem[\protect\citeauthoryear{{Liang}, {Li}, {Li}, {Zhou}, {Yan}, {Mo}  \&
  {Zhang}}{{Liang} et~al.}{2021}]{2021arXiv210103217L}
{Liang} F.-H.,  {Li} C.,  {Li} N.,  {Zhou} S.,  {Yan} R.,  {Mo} H.,   {Zhang}
  W.,  2021, arXiv e-prints, \href
  {https://ui.adsabs.harvard.edu/abs/2021arXiv210103217L} {p. arXiv:2101.03217}

\bibitem[\protect\citeauthoryear{{Lipovka}, {N{\'u}{\~n}ez-L{\'o}pez}  \&
  {Avila-Reese}}{{Lipovka} et~al.}{2005}]{2005MNRAS.361..850L}
{Lipovka} A.,  {N{\'u}{\~n}ez-L{\'o}pez} R.,   {Avila-Reese} V.,  2005, \mn@doi
  [\mnras] {10.1111/j.1365-2966.2005.09226.x}, \href
  {https://ui.adsabs.harvard.edu/abs/2005MNRAS.361..850L} {361, 850}

\bibitem[\protect\citeauthoryear{{Lique}, {Werfelli}, {Halvick}, {Stoecklin},
  {Faure}, {Wiesenfeld}  \& {Dagdigian}}{{Lique}
  et~al.}{2013}]{2013JChPh.138t4314L}
{Lique} F.,  {Werfelli} G.,  {Halvick} P.,  {Stoecklin} T.,  {Faure} A.,
  {Wiesenfeld} L.,   {Dagdigian} P.~J.,  2013, \mn@doi [\jcp]
  {10.1063/1.4807311}, \href
  {https://ui.adsabs.harvard.edu/abs/2013JChPh.138t4314L} {138, 204314}

\bibitem[\protect\citeauthoryear{{Lique}, {K{\l}os}, {Alexander}, {Le Picard}
  \& {Dagdigian}}{{Lique} et~al.}{2018}]{2018MNRAS.474.2313L}
{Lique} F.,  {K{\l}os} J.,  {Alexander} M.~H.,  {Le Picard} S.~D.,
  {Dagdigian} P.~J.,  2018, \mn@doi [\mnras] {10.1093/mnras/stx2907}, \href
  {https://ui.adsabs.harvard.edu/abs/2018MNRAS.474.2313L} {474, 2313}

\bibitem[\protect\citeauthoryear{{Liu} \& {Bromm}}{{Liu} \&
  {Bromm}}{2020}]{2020MNRAS.497.2839L}
{Liu} B.,  {Bromm} V.,  2020, \mn@doi [\mnras] {10.1093/mnras/staa2143}, \href
  {https://ui.adsabs.harvard.edu/abs/2020MNRAS.497.2839L} {497, 2839}

\bibitem[\protect\citeauthoryear{{Liu}, {Meynet}  \& {Bromm}}{{Liu}
  et~al.}{2021}]{2021MNRAS.501..643L}
{Liu} B.,  {Meynet} G.,   {Bromm} V.,  2021, \mn@doi [\mnras]
  {10.1093/mnras/staa3671}, \href
  {https://ui.adsabs.harvard.edu/abs/2021MNRAS.501..643L} {501, 643}

\bibitem[\protect\citeauthoryear{{Longmore} et~al.,}{{Longmore}
  et~al.}{2014}]{2014prpl.conf..291L}
{Longmore} S.~N.,  et~al., 2014, in {Beuther} H.,  {Klessen} R.~S.,
  {Dullemond} C.~P.,   {Henning} T.,  eds, Protostars and Planets VI. p.~291
  (\mn@eprint {arXiv} {1401.4175}),
  \mn@doi{10.2458/azu\_uapress\_9780816531240-ch013}

\bibitem[\protect\citeauthoryear{{Lopez}, {Mathur}, {Nguyen}, {Thompson}  \&
  {Olivier}}{{Lopez} et~al.}{2020}]{Lopez20a}
{Lopez} L.~A.,  {Mathur} S.,  {Nguyen} D.~D.,  {Thompson} T.~A.,   {Olivier}
  G.~M.,  2020, \mn@doi [\apj] {10.3847/1538-4357/abc010}, \href
  {https://ui.adsabs.harvard.edu/abs/2020ApJ...904..152L} {904, 152}

\bibitem[\protect\citeauthoryear{{Low} \& {Lynden-Bell}}{{Low} \&
  {Lynden-Bell}}{1976}]{Low76a}
{Low} C.,  {Lynden-Bell} D.,  1976, \mnras, \href
  {http://adsabs.harvard.edu/abs/1976MNRAS.176..367L} {176, 367}

\bibitem[\protect\citeauthoryear{{Machida} \& {Nakamura}}{{Machida} \&
  {Nakamura}}{2015}]{2015MNRAS.448.1405M}
{Machida} M.~N.,  {Nakamura} T.,  2015, \mn@doi [\mnras]
  {10.1093/mnras/stu2633}, \href
  {https://ui.adsabs.harvard.edu/abs/2015MNRAS.448.1405M} {448, 1405}

\bibitem[\protect\citeauthoryear{{Machida}, {Omukai}, {Matsumoto}  \&
  {Inutsuka}}{{Machida} et~al.}{2009}]{2009MNRAS.399.1255M}
{Machida} M.~N.,  {Omukai} K.,  {Matsumoto} T.,   {Inutsuka} S.-I.,  2009,
  \mn@doi [\mnras] {10.1111/j.1365-2966.2009.15394.x}, \href
  {https://ui.adsabs.harvard.edu/abs/2009MNRAS.399.1255M} {399, 1255}

\bibitem[\protect\citeauthoryear{{Madden} et~al.,}{{Madden}
  et~al.}{2020}]{2020A&A...643A.141M}
{Madden} S.~C.,  et~al., 2020, \mn@doi [\aap] {10.1051/0004-6361/202038860},
  \href {https://ui.adsabs.harvard.edu/abs/2020A&A...643A.141M} {643, A141}

\bibitem[\protect\citeauthoryear{{Marks}, {Kroupa}, {Dabringhausen}  \&
  {Pawlowski}}{{Marks} et~al.}{2012}]{2012MNRAS.422.2246M}
{Marks} M.,  {Kroupa} P.,  {Dabringhausen} J.,   {Pawlowski} M.~S.,  2012,
  \mn@doi [\mnras] {10.1111/j.1365-2966.2012.20767.x}, \href
  {https://ui.adsabs.harvard.edu/abs/2012MNRAS.422.2246M} {422, 2246}

\bibitem[\protect\citeauthoryear{{Mart{\'\i}n-Navarro}, {La Barbera},
  {Vazdekis}, {Falc{\'o}n-Barroso}  \& {Ferreras}}{{Mart{\'\i}n-Navarro}
  et~al.}{2015a}]{2015MNRAS.447.1033M}
{Mart{\'\i}n-Navarro} I.,  {La Barbera} F.,  {Vazdekis} A.,
  {Falc{\'o}n-Barroso} J.,   {Ferreras} I.,  2015a, \mn@doi [\mnras]
  {10.1093/mnras/stu2480}, \href
  {https://ui.adsabs.harvard.edu/abs/2015MNRAS.447.1033M} {447, 1033}

\bibitem[\protect\citeauthoryear{{Mart{\'\i}n-Navarro}
  et~al.,}{{Mart{\'\i}n-Navarro} et~al.}{2015b}]{2015ApJ...806L..31M}
{Mart{\'\i}n-Navarro} I.,  et~al., 2015b, \mn@doi [\apjl]
  {10.1088/2041-8205/806/2/L31}, \href
  {https://ui.adsabs.harvard.edu/abs/2015ApJ...806L..31M} {806, L31}

\bibitem[\protect\citeauthoryear{{Mart{\'\i}n-Navarro}
  et~al.,}{{Mart{\'\i}n-Navarro} et~al.}{2021}]{2021arXiv210714243M}
{Mart{\'\i}n-Navarro} I.,  et~al., 2021, arXiv e-prints, \href
  {https://ui.adsabs.harvard.edu/abs/2021arXiv210714243M} {p. arXiv:2107.14243}

\bibitem[\protect\citeauthoryear{{Mart{\'\i}n}, {Muller}, {Henkel}, {Meier},
  {Aladro}, {Sakamoto}  \& {van der Werf}}{{Mart{\'\i}n}
  et~al.}{2019}]{2019A&A...624A.125M}
{Mart{\'\i}n} S.,  {Muller} S.,  {Henkel} C.,  {Meier} D.~S.,  {Aladro} R.,
  {Sakamoto} K.,   {van der Werf} P.~P.,  2019, \mn@doi [\aap]
  {10.1051/0004-6361/201935106}, \href
  {https://ui.adsabs.harvard.edu/abs/2019A&A...624A.125M} {624, A125}

\bibitem[\protect\citeauthoryear{{Masunaga} \& {Inutsuka}}{{Masunaga} \&
  {Inutsuka}}{2000}]{Masunaga00a}
{Masunaga} H.,  {Inutsuka} S.-i.,  2000, \mn@doi [\apj] {10.1086/308439}, \href
  {http://adsabs.harvard.edu/abs/2000ApJ...531..350M} {531, 350}

\bibitem[\protect\citeauthoryear{{Masunaga}, {Miyama}  \&
  {Inutsuka}}{{Masunaga} et~al.}{1998}]{Masunaga98a}
{Masunaga} H.,  {Miyama} S.~M.,   {Inutsuka} S.-i.,  1998, \mn@doi [\apj]
  {10.1086/305281}, \href {http://adsabs.harvard.edu/abs/1998ApJ...495..346M}
  {495, 346}

\bibitem[\protect\citeauthoryear{{Mathew} \& {Federrath}}{{Mathew} \&
  {Federrath}}{2020}]{Mathew20a}
{Mathew} S.~S.,  {Federrath} C.,  2020, \mn@doi [\mnras]
  {10.1093/mnras/staa1931}, \href
  {https://ui.adsabs.harvard.edu/abs/2020MNRAS.496.5201M} {496, 5201}

\bibitem[\protect\citeauthoryear{{Mathew} \& {Federrath}}{{Mathew} \&
  {Federrath}}{2021}]{2021arXiv210606521M}
{Mathew} S.~S.,  {Federrath} C.,  2021, \mn@doi [\mnras]
  {10.1093/mnras/stab2338}, \href
  {https://ui.adsabs.harvard.edu/abs/2021MNRAS.tmp.2105M} {}

\bibitem[\protect\citeauthoryear{{Mattsson}}{{Mattsson}}{2010}]{2010A&A...515A..68M}
{Mattsson} L.,  2010, \mn@doi [\aap] {10.1051/0004-6361/200913315}, \href
  {https://ui.adsabs.harvard.edu/abs/2010A&A...515A..68M} {515, A68}

\bibitem[\protect\citeauthoryear{{Matzner} \& {McKee}}{{Matzner} \&
  {McKee}}{2000}]{2000ApJ...545..364M}
{Matzner} C.~D.,  {McKee} C.~F.,  2000, \mn@doi [\apj] {10.1086/317785}, \href
  {https://ui.adsabs.harvard.edu/abs/2000ApJ...545..364M} {545, 364}

\bibitem[\protect\citeauthoryear{{McDermid} et~al.,}{{McDermid}
  et~al.}{2014}]{2014ApJ...792L..37M}
{McDermid} R.~M.,  et~al., 2014, \mn@doi [\apjl] {10.1088/2041-8205/792/2/L37},
  \href {https://ui.adsabs.harvard.edu/abs/2014ApJ...792L..37M} {792, L37}

\bibitem[\protect\citeauthoryear{{McKee} \& {Krumholz}}{{McKee} \&
  {Krumholz}}{2010}]{2010ApJ...709..308M}
{McKee} C.~F.,  {Krumholz} M.~R.,  2010, \mn@doi [\apj]
  {10.1088/0004-637X/709/1/308}, \href
  {https://ui.adsabs.harvard.edu/abs/2010ApJ...709..308M} {709, 308}

\bibitem[\protect\citeauthoryear{{McKee} \& {Ostriker}}{{McKee} \&
  {Ostriker}}{2007}]{2007ARA&A..45..565M}
{McKee} C.~F.,  {Ostriker} E.~C.,  2007, \mn@doi [\araa]
  {10.1146/annurev.astro.45.051806.110602}, \href
  {https://ui.adsabs.harvard.edu/abs/2007ARA&A..45..565M} {45, 565}

\bibitem[\protect\citeauthoryear{{McKee} \& {Tan}}{{McKee} \&
  {Tan}}{2003}]{2003ApJ...585..850M}
{McKee} C.~F.,  {Tan} J.~C.,  2003, \mn@doi [\apj] {10.1086/346149}, \href
  {https://ui.adsabs.harvard.edu/abs/2003ApJ...585..850M} {585, 850}

\bibitem[\protect\citeauthoryear{{McQuinn} et~al.,}{{McQuinn}
  et~al.}{2015}]{2015ApJ...812..158M}
{McQuinn} K. B.~W.,  et~al., 2015, \mn@doi [\apj]
  {10.1088/0004-637X/812/2/158}, \href
  {https://ui.adsabs.harvard.edu/abs/2015ApJ...812..158M} {812, 158}

\bibitem[\protect\citeauthoryear{{Meece}, {Smith}  \& {O'Shea}}{{Meece}
  et~al.}{2014}]{2014ApJ...783...75M}
{Meece} G.~R.,  {Smith} B.~D.,   {O'Shea} B.~W.,  2014, \mn@doi [\apj]
  {10.1088/0004-637X/783/2/75}, \href
  {https://ui.adsabs.harvard.edu/abs/2014ApJ...783...75M} {783, 75}

\bibitem[\protect\citeauthoryear{{Meurer} et~al.,}{{Meurer}
  et~al.}{2009}]{2009ApJ...695..765M}
{Meurer} G.~R.,  et~al., 2009, \mn@doi [\apj] {10.1088/0004-637X/695/1/765},
  \href {https://ui.adsabs.harvard.edu/abs/2009ApJ...695..765M} {695, 765}

\bibitem[\protect\citeauthoryear{{Miville-Desch{\^e}nes}, {Murray}  \&
  {Lee}}{{Miville-Desch{\^e}nes} et~al.}{2017}]{2017ApJ...834...57M}
{Miville-Desch{\^e}nes} M.-A.,  {Murray} N.,   {Lee} E.~J.,  2017, \mn@doi
  [\apj] {10.3847/1538-4357/834/1/57}, \href
  {https://ui.adsabs.harvard.edu/abs/2017ApJ...834...57M} {834, 57}

\bibitem[\protect\citeauthoryear{{Mueller}, {Shirley}, {Evans}  \&
  {Jacobson}}{{Mueller} et~al.}{2002}]{2002ApJS..143..469M}
{Mueller} K.~E.,  {Shirley} Y.~L.,  {Evans} Neal~J. I.,   {Jacobson} H.~R.,
  2002, \mn@doi [\apjs] {10.1086/342881}, \href
  {https://ui.adsabs.harvard.edu/abs/2002ApJS..143..469M} {143, 469}

\bibitem[\protect\citeauthoryear{{Muratov}, {Gnedin}, {Gnedin}  \&
  {Zemp}}{{Muratov} et~al.}{2013}]{2013ApJ...773...19M}
{Muratov} A.~L.,  {Gnedin} O.~Y.,  {Gnedin} N.~Y.,   {Zemp} M.,  2013, \mn@doi
  [\apj] {10.1088/0004-637X/773/1/19}, \href
  {https://ui.adsabs.harvard.edu/abs/2013ApJ...773...19M} {773, 19}

\bibitem[\protect\citeauthoryear{{Myers}, {Krumholz}, {Klein}  \&
  {McKee}}{{Myers} et~al.}{2011}]{2011ApJ...735...49M}
{Myers} A.~T.,  {Krumholz} M.~R.,  {Klein} R.~I.,   {McKee} C.~F.,  2011,
  \mn@doi [\apj] {10.1088/0004-637X/735/1/49}, \href
  {https://ui.adsabs.harvard.edu/abs/2011ApJ...735...49M} {735, 49}

\bibitem[\protect\citeauthoryear{{Myers}, {Klein}, {Krumholz}  \&
  {McKee}}{{Myers} et~al.}{2014a}]{Myers14a}
{Myers} A.~T.,  {Klein} R.~I.,  {Krumholz} M.~R.,   {McKee} C.~F.,  2014a,
  \mn@doi [\mnras] {10.1093/mnras/stu190}, \href
  {https://ui.adsabs.harvard.edu/abs/2014MNRAS.439.3420M} {439, 3420}

\bibitem[\protect\citeauthoryear{{Myers}, {Klein}, {Krumholz}  \&
  {McKee}}{{Myers} et~al.}{2014b}]{2014MNRAS.439.3420M}
{Myers} A.~T.,  {Klein} R.~I.,  {Krumholz} M.~R.,   {McKee} C.~F.,  2014b,
  \mn@doi [\mnras] {10.1093/mnras/stu190}, \href
  {https://ui.adsabs.harvard.edu/abs/2014MNRAS.439.3420M} {439, 3420}

\bibitem[\protect\citeauthoryear{{Newman}, {Smith}, {Conroy}, {Villaume}  \&
  {van Dokkum}}{{Newman} et~al.}{2017}]{2017ApJ...845..157N}
{Newman} A.~B.,  {Smith} R.~J.,  {Conroy} C.,  {Villaume} A.,   {van Dokkum}
  P.,  2017, \mn@doi [\apj] {10.3847/1538-4357/aa816d}, \href
  {https://ui.adsabs.harvard.edu/abs/2017ApJ...845..157N} {845, 157}

\bibitem[\protect\citeauthoryear{{Nordlander} et~al.,}{{Nordlander}
  et~al.}{2019}]{2019MNRAS.488L.109N}
{Nordlander} T.,  et~al., 2019, \mn@doi [\mnras] {10.1093/mnrasl/slz109}, \href
  {https://ui.adsabs.harvard.edu/abs/2019MNRAS.488L.109N} {488, L109}

\bibitem[\protect\citeauthoryear{{Norman} \& {Spaans}}{{Norman} \&
  {Spaans}}{1997}]{1997ApJ...480..145N}
{Norman} C.~A.,  {Spaans} M.,  1997, \mn@doi [\apj] {10.1086/303940}, \href
  {https://ui.adsabs.harvard.edu/abs/1997ApJ...480..145N} {480, 145}

\bibitem[\protect\citeauthoryear{{Nozawa}, {Kozasa}, {Umeda}, {Maeda}  \&
  {Nomoto}}{{Nozawa} et~al.}{2003}]{2003ApJ...598..785N}
{Nozawa} T.,  {Kozasa} T.,  {Umeda} H.,  {Maeda} K.,   {Nomoto} K.,  2003,
  \mn@doi [\apj] {10.1086/379011}, \href
  {https://ui.adsabs.harvard.edu/abs/2003ApJ...598..785N} {598, 785}

\bibitem[\protect\citeauthoryear{{Nozawa}, {Kozasa}  \& {Nomoto}}{{Nozawa}
  et~al.}{2012}]{2012ApJ...756L..35N}
{Nozawa} T.,  {Kozasa} T.,   {Nomoto} K.,  2012, \mn@doi [\apjl]
  {10.1088/2041-8205/756/2/L35}, \href
  {https://ui.adsabs.harvard.edu/abs/2012ApJ...756L..35N} {756, L35}

\bibitem[\protect\citeauthoryear{{O'Meara}, {Tytler}, {Kirkman}, {Suzuki},
  {Prochaska}, {Lubin}  \& {Wolfe}}{{O'Meara}
  et~al.}{2001}]{2001ApJ...552..718O}
{O'Meara} J.~M.,  {Tytler} D.,  {Kirkman} D.,  {Suzuki} N.,  {Prochaska} J.~X.,
   {Lubin} D.,   {Wolfe} A.~M.,  2001, \mn@doi [\apj] {10.1086/320579}, \href
  {https://ui.adsabs.harvard.edu/abs/2001ApJ...552..718O} {552, 718}

\bibitem[\protect\citeauthoryear{{Offner}, {Klein}, {McKee}  \&
  {Krumholz}}{{Offner} et~al.}{2009}]{2009ApJ...703..131O}
{Offner} S. S.~R.,  {Klein} R.~I.,  {McKee} C.~F.,   {Krumholz} M.~R.,  2009,
  \mn@doi [\apj] {10.1088/0004-637X/703/1/131}, \href
  {https://ui.adsabs.harvard.edu/abs/2009ApJ...703..131O} {703, 131}

\bibitem[\protect\citeauthoryear{{Offner}, {Kratter}, {Matzner}, {Krumholz}  \&
  {Klein}}{{Offner} et~al.}{2010}]{2010ApJ...725.1485O}
{Offner} S. S.~R.,  {Kratter} K.~M.,  {Matzner} C.~D.,  {Krumholz} M.~R.,
  {Klein} R.~I.,  2010, \mn@doi [\apj] {10.1088/0004-637X/725/2/1485}, \href
  {https://ui.adsabs.harvard.edu/abs/2010ApJ...725.1485O} {725, 1485}

\bibitem[\protect\citeauthoryear{{Offner}, {Clark}, {Hennebelle}, {Bastian},
  {Bate}, {Hopkins}, {Moraux}  \& {Whitworth}}{{Offner}
  et~al.}{2014}]{2014prpl.conf...53O}
{Offner} S.~S.~R.,  {Clark} P.~C.,  {Hennebelle} P.,  {Bastian} N.,  {Bate}
  M.~R.,  {Hopkins} P.~F.,  {Moraux} E.,   {Whitworth} A.~P.,  2014, in
  {Beuther} H.,  {Klessen} R.~S.,  {Dullemond} C.~P.,   {Henning} T.,  eds,
  Protostars and Planets VI. p.~53 (\mn@eprint {arXiv} {1312.5326}),
  \mn@doi{10.2458/azu\_uapress\_9780816531240-ch003}

\bibitem[\protect\citeauthoryear{{Oldham} \& {Auger}}{{Oldham} \&
  {Auger}}{2018}]{2018MNRAS.474.4169O}
{Oldham} L.,  {Auger} M.,  2018, \mn@doi [\mnras] {10.1093/mnras/stx2969},
  \href {https://ui.adsabs.harvard.edu/abs/2018MNRAS.474.4169O} {474, 4169}

\bibitem[\protect\citeauthoryear{{Oliphant}}{{Oliphant}}{2006}]{oliphant2006guide}
{Oliphant} T.~E.,  2006, A guide to NumPy.
~ Vol. 1, Trelgol Publishing USA

\bibitem[\protect\citeauthoryear{{Omukai}}{{Omukai}}{2000}]{2000ApJ...534..809O}
{Omukai} K.,  2000, \mn@doi [\apj] {10.1086/308776}, \href
  {https://ui.adsabs.harvard.edu/abs/2000ApJ...534..809O} {534, 809}

\bibitem[\protect\citeauthoryear{{Omukai}, {Tsuribe}, {Schneider}  \&
  {Ferrara}}{{Omukai} et~al.}{2005}]{2005ApJ...626..627O}
{Omukai} K.,  {Tsuribe} T.,  {Schneider} R.,   {Ferrara} A.,  2005, \mn@doi
  [\apj] {10.1086/429955}, \href
  {https://ui.adsabs.harvard.edu/abs/2005ApJ...626..627O} {626, 627}

\bibitem[\protect\citeauthoryear{{Omukai}, {Hosokawa}  \& {Yoshida}}{{Omukai}
  et~al.}{2010}]{2010ApJ...722.1793O}
{Omukai} K.,  {Hosokawa} T.,   {Yoshida} N.,  2010, \mn@doi [\apj]
  {10.1088/0004-637X/722/2/1793}, \href
  {https://ui.adsabs.harvard.edu/abs/2010ApJ...722.1793O} {722, 1793}

\bibitem[\protect\citeauthoryear{{Ossenkopf} \& {Henning}}{{Ossenkopf} \&
  {Henning}}{1994}]{1994A&A...291..943O}
{Ossenkopf} V.,  {Henning} T.,  1994, \aap, \href
  {https://ui.adsabs.harvard.edu/abs/1994A&A...291..943O} {291, 943}

\bibitem[\protect\citeauthoryear{{Padoan}, {Nordlund}  \& {Jones}}{{Padoan}
  et~al.}{1997}]{1997MNRAS.288..145P}
{Padoan} P.,  {Nordlund} A.,   {Jones} B. J.~T.,  1997, \mn@doi [\mnras]
  {10.1093/mnras/288.1.145}, \href
  {https://ui.adsabs.harvard.edu/abs/1997MNRAS.288..145P} {288, 145}

\bibitem[\protect\citeauthoryear{{Padovani}, {Marcowith}, {Hennebelle}  \&
  {Ferri{\`e}re}}{{Padovani} et~al.}{2016}]{2016A&A...590A...8P}
{Padovani} M.,  {Marcowith} A.,  {Hennebelle} P.,   {Ferri{\`e}re} K.,  2016,
  \mn@doi [\aap] {10.1051/0004-6361/201628221}, \href
  {https://ui.adsabs.harvard.edu/abs/2016A&A...590A...8P} {590, A8}

\bibitem[\protect\citeauthoryear{{Padovani}, {Ivlev}, {Galli}  \&
  {Caselli}}{{Padovani} et~al.}{2018}]{2018A&A...614A.111P}
{Padovani} M.,  {Ivlev} A.~V.,  {Galli} D.,   {Caselli} P.,  2018, \mn@doi
  [\aap] {10.1051/0004-6361/201732202}, \href
  {https://ui.adsabs.harvard.edu/abs/2018A&A...614A.111P} {614, A111}

\bibitem[\protect\citeauthoryear{{Padovani} et~al.,}{{Padovani}
  et~al.}{2020}]{2020SSRv..216...29P}
{Padovani} M.,  et~al., 2020, \mn@doi [\ssr] {10.1007/s11214-020-00654-1},
  \href {https://ui.adsabs.harvard.edu/abs/2020SSRv..216...29P} {216, 29}

\bibitem[\protect\citeauthoryear{{Pandya}, {Mulchaey}  \& {Greene}}{{Pandya}
  et~al.}{2016}]{2016ApJ...819..162P}
{Pandya} V.,  {Mulchaey} J.,   {Greene} J.~E.,  2016, \mn@doi [\apj]
  {10.3847/0004-637X/819/2/162}, \href
  {https://ui.adsabs.harvard.edu/abs/2016ApJ...819..162P} {819, 162}

\bibitem[\protect\citeauthoryear{{Pantaleone}, {Enrique-Romero}, {Ceccarelli},
  {Ferrero}, {Balucani}, {Rimola}  \& {Ugliengo}}{{Pantaleone}
  et~al.}{2021}]{2021arXiv210506843P}
{Pantaleone} S.,  {Enrique-Romero} J.,  {Ceccarelli} C.,  {Ferrero} S.,
  {Balucani} N.,  {Rimola} A.,   {Ugliengo} P.,  2021, \mn@doi [\apj]
  {10.3847/1538-4357/ac0142}, \href
  {https://ui.adsabs.harvard.edu/abs/2021ApJ...917...49P} {917, 49}

\bibitem[\protect\citeauthoryear{{Papadopoulos}}{{Papadopoulos}}{2010}]{2010ApJ...720..226P}
{Papadopoulos} P.~P.,  2010, \mn@doi [\apj] {10.1088/0004-637X/720/1/226},
  \href {https://ui.adsabs.harvard.edu/abs/2010ApJ...720..226P} {720, 226}

\bibitem[\protect\citeauthoryear{{Papadopoulos}, {Thi}, {Miniati}  \&
  {Viti}}{{Papadopoulos} et~al.}{2011}]{2011MNRAS.414.1705P}
{Papadopoulos} P.~P.,  {Thi} W.-F.,  {Miniati} F.,   {Viti} S.,  2011, \mn@doi
  [\mnras] {10.1111/j.1365-2966.2011.18504.x}, \href
  {https://ui.adsabs.harvard.edu/abs/2011MNRAS.414.1705P} {414, 1705}

\bibitem[\protect\citeauthoryear{{Parikh} et~al.,}{{Parikh}
  et~al.}{2018}]{2018MNRAS.477.3954P}
{Parikh} T.,  et~al., 2018, \mn@doi [\mnras] {10.1093/mnras/sty785}, \href
  {https://ui.adsabs.harvard.edu/abs/2018MNRAS.477.3954P} {477, 3954}

\bibitem[\protect\citeauthoryear{{Peacock} et~al.,}{{Peacock}
  et~al.}{2017}]{2017ApJ...841...28P}
{Peacock} M.~B.,  et~al., 2017, \mn@doi [\apj] {10.3847/1538-4357/aa70eb},
  \href {https://ui.adsabs.harvard.edu/abs/2017ApJ...841...28P} {841, 28}

\bibitem[\protect\citeauthoryear{{Pflamm-Altenburg}, {Weidner}  \&
  {Kroupa}}{{Pflamm-Altenburg} et~al.}{2009}]{2009MNRAS.395..394P}
{Pflamm-Altenburg} J.,  {Weidner} C.,   {Kroupa} P.,  2009, \mn@doi [\mnras]
  {10.1111/j.1365-2966.2009.14522.x}, \href
  {https://ui.adsabs.harvard.edu/abs/2009MNRAS.395..394P} {395, 394}

\bibitem[\protect\citeauthoryear{{Phillipps}, {Young}, {Drinkwater}, {Gregg}
  \& {Karick}}{{Phillipps} et~al.}{2013}]{2013MNRAS.433.1444P}
{Phillipps} S.,  {Young} A.~J.,  {Drinkwater} M.~J.,  {Gregg} M.~D.,   {Karick}
  A.,  2013, \mn@doi [\mnras] {10.1093/mnras/stt820}, \href
  {https://ui.adsabs.harvard.edu/abs/2013MNRAS.433.1444P} {433, 1444}

\bibitem[\protect\citeauthoryear{{Pillai}, {Kauffmann}, {Wyrowski}, {Hatchell},
  {Gibb}  \& {Thompson}}{{Pillai} et~al.}{2011}]{2011A&A...530A.118P}
{Pillai} T.,  {Kauffmann} J.,  {Wyrowski} F.,  {Hatchell} J.,  {Gibb} A.~G.,
  {Thompson} M.~A.,  2011, \mn@doi [\aap] {10.1051/0004-6361/201015899}, \href
  {https://ui.adsabs.harvard.edu/abs/2011A&A...530A.118P} {530, A118}

\bibitem[\protect\citeauthoryear{{Pineda} et~al.,}{{Pineda}
  et~al.}{2017}]{2017ApJ...839..107P}
{Pineda} J.~L.,  et~al., 2017, \mn@doi [\apj] {10.3847/1538-4357/aa683a}, \href
  {https://ui.adsabs.harvard.edu/abs/2017ApJ...839..107P} {839, 107}

\bibitem[\protect\citeauthoryear{{Pirogov}}{{Pirogov}}{2009}]{2009ARep...53.1127P}
{Pirogov} L.~E.,  2009, \mn@doi [Astronomy Reports]
  {10.1134/S1063772909120051}, \href
  {https://ui.adsabs.harvard.edu/abs/2009ARep...53.1127P} {53, 1127}

\bibitem[\protect\citeauthoryear{{Placco} et~al.,}{{Placco}
  et~al.}{2021}]{2021ApJ...912L..32P}
{Placco} V.~M.,  et~al., 2021, \mn@doi [\apjl] {10.3847/2041-8213/abf93d},
  \href {https://ui.adsabs.harvard.edu/abs/2021ApJ...912L..32P} {912, L32}

\bibitem[\protect\citeauthoryear{{Pollack}, {Hollenbach}, {Beckwith},
  {Simonelli}, {Roush}  \& {Fong}}{{Pollack}
  et~al.}{1994}]{1994ApJ...421..615P}
{Pollack} J.~B.,  {Hollenbach} D.,  {Beckwith} S.,  {Simonelli} D.~P.,  {Roush}
  T.,   {Fong} W.,  1994, \mn@doi [\apj] {10.1086/173677}, \href
  {https://ui.adsabs.harvard.edu/abs/1994ApJ...421..615P} {421, 615}

\bibitem[\protect\citeauthoryear{{Prgomet}, {Rey}, {Andersson}, {Segovia
  Otero}, {Agertz}, {Renaud}, {Pontzen}  \& {Read}}{{Prgomet}
  et~al.}{2021}]{2021arXiv210700663P}
{Prgomet} M.,  {Rey} M.~P.,  {Andersson} E.~P.,  {Segovia Otero} A.,  {Agertz}
  O.,  {Renaud} F.,  {Pontzen} A.,   {Read} J.~I.,  2021, arXiv e-prints, \href
  {https://ui.adsabs.harvard.edu/abs/2021arXiv210700663P} {p. arXiv:2107.00663}

\bibitem[\protect\citeauthoryear{{Rees}}{{Rees}}{1976}]{Rees76a}
{Rees} M.~J.,  1976, \mnras, \href
  {http://adsabs.harvard.edu/abs/1976MNRAS.176..483R} {176, 483}

\bibitem[\protect\citeauthoryear{{R{\'e}my-Ruyer} et~al.,}{{R{\'e}my-Ruyer}
  et~al.}{2014}]{2014A&A...563A..31R}
{R{\'e}my-Ruyer} A.,  et~al., 2014, \mn@doi [\aap]
  {10.1051/0004-6361/201322803}, \href
  {https://ui.adsabs.harvard.edu/abs/2014A&A...563A..31R} {563, A31}

\bibitem[\protect\citeauthoryear{{Ripamonti} \& {Abel}}{{Ripamonti} \&
  {Abel}}{2004}]{2004MNRAS.348.1019R}
{Ripamonti} E.,  {Abel} T.,  2004, \mn@doi [\mnras]
  {10.1111/j.1365-2966.2004.07422.x}, \href
  {https://ui.adsabs.harvard.edu/abs/2004MNRAS.348.1019R} {348, 1019}

\bibitem[\protect\citeauthoryear{{Rodr{\'\i}guez-Fern{\'a}ndez},
  {Mart{\'\i}n-Pintado}, {de Vicente}, {Fuente}, {H{\"u}ttemeister}, {Wilson}
  \& {Kunze}}{{Rodr{\'\i}guez-Fern{\'a}ndez}
  et~al.}{2000}]{2000A&A...356..695R}
{Rodr{\'\i}guez-Fern{\'a}ndez} N.~J.,  {Mart{\'\i}n-Pintado} J.,  {de Vicente}
  P.,  {Fuente} A.,  {H{\"u}ttemeister} S.,  {Wilson} T.~L.,   {Kunze} D.,
  2000, \aap, \href {https://ui.adsabs.harvard.edu/abs/2000A&A...356..695R}
  {356, 695}

\bibitem[\protect\citeauthoryear{{Rossi}, {Salvadori}  \&
  {Sk{\'u}lad{\'o}ttir}}{{Rossi} et~al.}{2021}]{2021MNRAS.503.6026R}
{Rossi} M.,  {Salvadori} S.,   {Sk{\'u}lad{\'o}ttir} {\'A}.,  2021, \mn@doi
  [\mnras] {10.1093/mnras/stab821}, \href
  {https://ui.adsabs.harvard.edu/abs/2021MNRAS.503.6026R} {503, 6026}

\bibitem[\protect\citeauthoryear{{Rubio}, {Elmegreen}, {Hunter}, {Brinks},
  {Cort{\'e}s}  \& {Cigan}}{{Rubio} et~al.}{2015}]{Rubio15a}
{Rubio} M.,  {Elmegreen} B.~G.,  {Hunter} D.~A.,  {Brinks} E.,  {Cort{\'e}s}
  J.~R.,   {Cigan} P.,  2015, \mn@doi [\nat] {10.1038/nature14901}, \href
  {https://ui.adsabs.harvard.edu/abs/2015Natur.525..218R} {525, 218}

\bibitem[\protect\citeauthoryear{{Sarangi}, {Matsuura}  \&
  {Micelotta}}{{Sarangi} et~al.}{2018}]{2018SSRv..214...63S}
{Sarangi} A.,  {Matsuura} M.,   {Micelotta} E.~R.,  2018, \mn@doi [\ssr]
  {10.1007/s11214-018-0492-7}, \href
  {https://ui.adsabs.harvard.edu/abs/2018SSRv..214...63S} {214, 63}

\bibitem[\protect\citeauthoryear{{Sarzi}, {Spiniello}, {La Barbera},
  {Krajnovi{\'c}}  \& {van den Bosch}}{{Sarzi}
  et~al.}{2018}]{2018MNRAS.478.4084S}
{Sarzi} M.,  {Spiniello} C.,  {La Barbera} F.,  {Krajnovi{\'c}} D.,   {van den
  Bosch} R.,  2018, \mn@doi [\mnras] {10.1093/mnras/sty1092}, \href
  {https://ui.adsabs.harvard.edu/abs/2018MNRAS.478.4084S} {478, 4084}

\bibitem[\protect\citeauthoryear{{Schneider} \& {Omukai}}{{Schneider} \&
  {Omukai}}{2010}]{2010MNRAS.402..429S}
{Schneider} R.,  {Omukai} K.,  2010, \mn@doi [\mnras]
  {10.1111/j.1365-2966.2009.15891.x}, \href
  {https://ui.adsabs.harvard.edu/abs/2010MNRAS.402..429S} {402, 429}

\bibitem[\protect\citeauthoryear{{Schneider}, {Omukai}, {Inoue}  \&
  {Ferrara}}{{Schneider} et~al.}{2006}]{2006MNRAS.369.1437S}
{Schneider} R.,  {Omukai} K.,  {Inoue} A.~K.,   {Ferrara} A.,  2006, \mn@doi
  [\mnras] {10.1111/j.1365-2966.2006.10391.x}, \href
  {https://ui.adsabs.harvard.edu/abs/2006MNRAS.369.1437S} {369, 1437}

\bibitem[\protect\citeauthoryear{{Schneider}, {Omukai}, {Bianchi}  \&
  {Valiante}}{{Schneider} et~al.}{2012}]{2012MNRAS.419.1566S}
{Schneider} R.,  {Omukai} K.,  {Bianchi} S.,   {Valiante} R.,  2012, \mn@doi
  [\mnras] {10.1111/j.1365-2966.2011.19818.x}, \href
  {https://ui.adsabs.harvard.edu/abs/2012MNRAS.419.1566S} {419, 1566}

\bibitem[\protect\citeauthoryear{{Schneider} et~al.,}{{Schneider}
  et~al.}{2015}]{2015A&A...578A..29S}
{Schneider} N.,  et~al., 2015, \mn@doi [\aap] {10.1051/0004-6361/201424375},
  \href {https://ui.adsabs.harvard.edu/abs/2015A&A...578A..29S} {578, A29}

\bibitem[\protect\citeauthoryear{{Schneider}, {Hunt}  \&
  {Valiante}}{{Schneider} et~al.}{2016}]{2016MNRAS.457.1842S}
{Schneider} R.,  {Hunt} L.,   {Valiante} R.,  2016, \mn@doi [\mnras]
  {10.1093/mnras/stw114}, \href
  {https://ui.adsabs.harvard.edu/abs/2016MNRAS.457.1842S} {457, 1842}

\bibitem[\protect\citeauthoryear{{Schneider} et~al.,}{{Schneider}
  et~al.}{2018}]{2018Sci...359...69S}
{Schneider} F.~R.~N.,  et~al., 2018, \mn@doi [Science]
  {10.1126/science.aan0106}, \href
  {https://ui.adsabs.harvard.edu/abs/2018Sci...359...69S} {359, 69}

\bibitem[\protect\citeauthoryear{{Sch{\"o}ier}, {van der Tak}, {van Dishoeck}
  \& {Black}}{{Sch{\"o}ier} et~al.}{2005}]{2005A&A...432..369S}
{Sch{\"o}ier} F.~L.,  {van der Tak} F.~F.~S.,  {van Dishoeck} E.~F.,   {Black}
  J.~H.,  2005, \mn@doi [\aap] {10.1051/0004-6361:20041729}, \href
  {https://ui.adsabs.harvard.edu/abs/2005A&A...432..369S} {432, 369}

\bibitem[\protect\citeauthoryear{{Schroder}, {Staemmler}, {Smith}, {Flower}  \&
  {Jaquet}}{{Schroder} et~al.}{1991}]{1991JPhB...24.2487S}
{Schroder} K.,  {Staemmler} V.,  {Smith} M.~D.,  {Flower} D.~R.,   {Jaquet} R.,
   1991, \mn@doi [Journal of Physics B Atomic Molecular Physics]
  {10.1088/0953-4075/24/10/007}, \href
  {https://ui.adsabs.harvard.edu/abs/1991JPhB...24.2487S} {24, 2487}

\bibitem[\protect\citeauthoryear{{Schruba} et~al.,}{{Schruba}
  et~al.}{2017}]{2017ApJ...835..278S}
{Schruba} A.,  et~al., 2017, \mn@doi [\apj] {10.3847/1538-4357/835/2/278},
  \href {https://ui.adsabs.harvard.edu/abs/2017ApJ...835..278S} {835, 278}

\bibitem[\protect\citeauthoryear{{Semenov}, {Henning}, {Helling}, {Ilgner}  \&
  {Sedlmayr}}{{Semenov} et~al.}{2003}]{2003A&A...410..611S}
{Semenov} D.,  {Henning} T.,  {Helling} C.,  {Ilgner} M.,   {Sedlmayr} E.,
  2003, \mn@doi [\aap] {10.1051/0004-6361:20031279}, \href
  {https://ui.adsabs.harvard.edu/abs/2003A&A...410..611S} {410, 611}

\bibitem[\protect\citeauthoryear{{Sharda}, {Federrath}, {da Cunha}, {Swinbank}
  \& {Dye}}{{Sharda} et~al.}{2018}]{2018MNRAS.477.4380S}
{Sharda} P.,  {Federrath} C.,  {da Cunha} E.,  {Swinbank} A.~M.,   {Dye} S.,
  2018, \mn@doi [\mnras] {10.1093/mnras/sty886}, \href
  {https://ui.adsabs.harvard.edu/abs/2018MNRAS.477.4380S} {477, 4380}

\bibitem[\protect\citeauthoryear{{Sharda} et~al.,}{{Sharda}
  et~al.}{2019a}]{2019MNRAS.487.4305S}
{Sharda} P.,  et~al., 2019a, \mn@doi [\mnras] {10.1093/mnras/stz1543}, \href
  {https://ui.adsabs.harvard.edu/abs/2019MNRAS.487.4305S} {487, 4305}

\bibitem[\protect\citeauthoryear{{Sharda}, {Krumholz}  \& {Federrath}}{{Sharda}
  et~al.}{2019b}]{2019MNRAS.490..513S}
{Sharda} P.,  {Krumholz} M.~R.,   {Federrath} C.,  2019b, \mn@doi [\mnras]
  {10.1093/mnras/stz2618}, \href
  {https://ui.adsabs.harvard.edu/abs/2019MNRAS.490..513S} {490, 513}

\bibitem[\protect\citeauthoryear{{Sharda}, {Federrath}  \& {Krumholz}}{{Sharda}
  et~al.}{2020}]{2020MNRAS.497..336S}
{Sharda} P.,  {Federrath} C.,   {Krumholz} M.~R.,  2020, \mn@doi [\mnras]
  {10.1093/mnras/staa1926}, \href
  {https://ui.adsabs.harvard.edu/abs/2020MNRAS.497..336S} {497, 336}

\bibitem[\protect\citeauthoryear{{Sharda}, {Krumholz}, {Wisnioski}, {Forbes},
  {Federrath}  \& {Acharyya}}{{Sharda} et~al.}{2021a}]{2021MNRAS.502.5935S}
{Sharda} P.,  {Krumholz} M.~R.,  {Wisnioski} E.,  {Forbes} J.~C.,  {Federrath}
  C.,   {Acharyya} A.,  2021a, \mn@doi [\mnras] {10.1093/mnras/stab252}, \href
  {https://ui.adsabs.harvard.edu/abs/2021MNRAS.502.5935S} {502, 5935}

\bibitem[\protect\citeauthoryear{{Sharda}, {Federrath}, {Krumholz}  \&
  {Schleicher}}{{Sharda} et~al.}{2021b}]{2021MNRAS.503.2014S}
{Sharda} P.,  {Federrath} C.,  {Krumholz} M.~R.,   {Schleicher} D. R.~G.,
  2021b, \mn@doi [\mnras] {10.1093/mnras/stab531}, \href
  {https://ui.adsabs.harvard.edu/abs/2021MNRAS.503.2014S} {503, 2014}

\bibitem[\protect\citeauthoryear{{Sharda}, {Krumholz}, {Wisnioski}, {Acharyya},
  {Federrath}  \& {Forbes}}{{Sharda} et~al.}{2021c}]{2021MNRAS.504...53S}
{Sharda} P.,  {Krumholz} M.~R.,  {Wisnioski} E.,  {Acharyya} A.,  {Federrath}
  C.,   {Forbes} J.~C.,  2021c, \mn@doi [\mnras] {10.1093/mnras/stab868}, \href
  {https://ui.adsabs.harvard.edu/abs/2021MNRAS.504...53S} {504, 53}

\bibitem[\protect\citeauthoryear{{Sharda}, {Wisnioski}, {Krumholz}  \&
  {Federrath}}{{Sharda} et~al.}{2021d}]{2021MNRAS.506.1295S}
{Sharda} P.,  {Wisnioski} E.,  {Krumholz} M.~R.,   {Federrath} C.,  2021d,
  \mn@doi [\mnras] {10.1093/mnras/stab1836}, \href
  {https://ui.adsabs.harvard.edu/abs/2021MNRAS.506.1295S} {506, 1295}

\bibitem[\protect\citeauthoryear{{Shi}, {Wang}, {Zhang}, {Gao}, {Hao}, {Xia}
  \& {Gu}}{{Shi} et~al.}{2016}]{Shi16a}
{Shi} Y.,  {Wang} J.,  {Zhang} Z.-Y.,  {Gao} Y.,  {Hao} C.-N.,  {Xia} X.-Y.,
  {Gu} Q.,  2016, \mn@doi [Nature Communications] {10.1038/ncomms13789}, \href
  {https://ui.adsabs.harvard.edu/abs/2016NatCo...713789S} {7, 13789}

\bibitem[\protect\citeauthoryear{{Shima} \& {Hosokawa}}{{Shima} \&
  {Hosokawa}}{2021}]{2021arXiv210206312S}
{Shima} K.,  {Hosokawa} T.,  2021, arXiv e-prints, \href
  {https://ui.adsabs.harvard.edu/abs/2021arXiv210206312S} {p. arXiv:2102.06312}

\bibitem[\protect\citeauthoryear{{Shu}, {Li}  \& {Allen}}{{Shu}
  et~al.}{2004}]{Shu04a}
{Shu} F.~H.,  {Li} Z.-Y.,   {Allen} A.,  2004, \mn@doi [\apj] {10.1086/380602},
  \href {http://adsabs.harvard.edu/abs/2004ApJ...601..930S} {601, 930}

\bibitem[\protect\citeauthoryear{{Skillman}}{{Skillman}}{2008}]{2008IAUS..255..285S}
{Skillman} E.~D.,  2008, in {Hunt} L.~K.,  {Madden} S.~C.,   {Schneider} R.,
  eds, ~ Vol. 255, Low-Metallicity Star Formation: From the First Stars to
  Dwarf Galaxies. IAU, pp 285--296, \mn@doi{10.1017/S1743921308024964}

\bibitem[\protect\citeauthoryear{{Sk{\'u}lad{\'o}ttir}
  et~al.,}{{Sk{\'u}lad{\'o}ttir} et~al.}{2021}]{2021arXiv210611592S}
{Sk{\'u}lad{\'o}ttir} {\'A}.,  et~al., 2021, \mn@doi [\apjl]
  {10.3847/2041-8213/ac0dc2}, \href
  {https://ui.adsabs.harvard.edu/abs/2021ApJ...915L..30S} {915, L30}

\bibitem[\protect\citeauthoryear{{Smith}}{{Smith}}{2014}]{2014MNRAS.443L..69S}
{Smith} R.~J.,  2014, \mn@doi [\mnras] {10.1093/mnrasl/slu082}, \href
  {https://ui.adsabs.harvard.edu/abs/2014MNRAS.443L..69S} {443, L69}

\bibitem[\protect\citeauthoryear{{Smith}}{{Smith}}{2020}]{2020ARA&A..58..577S}
{Smith} R.~J.,  2020, \mn@doi [\araa] {10.1146/annurev-astro-032620-020217},
  \href {https://ui.adsabs.harvard.edu/abs/2020ARA&A..58..577S} {58, 577}

\bibitem[\protect\citeauthoryear{{Smith}, {Lucey}  \& {Conroy}}{{Smith}
  et~al.}{2015}]{2015MNRAS.449.3441S}
{Smith} R.~J.,  {Lucey} J.~R.,   {Conroy} C.,  2015, \mn@doi [\mnras]
  {10.1093/mnras/stv518}, \href
  {https://ui.adsabs.harvard.edu/abs/2015MNRAS.449.3441S} {449, 3441}

\bibitem[\protect\citeauthoryear{{Spaans} \& {Norman}}{{Spaans} \&
  {Norman}}{1997}]{1997ApJ...483...87S}
{Spaans} M.,  {Norman} C.~A.,  1997, \mn@doi [\apj] {10.1086/304231}, \href
  {https://ui.adsabs.harvard.edu/abs/1997ApJ...483...87S} {483, 87}

\bibitem[\protect\citeauthoryear{{Spitzer}}{{Spitzer}}{1987}]{1987degc.book.....S}
{Spitzer} L.,  1987, {Dynamical evolution of globular clusters}.
Princeton University Press

\bibitem[\protect\citeauthoryear{{Stacy} \& {Bromm}}{{Stacy} \&
  {Bromm}}{2007}]{2007MNRAS.382..229S}
{Stacy} A.,  {Bromm} V.,  2007, \mn@doi [\mnras]
  {10.1111/j.1365-2966.2007.12247.x}, \href
  {https://ui.adsabs.harvard.edu/abs/2007MNRAS.382..229S} {382, 229}

\bibitem[\protect\citeauthoryear{{Stacy}, {Greif}  \& {Bromm}}{{Stacy}
  et~al.}{2012}]{2012MNRAS.422..290S}
{Stacy} A.,  {Greif} T.~H.,   {Bromm} V.,  2012, \mn@doi [\mnras]
  {10.1111/j.1365-2966.2012.20605.x}, \href
  {https://ui.adsabs.harvard.edu/abs/2012MNRAS.422..290S} {422, 290}

\bibitem[\protect\citeauthoryear{{Staemmler} \& {Flower}}{{Staemmler} \&
  {Flower}}{1991}]{1991JPhB...24.2343S}
{Staemmler} V.,  {Flower} D.~R.,  1991, \mn@doi [Journal of Physics B Atomic
  Molecular Physics] {10.1088/0953-4075/24/9/013}, \href
  {https://ui.adsabs.harvard.edu/abs/1991JPhB...24.2343S} {24, 2343}

\bibitem[\protect\citeauthoryear{{Stahler}, {Shu}  \& {Taam}}{{Stahler}
  et~al.}{1980}]{1980ApJ...241..637S}
{Stahler} S.~W.,  {Shu} F.~H.,   {Taam} R.~E.,  1980, \mn@doi [\apj]
  {10.1086/158377}, \href
  {https://ui.adsabs.harvard.edu/abs/1980ApJ...241..637S} {241, 637}

\bibitem[\protect\citeauthoryear{{Starkenburg} et~al.,}{{Starkenburg}
  et~al.}{2017}]{2017MNRAS.471.2587S}
{Starkenburg} E.,  et~al., 2017, \mn@doi [\mnras] {10.1093/mnras/stx1068},
  \href {https://ui.adsabs.harvard.edu/abs/2017MNRAS.471.2587S} {471, 2587}

\bibitem[\protect\citeauthoryear{{Starkenburg} et~al.,}{{Starkenburg}
  et~al.}{2018}]{2018MNRAS.481.3838S}
{Starkenburg} E.,  et~al., 2018, \mn@doi [\mnras] {10.1093/mnras/sty2276},
  \href {https://ui.adsabs.harvard.edu/abs/2018MNRAS.481.3838S} {481, 3838}

\bibitem[\protect\citeauthoryear{{Sternberg} \& {Neufeld}}{{Sternberg} \&
  {Neufeld}}{1999}]{1999ApJ...516..371S}
{Sternberg} A.,  {Neufeld} D.~A.,  1999, \mn@doi [\apj] {10.1086/307115}, \href
  {https://ui.adsabs.harvard.edu/abs/1999ApJ...516..371S} {516, 371}

\bibitem[\protect\citeauthoryear{{Sternberg}, {Le Petit}, {Roueff}  \& {Le
  Bourlot}}{{Sternberg} et~al.}{2014}]{2014ApJ...790...10S}
{Sternberg} A.,  {Le Petit} F.,  {Roueff} E.,   {Le Bourlot} J.,  2014, \mn@doi
  [\apj] {10.1088/0004-637X/790/1/10}, \href
  {https://ui.adsabs.harvard.edu/abs/2014ApJ...790...10S} {790, 10}

\bibitem[\protect\citeauthoryear{{Sternberg}, {Gurman}  \& {Bialy}}{{Sternberg}
  et~al.}{2021}]{2021arXiv210501681S}
{Sternberg} A.,  {Gurman} A.,   {Bialy} S.,  2021, arXiv e-prints, \href
  {https://ui.adsabs.harvard.edu/abs/2021arXiv210501681S} {p. arXiv:2105.01681}

\bibitem[\protect\citeauthoryear{{Suda}, {Yamada}, {Katsuta}, {Komiya},
  {Ishizuka}, {Aoki}  \& {Fujimoto}}{{Suda} et~al.}{2011}]{2011MNRAS.412..843S}
{Suda} T.,  {Yamada} S.,  {Katsuta} Y.,  {Komiya} Y.,  {Ishizuka} C.,  {Aoki}
  W.,   {Fujimoto} M.~Y.,  2011, \mn@doi [\mnras]
  {10.1111/j.1365-2966.2011.17943.x}, \href
  {https://ui.adsabs.harvard.edu/abs/2011MNRAS.412..843S} {412, 843}

\bibitem[\protect\citeauthoryear{{Suda} et~al.,}{{Suda}
  et~al.}{2013}]{2013MNRAS.432L..46S}
{Suda} T.,  et~al., 2013, \mn@doi [\mnras] {10.1093/mnrasl/slt033}, \href
  {https://ui.adsabs.harvard.edu/abs/2013MNRAS.432L..46S} {432, L46}

\bibitem[\protect\citeauthoryear{{Sugimura}, {Matsumoto}, {Hosokawa}, {Hirano}
  \& {Omukai}}{{Sugimura} et~al.}{2020}]{2020ApJ...892L..14S}
{Sugimura} K.,  {Matsumoto} T.,  {Hosokawa} T.,  {Hirano} S.,   {Omukai} K.,
  2020, \mn@doi [\apjl] {10.3847/2041-8213/ab7d37}, \href
  {https://ui.adsabs.harvard.edu/abs/2020ApJ...892L..14S} {892, L14}

\bibitem[\protect\citeauthoryear{{Susa}}{{Susa}}{2019}]{2019ApJ...877...99S}
{Susa} H.,  2019, \mn@doi [\apj] {10.3847/1538-4357/ab1b6f}, \href
  {https://ui.adsabs.harvard.edu/abs/2019ApJ...877...99S} {877, 99}

\bibitem[\protect\citeauthoryear{{Susa}, {Hasegawa}  \& {Tominaga}}{{Susa}
  et~al.}{2014}]{2014ApJ...792...32S}
{Susa} H.,  {Hasegawa} K.,   {Tominaga} N.,  2014, \mn@doi [\apj]
  {10.1088/0004-637X/792/1/32}, \href
  {https://ui.adsabs.harvard.edu/abs/2014ApJ...792...32S} {792, 32}

\bibitem[\protect\citeauthoryear{{Thies}, {Kroupa}, {Goodwin}, {Stamatellos}
  \& {Whitworth}}{{Thies} et~al.}{2010}]{2010ApJ...717..577T}
{Thies} I.,  {Kroupa} P.,  {Goodwin} S.~P.,  {Stamatellos} D.,   {Whitworth}
  A.~P.,  2010, \mn@doi [\apj] {10.1088/0004-637X/717/1/577}, \href
  {https://ui.adsabs.harvard.edu/abs/2010ApJ...717..577T} {717, 577}

\bibitem[\protect\citeauthoryear{{Thies}, {Pflamm-Altenburg}, {Kroupa}  \&
  {Marks}}{{Thies} et~al.}{2015}]{2015ApJ...800...72T}
{Thies} I.,  {Pflamm-Altenburg} J.,  {Kroupa} P.,   {Marks} M.,  2015, \mn@doi
  [\apj] {10.1088/0004-637X/800/1/72}, \href
  {https://ui.adsabs.harvard.edu/abs/2015ApJ...800...72T} {800, 72}

\bibitem[\protect\citeauthoryear{{Thomas}}{{Thomas}}{1967}]{1967rgd1.conf..155T}
{Thomas} L.~B.,  1967, in Rarefied Gas Dynamics, Volume 1. p.~155

\bibitem[\protect\citeauthoryear{{Thorne} et~al.,}{{Thorne}
  et~al.}{2021}]{2021MNRAS.tmp.1310T}
{Thorne} J.~E.,  et~al., 2021, \mn@doi [\mnras] {10.1093/mnras/stab1294}, \href
  {https://ui.adsabs.harvard.edu/abs/2021MNRAS.505..540T} {505, 540}

\bibitem[\protect\citeauthoryear{{Tobin} et~al.,}{{Tobin}
  et~al.}{2016}]{2016ApJ...818...73T}
{Tobin} J.~J.,  et~al., 2016, \mn@doi [\apj] {10.3847/0004-637X/818/1/73},
  \href {https://ui.adsabs.harvard.edu/abs/2016ApJ...818...73T} {818, 73}

\bibitem[\protect\citeauthoryear{{Todini} \& {Ferrara}}{{Todini} \&
  {Ferrara}}{2001}]{2001MNRAS.325..726T}
{Todini} P.,  {Ferrara} A.,  2001, \mn@doi [\mnras]
  {10.1046/j.1365-8711.2001.04486.x}, \href
  {https://ui.adsabs.harvard.edu/abs/2001MNRAS.325..726T} {325, 726}

\bibitem[\protect\citeauthoryear{{Tokovinin} \& {Moe}}{{Tokovinin} \&
  {Moe}}{2020}]{2020MNRAS.491.5158T}
{Tokovinin} A.,  {Moe} M.,  2020, \mn@doi [\mnras] {10.1093/mnras/stz3299},
  \href {https://ui.adsabs.harvard.edu/abs/2020MNRAS.491.5158T} {491, 5158}

\bibitem[\protect\citeauthoryear{{Treu}, {Auger}, {Koopmans}, {Gavazzi},
  {Marshall}  \& {Bolton}}{{Treu} et~al.}{2010}]{2010ApJ...709.1195T}
{Treu} T.,  {Auger} M.~W.,  {Koopmans} L. V.~E.,  {Gavazzi} R.,  {Marshall}
  P.~J.,   {Bolton} A.~S.,  2010, \mn@doi [\apj]
  {10.1088/0004-637X/709/2/1195}, \href
  {https://ui.adsabs.harvard.edu/abs/2010ApJ...709.1195T} {709, 1195}

\bibitem[\protect\citeauthoryear{{Tricco}, {Price}  \& {Laibe}}{{Tricco}
  et~al.}{2017}]{2017MNRAS.471L..52T}
{Tricco} T.~S.,  {Price} D.~J.,   {Laibe} G.,  2017, \mn@doi [\mnras]
  {10.1093/mnrasl/slx096}, \href
  {https://ui.adsabs.harvard.edu/abs/2017MNRAS.471L..52T} {471, L52}

\bibitem[\protect\citeauthoryear{{Tumlinson}}{{Tumlinson}}{2006}]{2006ApJ...641....1T}
{Tumlinson} J.,  2006, \mn@doi [\apj] {10.1086/500383}, \href
  {https://ui.adsabs.harvard.edu/abs/2006ApJ...641....1T} {641, 1}

\bibitem[\protect\citeauthoryear{{Tumlinson}}{{Tumlinson}}{2007}]{2007ApJ...665.1361T}
{Tumlinson} J.,  2007, \mn@doi [\apj] {10.1086/519917}, \href
  {https://ui.adsabs.harvard.edu/abs/2007ApJ...665.1361T} {665, 1361}

\bibitem[\protect\citeauthoryear{{Turner}, {Beck}  \& {Ho}}{{Turner}
  et~al.}{2000}]{2000ApJ...532L.109T}
{Turner} J.~L.,  {Beck} S.~C.,   {Ho} P. T.~P.,  2000, \mn@doi [\apjl]
  {10.1086/312586}, \href
  {https://ui.adsabs.harvard.edu/abs/2000ApJ...532L.109T} {532, L109}

\bibitem[\protect\citeauthoryear{{Tytler}, {Fan}  \& {Burles}}{{Tytler}
  et~al.}{1996}]{1996Natur.381..207T}
{Tytler} D.,  {Fan} X.-M.,   {Burles} S.,  1996, \mn@doi [\nat]
  {10.1038/381207a0}, \href
  {https://ui.adsabs.harvard.edu/abs/1996Natur.381..207T} {381, 207}

\bibitem[\protect\citeauthoryear{{Urban}, {Martel}  \& {Evans}}{{Urban}
  et~al.}{2010}]{2010ApJ...710.1343U}
{Urban} A.,  {Martel} H.,   {Evans} Neal~J. I.,  2010, \mn@doi [\apj]
  {10.1088/0004-637X/710/2/1343}, \href
  {https://ui.adsabs.harvard.edu/abs/2010ApJ...710.1343U} {710, 1343}

\bibitem[\protect\citeauthoryear{{Ventura} et~al.,}{{Ventura}
  et~al.}{2021}]{2021arXiv210804471V}
{Ventura} P.,  et~al., 2021, arXiv e-prints, \href
  {https://ui.adsabs.harvard.edu/abs/2021arXiv210804471V} {p. arXiv:2108.04471}

\bibitem[\protect\citeauthoryear{{Virtanen} et~al.,}{{Virtanen}
  et~al.}{2020}]{2020NatMe..17..261V}
{Virtanen} P.,  et~al., 2020, \mn@doi [Nature Methods]
  {10.1038/s41592-019-0686-2}, \href
  {https://ui.adsabs.harvard.edu/abs/2020NatMe..17..261V} {17, 261}

\bibitem[\protect\citeauthoryear{{Wang} \& {Dai}}{{Wang} \&
  {Dai}}{2011}]{2011ApJ...727L..34W}
{Wang} F.~Y.,  {Dai} Z.~G.,  2011, \mn@doi [\apjl]
  {10.1088/2041-8205/727/2/L34}, \href
  {https://ui.adsabs.harvard.edu/abs/2011ApJ...727L..34W} {727, L34}

\bibitem[\protect\citeauthoryear{{Wang}, {Kroupa}, {Takahashi}  \&
  {Je{\v{r}}{\'a}bkov{\'a}}}{{Wang} et~al.}{2020}]{2020MNRAS.491..440W}
{Wang} L.,  {Kroupa} P.,  {Takahashi} K.,   {Je{\v{r}}{\'a}bkov{\'a}} T.,
  2020, \mn@doi [\mnras] {10.1093/mnras/stz3033}, \href
  {https://ui.adsabs.harvard.edu/abs/2020MNRAS.491..440W} {491, 440}

\bibitem[\protect\citeauthoryear{{Watson} \& {Salpeter}}{{Watson} \&
  {Salpeter}}{1972}]{1972ApJ...174..321W}
{Watson} W.~D.,  {Salpeter} E.~E.,  1972, \mn@doi [\apj] {10.1086/151492},
  \href {https://ui.adsabs.harvard.edu/abs/1972ApJ...174..321W} {174, 321}

\bibitem[\protect\citeauthoryear{{Webb} \& {Leigh}}{{Webb} \&
  {Leigh}}{2015}]{2015MNRAS.453.3278W}
{Webb} J.~J.,  {Leigh} N. W.~C.,  2015, \mn@doi [\mnras]
  {10.1093/mnras/stv1780}, \href
  {https://ui.adsabs.harvard.edu/abs/2015MNRAS.453.3278W} {453, 3278}

\bibitem[\protect\citeauthoryear{{Webb}, {Vesperini}, {Dalessandro}, {Beccari},
  {Ferraro}  \& {Lanzoni}}{{Webb} et~al.}{2017}]{2017MNRAS.471.3845W}
{Webb} J.~J.,  {Vesperini} E.,  {Dalessandro} E.,  {Beccari} G.,  {Ferraro}
  F.~R.,   {Lanzoni} B.,  2017, \mn@doi [\mnras] {10.1093/mnras/stx1874}, \href
  {https://ui.adsabs.harvard.edu/abs/2017MNRAS.471.3845W} {471, 3845}

\bibitem[\protect\citeauthoryear{{Weidner}, {Kroupa}  \&
  {Pflamm-Altenburg}}{{Weidner} et~al.}{2013}]{2013MNRAS.434...84W}
{Weidner} C.,  {Kroupa} P.,   {Pflamm-Altenburg} J.,  2013, \mn@doi [\mnras]
  {10.1093/mnras/stt1002}, \href
  {https://ui.adsabs.harvard.edu/abs/2013MNRAS.434...84W} {434, 84}

\bibitem[\protect\citeauthoryear{{Weidner}, {Kroupa}  \&
  {Pflamm-Altenburg}}{{Weidner} et~al.}{2014}]{2014MNRAS.441.3348W}
{Weidner} C.,  {Kroupa} P.,   {Pflamm-Altenburg} J.,  2014, \mn@doi [\mnras]
  {10.1093/mnras/stu640}, \href
  {https://ui.adsabs.harvard.edu/abs/2014MNRAS.441.3348W} {441, 3348}

\bibitem[\protect\citeauthoryear{{Weingartner} \& {Draine}}{{Weingartner} \&
  {Draine}}{2001}]{2001ApJ...548..296W}
{Weingartner} J.~C.,  {Draine} B.~T.,  2001, \mn@doi [\apj] {10.1086/318651},
  \href {https://ui.adsabs.harvard.edu/abs/2001ApJ...548..296W} {548, 296}

\bibitem[\protect\citeauthoryear{{Weisz} et~al.,}{{Weisz}
  et~al.}{2012}]{Weisz12a}
{Weisz} D.~R.,  et~al., 2012, \mn@doi [\apj] {10.1088/0004-637X/744/1/44},
  \href {http://adsabs.harvard.edu/abs/2012ApJ...744...44W} {744, 44}

\bibitem[\protect\citeauthoryear{{Wiesenfeld} \& {Goldsmith}}{{Wiesenfeld} \&
  {Goldsmith}}{2014}]{2014ApJ...780..183W}
{Wiesenfeld} L.,  {Goldsmith} P.~F.,  2014, \mn@doi [\apj]
  {10.1088/0004-637X/780/2/183}, \href
  {https://ui.adsabs.harvard.edu/abs/2014ApJ...780..183W} {780, 183}

\bibitem[\protect\citeauthoryear{{Yan}, {Je{\v{r}}{\'a}bkov{\'a}}  \&
  {Kroupa}}{{Yan} et~al.}{2020}]{2020A&A...637A..68Y}
{Yan} Z.,  {Je{\v{r}}{\'a}bkov{\'a}} T.,   {Kroupa} P.,  2020, \mn@doi [\aap]
  {10.1051/0004-6361/202037567}, \href
  {https://ui.adsabs.harvard.edu/abs/2020A&A...637A..68Y} {637, A68}

\bibitem[\protect\citeauthoryear{{Yan}, {Jerabkova}  \& {Kroupa}}{{Yan}
  et~al.}{2021}]{2021arXiv210703388Y}
{Yan} Z.,  {Jerabkova} T.,   {Kroupa} P.,  2021, arXiv e-prints, \href
  {https://ui.adsabs.harvard.edu/abs/2021arXiv210703388Y} {p. arXiv:2107.03388}

\bibitem[\protect\citeauthoryear{{Yong} et~al.,}{{Yong}
  et~al.}{2013}]{2013ApJ...762...26Y}
{Yong} D.,  et~al., 2013, \mn@doi [\apj] {10.1088/0004-637X/762/1/26}, \href
  {https://ui.adsabs.harvard.edu/abs/2013ApJ...762...26Y} {762, 26}

\bibitem[\protect\citeauthoryear{{Yong} et~al.,}{{Yong}
  et~al.}{2021a}]{2021arXiv210706430Y}
{Yong} D.,  et~al., 2021a, \mn@doi [\mnras] {10.1093/mnras/stab2001}, \href
  {https://ui.adsabs.harvard.edu/abs/2021MNRAS.507.4102Y} {507, 4102}

\bibitem[\protect\citeauthoryear{{Yong} et~al.,}{{Yong}
  et~al.}{2021b}]{2021Natur.595..223Y}
{Yong} D.,  et~al., 2021b, \mn@doi [\nat] {10.1038/s41586-021-03611-2}, \href
  {https://ui.adsabs.harvard.edu/abs/2021Natur.595..223Y} {595, 223}

\bibitem[\protect\citeauthoryear{{Yoshida}, {Omukai}, {Hernquist}  \&
  {Abel}}{{Yoshida} et~al.}{2006}]{2006ApJ...652....6Y}
{Yoshida} N.,  {Omukai} K.,  {Hernquist} L.,   {Abel} T.,  2006, \mn@doi [\apj]
  {10.1086/507978}, \href
  {https://ui.adsabs.harvard.edu/abs/2006ApJ...652....6Y} {652, 6}

\bibitem[\protect\citeauthoryear{{Youakim} et~al.,}{{Youakim}
  et~al.}{2020}]{2020MNRAS.492.4986Y}
{Youakim} K.,  et~al., 2020, \mn@doi [\mnras] {10.1093/mnras/stz3619}, \href
  {https://ui.adsabs.harvard.edu/abs/2020MNRAS.492.4986Y} {492, 4986}

\bibitem[\protect\citeauthoryear{{Yusef-Zadeh}, {Law}  \&
  {Wardle}}{{Yusef-Zadeh} et~al.}{2002}]{2002ApJ...568L.121Y}
{Yusef-Zadeh} F.,  {Law} C.,   {Wardle} M.,  2002, \mn@doi [\apjl]
  {10.1086/340379}, \href
  {https://ui.adsabs.harvard.edu/abs/2002ApJ...568L.121Y} {568, L121}

\bibitem[\protect\citeauthoryear{{Yusef-Zadeh}, {Wardle}  \&
  {Roy}}{{Yusef-Zadeh} et~al.}{2007}]{2007ApJ...665L.123Y}
{Yusef-Zadeh} F.,  {Wardle} M.,   {Roy} S.,  2007, \mn@doi [\apjl]
  {10.1086/521359}, \href
  {https://ui.adsabs.harvard.edu/abs/2007ApJ...665L.123Y} {665, L123}

\bibitem[\protect\citeauthoryear{{Zaritsky}, {Colucci}, {Pessev}, {Bernstein}
  \& {Chandar}}{{Zaritsky} et~al.}{2014}]{2014ApJ...796...71Z}
{Zaritsky} D.,  {Colucci} J.~E.,  {Pessev} P.~M.,  {Bernstein} R.~A.,
  {Chandar} R.,  2014, \mn@doi [\apj] {10.1088/0004-637X/796/2/71}, \href
  {https://ui.adsabs.harvard.edu/abs/2014ApJ...796...71Z} {796, 71}

\bibitem[\protect\citeauthoryear{{Zhang}, {Romano}, {Ivison}, {Papadopoulos}
  \& {Matteucci}}{{Zhang} et~al.}{2018}]{2018Natur.558..260Z}
{Zhang} Z.-Y.,  {Romano} D.,  {Ivison} R.~J.,  {Papadopoulos} P.~P.,
  {Matteucci} F.,  2018, \mn@doi [\nat] {10.1038/s41586-018-0196-x}, \href
  {https://ui.adsabs.harvard.edu/abs/2018Natur.558..260Z} {558, 260}

\bibitem[\protect\citeauthoryear{{Zinn}}{{Zinn}}{1985}]{1985ApJ...293..424Z}
{Zinn} R.,  1985, \mn@doi [\apj] {10.1086/163249}, \href
  {https://ui.adsabs.harvard.edu/abs/1985ApJ...293..424Z} {293, 424}

\bibitem[\protect\citeauthoryear{{Zubko}, {Dwek}  \& {Arendt}}{{Zubko}
  et~al.}{2004}]{2004ApJS..152..211Z}
{Zubko} V.,  {Dwek} E.,   {Arendt} R.~G.,  2004, \mn@doi [\apjs]
  {10.1086/382351}, \href
  {https://ui.adsabs.harvard.edu/abs/2004ApJS..152..211Z} {152, 211}

\bibitem[\protect\citeauthoryear{{van Dokkum} \& {Conroy}}{{van Dokkum} \&
  {Conroy}}{2010}]{2010Natur.468..940V}
{van Dokkum} P.~G.,  {Conroy} C.,  2010, \mn@doi [\nat] {10.1038/nature09578},
  \href {https://ui.adsabs.harvard.edu/abs/2010Natur.468..940V} {468, 940}

\bibitem[\protect\citeauthoryear{{van Dokkum} et~al.,}{{van Dokkum}
  et~al.}{2008}]{2008ApJ...677L...5V}
{van Dokkum} P.~G.,  et~al., 2008, \mn@doi [\apjl] {10.1086/587874}, \href
  {https://ui.adsabs.harvard.edu/abs/2008ApJ...677L...5V} {677, L5}

\bibitem[\protect\citeauthoryear{{van der Tak}, {van Dishoeck}, {Evans}  \&
  {Blake}}{{van der Tak} et~al.}{2000}]{2000ApJ...537..283V}
{van der Tak} F. F.~S.,  {van Dishoeck} E.~F.,  {Evans} Neal~J. I.,   {Blake}
  G.~A.,  2000, \mn@doi [\apj] {10.1086/309011}, \href
  {https://ui.adsabs.harvard.edu/abs/2000ApJ...537..283V} {537, 283}

\bibitem[\protect\citeauthoryear{{van der Tak}, {Lique}, {Faure}, {Black}  \&
  {van Dishoeck}}{{van der Tak} et~al.}{2020}]{2020Atoms...8...15V}
{van der Tak} F. F.~S.,  {Lique} F.,  {Faure} A.,  {Black} J.~H.,   {van
  Dishoeck} E.~F.,  2020, \mn@doi [Atoms] {10.3390/atoms8020015}, \href
  {https://ui.adsabs.harvard.edu/abs/2020Atoms...8...15V} {8, 15}

\makeatother
\end{thebibliography}

\bsp	
\label{lastpage}
\end{document}